\documentclass[12pt]{article}

\usepackage{graphicx}
\usepackage{float}
\usepackage{amsmath}
\usepackage{amssymb}
\usepackage{subfigure}
\usepackage{cases}
\usepackage{mathrsfs}
\usepackage[flushleft]{threeparttable}
\usepackage{amssymb}
\usepackage{pifont}
\usepackage{epstopdf}
\usepackage{xcolor}
\usepackage{hyperref}
\usepackage[all]{hypcap}
\usepackage{mwe,tikz}
\usepackage{bm}
\usepackage{multirow}
\usepackage{longtable}
\usepackage{xspace}
\usepackage{rotating}
\usepackage{CJK}
\usepackage{url}
\usepackage{upgreek}
\usepackage{booktabs}
\usepackage{textcomp}
\usepackage{makecell}
\usepackage{enumitem}
\usepackage{csquotes}
\usepackage{soul}
\usepackage{tabularx}
\usepackage{overpic}
\usepackage{caption}
\usepackage{float}
\usepackage{threeparttable}
\usepackage{threeparttablex}
\usepackage{subfiles}
\usepackage{array}
\usepackage{comment}
\usepackage{multibib}
\newcites{output}{References}

\usepackage{newtxtext,newtxmath}
\usepackage{graphicx}
\usepackage[letterpaper,margin=1in]{geometry}
\renewenvironment{abstract}
	{\quotation}
	{\endquotation}

\date{}

\newcommand{\zjp}[1] {{\textcolor{red}{\textbf{{#1}}}}}

\makeatletter
\renewcommand{\fnum@figure}{\textbf{Figure \thefigure}}
\renewcommand{\fnum@table}{\textbf{Table \thetable}}
\makeatother
\usepackage{scicite}
\usepackage{url}

\def\scititle{
X-rays breaking out of pre-explosion ejecta mark a supernova's first light
}

\title{\bfseries \boldmath \scititle}

\author{
    Weimin~Yuan$^{{1,2},\dagger,\ast}$,
    Qiu-Ju~Huang$^{{3,4},\dagger}$,
    Jin-Ping~Zhu$^{{5,6},\dagger}$
    Yun-Wei~Yu$^{{7},\ast}$,
    Dong~Xu$^{{1,8},\ast}$,\and
    Chen~Zhang$^{{1},\ast}$
    Zhuo~Li$^{{9}}$,
    Yuan~Liu$^{{1}}$,
    Tao~An$^{{10}}$,
    Giulia~Gianfagna$^{{11}}$
% ...
\and
    \small$^{1}$National Astronomical Observatories, Chinese Academy of Sciences, Beijing 100101, China.\and
    \small$^{2}$School of Astronomy and Space Science, University of Chinese Academy of Sciences, \and
    \small Chinese Academy of Sciences, Beijing 100049, China.\and
    \small$^{3}$Purple Mountain Observatory, Chinese Academy of Sciences, Nanjing 210023, China.\and
    \small$^{4}$School of Astronomy and Space Sciences, University of Science and Technology of China, Hefei, 230026, China.\and
    \small$^{5}$The Hong Kong Institute for Astronomy and Astrophysics, The University of Hong Kong, \and
    \small Pokfulam Road, Hong Kong, People's Republic of China.\and
    \small$^{6}$Department of Physics, The University of Hong Kong, Pokfulam Road, Hong Kong, People's Republic of China.\and
    \small$^{7}$Institute of Astrophysics, Central China Normal University, Wuhan 430079, China.\and
    \small$^{8}$Altay Astronomical Observatory, Altay, Xinjiang 836500, China.\and
    \small$^{9}$Department of Astronomy, School of Physics, Peking University, Beijing 100871, People's Republic of China.\and
    \small$^{10}$Department of Astronomy, University of Science and Technology of China, Hefei 230026, China.\and
    \small$^{11}$INAF -- Istituto di Astrofisica e Planetologia Spaziali, via Fosso del Cavaliere 100, I-00133 Rome, Italy.\and
    \small$^{12}$Department of Astronomy, University of California, Berkeley, CA 94720-3411, USA.\and
    \small$^{13}$South-Western Institute for Astronomy Research, Yunnan Key Laboratory of Survey Science, \and
    \small Yunnan University, Kunming, Yunnan 650504, China.\and
    \small$^{14}$National Astronomical Observatory of Japan, 2-21-1 Osawa, Mitaka, Tokyo 181-8588, Japan.\and
    \small$^{15}$Cosmic Dawn Center (DAWN), Copenhagen 2200, Denmark.\and
    \small$^{16}$Niels Bohr Institute, University of Copenhagen, Copenhagen 2200, Denmark.\and
    \small$^{17}$Space Research Institute, Russian Academy of Sciences, Moscow 117997, Russia.\and
    \small$^{18}$Faculty of Physics, Higher School of Economics, Moscow 101000, Russia.\and
    \small$^{19}$Physics Department, Tsinghua University, Beijing 100084, China.\and
    \small$^{20}$ICRA, Dip. di Fisica, Sapienza Università di Roma, Piazzale Aldo Moro 5, Roma, I-00185, Italy.\and
    \small$^{21}$ICRANet, Piazza della Repubblica 10, Pescara, 65122, Italy.\and
    \small$^{22}$INAF - Osservatorio Astronomico d'Abruzzo, Via M. Maggini snc, Teramo, I-64100, Italy.\and
    \small$^{23}$School of Astronomy and Space Science, Nanjing University, Nanjing 210023, China.\and
    \small$^{24}$Key Laboratory of Modern Astronomy and Astrophysics (Nanjing University), \and
    \small Ministry of Education, Nanjing, 210023, China.\and
    \small$^{25}$State Key Laboratory of Particle Astrophysics, Institute of High Energy Physics, \and
    \small Chinese Academy of Sciences, Chinese Academy of Sciences, Beijing 100049, China.\and
    \small$^{26}$Key Laboratory for Satellite Digitalization Technology, Innovation Academy for Microsatellite, \and
    \small Chinese Academy of Sciences, Shanghai, 201304, China.\and
    \small$^{27}$Yunnan Observatories, Chinese Academy of Sciences, Kunming 650216, China.\and
    \small$^{28}$International Centre of Supernovae (ICESUN), Yunnan Key Laboratory of Supernova Research, \and
    \small Yunnan Observatories, Chinese Academy of Sciences (CAS), Kunming 650216, China.\and
    \small$^{29}$State Key Laboratory of Radio Astronomy and Technology, Purple Mountain Observatory, \and
    \small Chinese Academy of Sciences, 10 Yuanhua Road, Nanjing 210023, China.\and
    \small$^{30}$Indian Institute of Astrophysics, II Block Koramangala, Bengaluru 560034, India.\and
    \small$^{31}$School of Physics and Astronomy, University of Southampton, Southampton, SO17 1BJ, UK.\and
    \small$^{32}$Department of Physics, Indian Institute of Technology Bombay, Powai, Mumbai 400076, India.\and
    \small$^{33}$National Astronomical Research Institute of Thailand, 260 Moo 4, Donkaew, Maerim, Chiang Mai 50180, Thailand.\and
    \small$^{34}$CEA Paris-Saclay, IRFU/Département d'Astrophysique-AIM, 91191 Gif-sur-Yvette, France.\and
    \small$^{35}$Institute of Solar-Terrestrial Physics, Russian Academy of Sciences (Siberian Branch), Irkutsk 664033, Russia.\and
    \small$^{36}$Department of Astronomy, School of Physics and Technology, Wuhan University, \and
    \small Wuhan 430072, People's Republic of China.\and
    \small$^{37}$Hagler Institute for Advanced Study, Texas A\&M University, 3572 TAMU, College Station, TX 77843, USA.\and
    \small$^{38}$Department of Astronomy, School of Physics, Huazhong University of Science and Technology, \and
    \small Wuhan, 430074, People's Republic of China.\and
    \small$^{39}$Institute for Frontiers in Astronomy and Astrophysics, Beijing Normal University, Beijing 102206, China.\and
    \small$^{40}$Department of Astronomy, Beijing Normal University, Beijing 100875, China.\and
    \small$^{41}$Sternberg Astronomical Institute, Moscow State University, Universitetski pr., 13, Moscow 119992, Russia.\and
    \small$^{42}$Special Astrophysical Observatory, Russian Academy of Sciences, Nizhnij Arkhyz, 369167, Russia.\and
    \small$^{43}$School of Astronomy and Space Science, University of Chinese Academy of Sciences,\and
    \small 19A Yuquan Road, Beijing 100049, China.\and
    \small$^{44}$North Night Vision Technology Co., LTD, Nanjing 210008, People's Republic of China.\and
    \small$^{45}$Department of Astrophysics/IMAPP, Radboud University, 6525 AJ Nijmegen, The Netherlands\and
    \small$^{46}$Institute of Astronomy, National Tsing Hua University, Hsinchu 300044, Taiwan.\and
    \small$^{47}$University of Chinese Academy of Sciences, Beijing 101408, China.\and
    \small$^{48}$Innovation Academy for Microsatellites of the Chinese Academy of Sciences.\and
    \small$^{49}$Shanghai Astronomical Observatory, Chinese Academy of Sciences, 80 Nandan Road, Shanghai 200030, China.\and
    \small$^{50}$Center for Astronomical Mega-Science, Chinese Academy of Sciences, Beijing 100012, China.\and
    \small$^{51}$Key Laboratory for the Structure and Evolution of Celestial Objects, \and
    \small Chinese Academy of Sciences, Kunming 650216, China.\and
    \small$^{52}$Max-Planck-Institut für extraterrestrische Physik, Giessenbachstrasse 1, 85748 Garching, Germany.\and
    \small$^{53}$European Space Astronomy Centre, European Space Agency, Villanueva de la Cañada, Spain.\and
    \small$^{54}$School of Physics and Astronomy, University of Leicester, LE1 7RH, UK.\and
    \small$^{55}$Department of Astrophysics/IMAPP, Radboud University, PO Box 9010, 6500 GL, The Netherlands.\and
    \small$^{56}$Institute of Space Sciences (ICE), Consejo Superior de Investigaciones Científicas (CSIC), Barcelona, Spain.\and
    \small$^{57}$Institut d'Estudis Espacials de Catalunya (IEEC), Barcelona, Spain.\and
    \small$^{58}$The Shanghai Institute of Technical Physics of the Chinese Academy of Sciences.\and
    \small$^{59}$National Space Science Center, Chinese Academy of Sciences, Beijing, 100190, People's Republic of China.\and
    \small$^\ast$Corresponding author: W. Yuan (wmy@nao.cas.cn), Y. Yu (yuyw@ccnu.edu.cn),\and
    \small D. Xu (dxu@nao.cas.cn), C. Zhang (chzhang@bao.ac.cn).\and
    \small$^\dagger$These authors contributed equally to this work.\and
}

\begin{document} 

\maketitle

\begin{abstract} \bfseries \boldmath
Massive stars die as core-collapse supernovae, whose optical light emerges days after the implosion. Theory predicts that the initial collapse-driven shock, upon breaking through the star and dense circumstellar medium, emits a brief thermal flash of soft X-rays and ultraviolet. Yet these elusive first signals have remained largely undetected, owing to limited wide-field soft X-ray monitoring. Here we report the discovery of a soft X-ray flash, EP260321a, followed days later by a broad-lined supernova from an envelope-stripped progenitor. Its X-ray spectrum, best modeled with blackbody, establishes it as the long-sought archetypal shock breakout. The burst's duration and energetics place the breakout at a radius of 300 solar radii, tracing a dense surrounding shell and revealing abrupt mass ejection within the final month before collapse.

\end{abstract}

\textbf{One sentence summary:} Einstein Probe detected the archetypal thermal soft X-ray shock breakout of a core-collapse supernova from a pre-explosion shell.


Massive stars end their lives as energetic supernova explosions, which have been discovered and studied invariably via their optical light that emerges days after the core implosion. Yet the physical processes during and before such explosions are poorly understood. This is due largely to the lack of detection of neutrino and electromagnetic signals in the prompt and very early stages. Theory predicts that a neutrino-driven shock wave is launched immediately following the stellar core-collapse. 
The explosion shock, propagating outward through the dense stellar envelope, is radiation-mediated, wherein photons are effectively trapped, governing momentum and energy transport. 
At the moment the shock breaks out of the stellar envelope or a dense circumstellar medium (CSM), photons of this trapped radiation escape from the shock front, producing the first light of the supernova. Such shock breakouts (SBO) are predicted to emit a burst of thermal spectrum in the soft X-ray to ultraviolet (UV) band, lasting from tens to thousands of seconds, depending on the size of the enveloping material and total energy output \cite{Colgate1974ApJ...187..333C,Waxman2017hsn..book..967W,Chevalier2011ApJ...729L...6C}.
Supernova SBO events provide a direct probe of the explosion energetics, the progenitor stars, and their immediate environment. Furthermore, they offer more precise timing of core-collapse events than optical supernova light does, allowing for more efficient searches for associated neutrinos and possible gravitational-wave signals.

Supernova SBOs are elusive and remain largely undetected, owing to their fleeting and faint nature and a historical lack of sensitive soft X-ray and UV wide-field monitors. To date, despite a few claimed candidates \cite{Alp2020Blast,Eappachen2024}, only two X-ray SBO events have been reported, both by {\it Swift} over the past two decades \cite{Campana2006,Mazzali2006Natur.442.1018M,Soderberg2008Natur.453..469S,Mazzali2008}; however, neither matches theoretical predictions: their spectra deviate largely from the simple thermal form because of complex physical processes, and one is associated with a relativistic jet-driven gamma-ray burst. 
A systematic search for SBOs is now within reach, enabled by the Einstein Probe (EP) \cite{Yuan2022,Yuan2025}, which offers a large field of view, excellent soft-X-ray sensitivity, and rapid autonomous follow-up. Here we report the discovery by EP of a fast X-ray transient, followed days later by a broad-lined Type Ic supernova from an envelop-stripped progenitor. Its X-ray emission is soft and matches a single-temperature blackbody, establishes it as a long-sought bona fide supernova SBO.

\noindent
\section*{Discovery of the fast X-ray transient EP260321a}

During a survey observation, the Wide-field X-ray Telescope (WXT) onboard the Einstein Probe detected an X-ray transient designated EP260321a\cite{Huang2026GCN.44068....1H} on 21 March 2026 at 12:30:18 UTC ($T_{\rm{det}}$).
An autonomous observation by EP's Follow-up X-ray Telescope (FXT) was triggered and started from $T_{\rm{det}}+215$~s, further localizing this uncataloged source to a position consistent with the galaxy SDSS J095942.88+002506.2 
at a redshift of $z=0.0344$ (see Ref. \cite{Lee2026GCN.44070....1L}). A General Coordinates Network (GCN) Notice was issued automatically at $T_{\rm{det}}+329$~s. Two GCN Circulars were issued at $T_{\rm{det}}+31.4$~min and $T_{\rm{det}}+16$~hr, which identified EP260321a as a supernova shock breakout candidate\cite{EPGCN1,EPGCN2} and induced extensive multi-wavelength follow-up observations. WXT had in fact captured the onset of the transient at $T_0=$2026-03-21T12:16:08 ($T_{\rm{det}}-850$~s) in the preceding survey observation. The X-ray light curve increased linearly over $\sim$700 s, followed by an exponential decay (see the lower panel of Figure \ref{fig:Xraylc}). The entire evolution of the decay stage was covered by FXT up to $T_0+2300~\rm{s}$. Generally, the X-ray light curve evolved smoothly, with no significant substructure. Several monitoring observations with FXT were also conducted from $T_0+0.4$ days to $T_0+20$ days, and the Chandra X-ray Observatory observed EP260321a on $T_0+15$ days\cite{OConnor2026GCN.44250....1O}. All these subsequent observations resulted in nondetections. Meanwhile, although the source was covered by several gamma-ray monitors around the time of the outburst, no gamma-ray burst (GRB) counterpart to EP260321a was reported.  

The X-ray spectra of WXT ($T_0$ to $T_0+2200$ s)  and FXT ($T_0+1065$ s to $T_0+2359$ s) are well described by an absorbed blackbody function with an intrinsic absorption column density of $N_{\rm H} = 8.4^{+1.6}_{-1.5} \times 10^{20}$\,$\rm cm^{-2}$ (in excess of the Galactic absorption of \(2.64 \times 10^{20}\,\mathrm{cm}^{-2}\))\cite{Willingale2013} and blackbody temperatures of $kT_{\rm obs} = 124_{-6}^{+7}$ eV and $112.7_{-2.1}^{+2.1}$ eV for WXT and FXT, respectively, where $k$ is the Boltzmann constant (see Figure \ref{fig:Xrayspectra}). Although a cutoff power-law model also provides a statistically acceptable fit, the best-fitting parameters effectively mimic a blackbody spectrum, whereas a simple power-law model is strongly disfavored (Methods). 
From the blackbody spectral fits, it can be calculated that the $0.3$--$10$ keV luminosity of EP260321a reached its peak value of $L_{\rm p,X}=(1.0\pm0.3)\times10^{45}~\rm{erg~s^{-1}}$ at $t_{\rm p,X}=690\pm60~{\rm s}$, which yields an apparent blackbody radius of $(L_{\rm p,X}/4\pi \sigma T_{\rm obs}^4)^{1/2}\sim 10\,R_\odot$. 
The total radiated energy in the X-ray band ($0.3$--$10$ keV) is $E_{\rm th,X}=(8.7\pm0.6)\times10^{47}~\rm{erg}$, with a duration of $T_{90}\sim1750\,{\rm s}$. Accompanied by a decrease in flux after the peak, the blackbody temperature decreases slightly from a maximum value of $\sim$135 eV to $\sim$110 eV. The temperature and total radiated energy of EP260321a indicate that it is the softest and least luminous among all extragalactic fast X-ray transients detected by EP to date\cite{Sun2025,Li2025}. The blackbody spectrum, smooth light curve, and extremely low luminosity of EP260321a all suggest that it is a canonical SBO event. In the left panel of Figure \ref{fig:ComXrayLC}, we compare the X-ray light curve of EP260321a with two previously reported SBO events: XRF 060218\cite{Campana2006,Mazzali2006Natur.442.1018M} and XRO 080109\cite{Mazzali2008,Soderberg2008Natur.453..469S}, both of which are dominated by a nonthermal component and have lasting tail emission. In comparison, EP260321a was no longer detected after the second orbit of the autonomous FXT observation from $T_{0} + 5137$ s to $T_{0} + 8111$ s, which may simply result from the absence of the nonthermal component.

\section*{Multi-wavelength follow-up and association with SN\,2026gzf}

Our optical and near-infrared follow-up campaign was carried out using multiple ground-based facilities. Only 40 min after $T_0$, the 0.7\,m Thai Robotic Telescope began detecting the optical counterpart of EP260321a in the $R$ band, and shortly thereafter the 1.6\,m Mephisto telescope in either the $ugi$ or $vrz$ bands\cite{Yang2024ApJ...969..126Y,Chen2024ApJ...971L...2C}, showing that the counterpart appears on top of a north-western bright blue knot of the host galaxy SDSS J095942.88+002506.2 (see the upper panel of Figure 1) and that there is only slight flux variations within the first few hours. Approximately one day later, the fluxes across all filters started to increase with a temporal index less than 1, peaking around 12 days, as presented in Figure \ref{fig:SNlightcurves}. 
High-quality spectra were obtained for the optical counterpart of EP260321a, spanning the phase from $T_{0}$ + 2.7 days to $T_{0}$ + 56.4 days, 
which allow us to identify it as a broad-lined Type Ic supernova (SN Ic-BL), SN 2026gzf (see also Refs. \cite{Chen2026,MartinCarrillo2026,Rastinejad2026,OConnor2026}) in the galaxy, surprisingly in a local environment of extremely low metallicity. Furthermore, the bolometric light curve of SN 2026gzf shows a close resemblance to that of SN 2006aj\cite{Pian2006Natur.442.1011P}, SN 1998bw, and SN 2025kg\cite{Li2025}, except for its relatively weak presupernova optical bump. In summary, the association of EP260321a with SN 2026gzf provides further support for its identity as a supernova SBO event.

We performed radio observations of EP260321a/SN~2026gzf at multiple frequencies in the range 0.7--23 GHz from $T_0+5$ days to $T_0+$31 days using MeerKAT, ATCA, e-MERLIN, and uGMRT. We also compiled radio observations from GCN Circulars. No reliable radio counterpart is detected (Methods). These upper limits lie well below the radio afterglow fluxes of the majority of GRBs\cite{Weiler2002,Chandra2012},  disfavoring the existence of an on-axis relativistic jet (Methods; see also Ref. \cite{MartinCarrillo2026}). This is consistent with the nondetection of EP260321a in gamma rays. Furthermore, even for an off-axis jet or a choked jet, higher energy emission than EP260321a could also be expected to arise from the inner cocoon or the cocoon breakout\cite{Nakar2015,Zhu2025MNRAS.544L.139Z,Zheng2026}. Overall, these radio and gamma-ray constraints strengthen our claim that the soft X-ray emission of EP260321a arises from supernova SBO.

\section*{Supernova shock breakout origin of EP260321a}

Supernova SBO occurs when the explosion-driven shock reaches a photon diffusion surface at an optical depth of $\tau_{\rm SBO} \approx c/v_{\rm sh}$ (the speed ratio of light to shock wave), before which the heat converted from the kinetic energy of the shock cannot diffuse out efficiently, and thus the shock is radiation-mediated. After SBO, the heat trapped behind the shock starts to be released to produce thermal emission in the UV and soft X-ray bands. However, specifically, a radiation-mediated shock moving at a much faster velocity than $v_{\rm sh}\gtrsim0.1\,c$ would deviate substantially from local thermal equilibrium and lead to a characteristic color temperature much higher than $0.1\,{\rm keV}$ (see Ref. \cite{Katz2010}). Therefore, the observed thermal spectra and relatively low temperature of EP260321a strongly suggest that the shock is nonrelativistic, with a velocity \(v_{\rm sh}\lesssim0.1\,c\). Meanwhile, the shock velocity is also required to be not much lower than $\sim0.1\,c$, such that the kinetic energy of the supernova ejecta moving faster than $v_{\rm sh}$ can be matched to the observed X-ray energy.

According to the peak time of EP260321a, the breakout radius $r_{\rm SBO}$ can be inferred to be around $300\,R_\odot$, from either photon diffusion or the light-travel-time effect (Methods). This is much larger than the progenitor's radius, as it should be a compact Wolf-Rayet star for SNe Ic-BL (i.e., $r_\ast\lesssim R_\odot$)\cite{Crowther2007}. Therefore, the EP260321a SBO event is very unlikely to occur at the stellar surface and, instead, is more likely to be driven by the interaction of the supernova ejecta with dense and optically thick CSM. More notably, $r_{\rm SBO}$ is also considerably larger than the apparent blackbody radius directly inferred from observations. The primary reason for this difference is that the observed spectra, which appear as blackbody, are actually Comptonized spectra determined by a scattering equilibrium between electrons and photons, due to the high speed of the shock. In this case, the effective temperature $T_{\rm th}$ defined from the Stefan-Boltzmann law for the radiation energy density as $u=aT_{\rm th}^4$ is a few times lower than the color temperature $T_{\rm obs}$ of the observed spectra\cite{Katz2010,Nakar2010}, where $u$ is the energy density behind the radiation-mediated shock and $a$ is the radiation constant. 
Furthermore, the light-propagation effect would also cause the SBO emission at a given time to be dominated by a ring-like region instead of the entire sphere.

Since the post-shock energy density can also be related to its upstream kinetic energy density by $u=(6/7)\rho_{\rm SBO} v_{\rm sh}^2$, we can derive the density of the upstream CSM to be $\rho_{\rm SBO}= 7aT_{\rm obs}^4/(18f_{\rm col}^4v_{\rm sh}^2)\sim 10^{-11} \,{\rm g}\,{\rm cm}^{-3}$, 
where the color correction factor $f_{\rm col}\equiv T_{\rm obs}/T_{\rm th}\sim 3.5$ is introduced to represent the deviation from complete local thermal equilibrium. 
Then, according to the SBO condition as $\tau_{\rm SBO}\approx\kappa_{\rm X}\rho_{\rm SBO}(r_{\rm out}-r_{\rm SBO})\approx c/v_{\rm sh}$, we find that the outer radius of the CSM is only slightly larger than the breakout radius as $(r_{\rm out}-r_{\rm SBO})\sim50\,R_\odot$, 
where $\kappa_{\rm X}=0.2\,\rm cm^2g^{-1}$ is the opacity of fully ionized He/C/O-rich material for soft X-rays.
Furthermore, being constrained by the energy budget of X-ray emission, the total mass of the shocked CSM that contributes to the X-ray emission can be derived as $M_{\rm CSM}\sim 10^{-4}\,M_\odot$ 
from $M_{\rm CSM}v_{\rm sh}^2/2\sim E_{\rm th,X}$. The constraints on $M_{\rm CSM}$ as well as $\rho_{\rm SBO}$ indicate that the shocked CSM is accumulated by the shock from a narrow range around the breakout radius with a width of $\Delta r\approx M_{\rm CSM}/( 4\pi r_{\rm SBO}^2\rho_{\rm SBO})\sim 50\,R_\odot$, which is comparable to the value of $(r_{\rm out}-r_{\rm SBO})$. Thus, we suggest that the SBO emission does not originate from the CSM at radii much smaller than $r_{\rm SBO}$. This implies that the CSM is most likely confined within a thin shell, as any material located at smaller radii would be compressed to near $r_{\rm SBO}$ if present.

Detailed modeling of the X-ray light curve and temperature evolution of EP260321a is presented by the solid line in the lower panel of Figure \ref{fig:Xraylc}, where a confined CSM shell with a truncated wind density profile is adopted for $r_{\rm in}\approx260
\,R_\odot$ and $r_{\rm out}\approx 320
\,R_\odot$ and the shock velocity is set at $v_{\rm sh}\approx0.089
~c$. Here, it is considered that all heat accumulated by the shock contributed to the SBO X-ray emission. The most promising origin of such a shell-like CSM is an episodic mass ejection by the progenitor star before the supernova explosion\cite{Chen2026}, which could be a wave-driven outburst and powered by late-stage nuclear burning\cite{Quataert2012,Fuller2018,Leung2021,Wu2021,Wu2022}. The inferred CSM mass and radius of EP260321a also appear to be broadly consistent with the typical properties expected from wave-driven outbursts\cite{Leung2021}. The corresponding final mass ejection likely occurred at $t_{\rm CSM}\approx r_{\rm SBO}/v_{\rm CSM}\lesssim20\,{\rm days}$ 
before the supernova explosion, where the velocity of wave-driven outbursts for a stripped-envelope star is taken as a few $100\,{\rm km}\,{\rm s}^{-1}$ \cite{Fuller2018,Leung2021}, 
suggesting a possible connection to oxygen/neon burning inside the progenitor. In principle, such episodic wave-driven outbursts can take place multiple times for a stripped-envelope star, thereby forming a series of distinct CSM shells at different radii. However, assuming that the mass losses during each ejection are comparable, the outer CSM shells at larger radii are likely optically thin, and thus the interaction of the supernova ejecta with these outer shells could not produce a significant observational signal. 

\section*{The optical supernova SN\,2026gzf}

As a baseline test, we fit the multiband light curves of SN~2026gzf with a purely radioactive model powered by the decay chain of $^{56}$Ni using \texttt{TransFit}, a time-dependent radiative-diffusion code for supernova light-curve inference\cite{Liu2025}. The best-fit model, shown in Figure~\ref{fig:SNlightcurves} , requires a $^{56}$Ni mass of $0.45
M_{\odot}$ and an ejecta mass of $1.85
M_{\odot}$. To reproduce the shallow rise and the $\sim 10$-day peak with radioactive heating alone, the model further requires nearly complete $^{56}$Ni mixing, with $x_{\rm Ni}=0.993$, which corresponds to an almost homogeneous distribution of radioactive material throughout the ejecta. Such strong mixing could in principle be enhanced by explosion-driven hydrodynamic instabilities in a compact Wolf--Rayet progenitor that experienced vigorous late-stage mass loss\cite{Hammer2010,Wongwathanarat2015,AguileraDena2018}. However, this level of $^{56}$Ni homogenization is difficult to obtain in multidimensional explosion simulations, which generally produce asymmetric, clumpy, or finger-like distributions of iron-group material rather than a nearly uniform radioactive component\cite{Hammer2010,Wongwathanarat2015}. If weaker mixing is imposed, the early emission can be reproduced only by reducing the ejecta mass to below $\sim 1\,M_{\odot}$, which is unusually low for a stripped-envelope supernova\cite{Lyman2016,Taddia2018}. The relatively high nickel-to-ejecta mass ratio ($\gtrsim 25\%$) suggests that other energy sources may play an important role in powering SN~2026gzf (Methods). 

\section*{Event rates and implications}

Finally, from the single detection of EP260321a in the systematic survey with EP-WXT, we derive a local event rate density of X-ray SBO transients of $\rho_{0, \rm  SBO}({\rm EP260321a})  \geq 170$ $\rm{Gpc}^{-3}\ {yr}^{-1} $ above a peak luminosity of $1.0\times10^{45}~\rm{erg~s^{-1}}$ (see Methods). 
This event-rate density sits between the more common XRO 080109, associated with an SN Ib/c, and the more luminous and rare event XRF 060218, associated with an SN Ic-BL, forming a power-law luminosity function for X-ray-detected SBOs with an index of $1.3 \pm 0.4$ as shown in the right panel of Figure~\ref{fig:ComXrayLC}.
This fact, together with more common SBOs observed in the optical band in Type II SNe\cite{Garnavich2016,Bersten2018,Li2024}, implies a unified scheme for stellar shock breakouts, with rarer, more luminous events arising from more extreme stellar explosions involving higher angular momentum and more envelope stripping. 
Furthermore, the detection of supernova SBOs yields more precise timing of core-collapse events compared to optical supernovae. This allows for more efficient searches for associated neutrinos and potential gravitational-wave signals in the nearby universe. 
Future continued observations with Einstein Probe and other wide-field monitors, together with simultaneous multi-messenger observations in gravitational waves and neutrinos\cite{Kotake2006RPPh...69..971K} at the time of SBOs, will further constrain SBO physics and may uncover how massive stars end their lives.


\begin{figure}[h]
\centering
\begin{overpic}[width=0.9\textwidth]{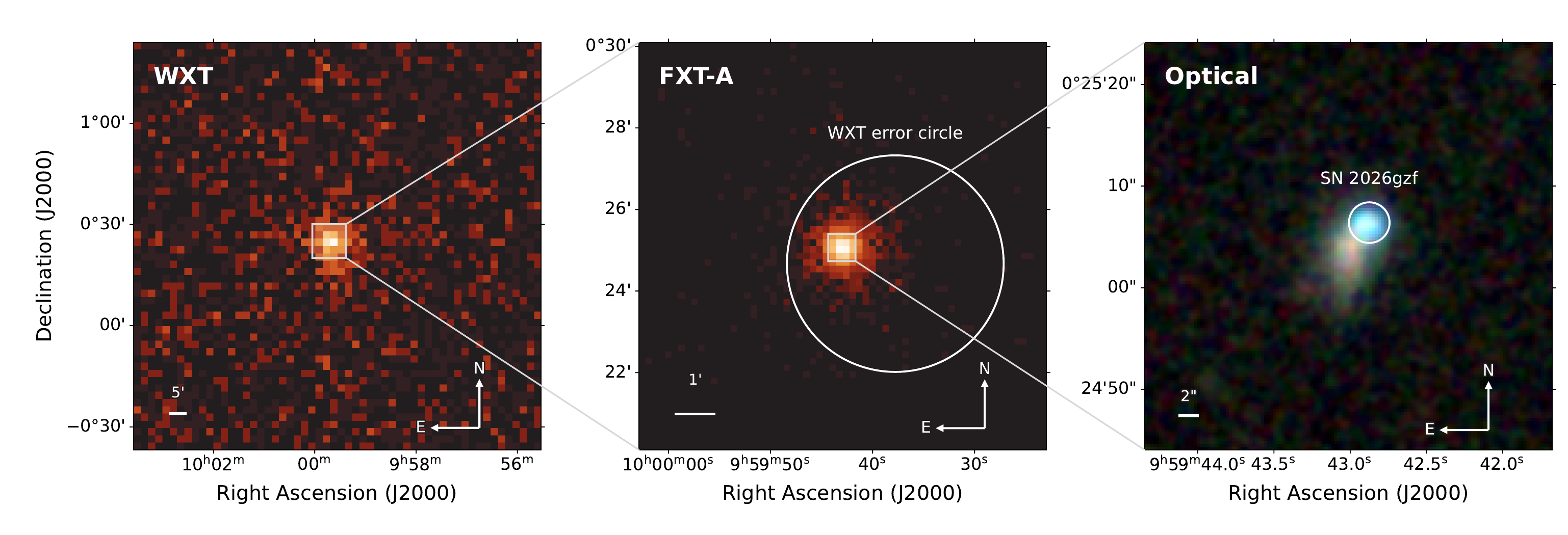}\put(0,30){\textbf{(A)}}
\end{overpic}
\begin{overpic}[width=0.5\textwidth]{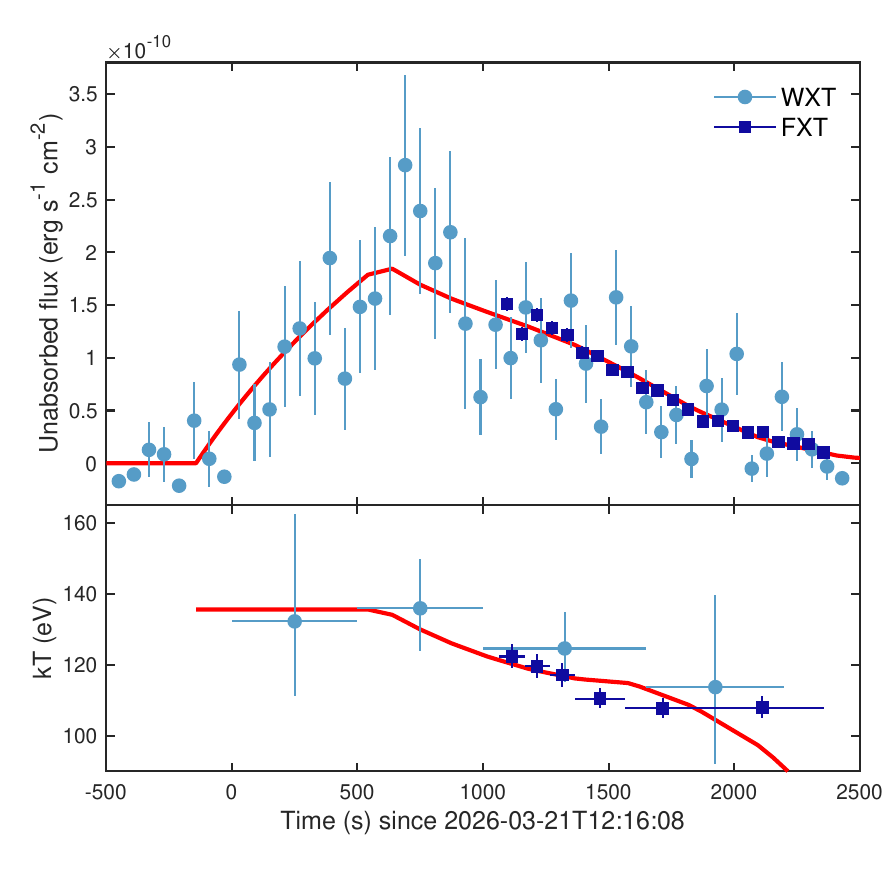}\put(0,92){\textbf{(B)}}
\end{overpic}
\caption{{\bf X-ray/optical images and X-ray evolution of EP260321a.} {{(A)}}  WXT and FXT-A image in 0.4--2.0 keV. The white circle shows the positional error circle of WXT with a radius of $2.7'$ (90\% confidence level). The optical image is the composite Mephisto-\textit{g}\textit{r}\textit{i} band images taken at  $\sim$$T_0 + 2.3$ hr.  {{(B)}} The evolution of unabsorbed X-ray flux in 0.4--2.0 keV (upper) and temperature (lower) under the absorbed blackbody model. The transient began at $T_0 = \mathrm{2026\mbox{-}03\mbox{-}21T12{:}16{:}08}$ UTC (corresponding to the time of a $3\sigma$ excess above the background). Raw count rates were converted to the unabsorbed flux using the factors determined by the time-resolved spectral fitting (Methods). The vertical error bars show the $1\sigma$ uncertainty and the horizontal error bars indicate the corresponding time intervals. The solid lines represent the best fit with a supernova SBO model (Methods).}
\label{fig:Xraylc}
\end{figure}

\begin{figure}[h]
\centering
\begin{overpic}[width=0.45\textwidth]{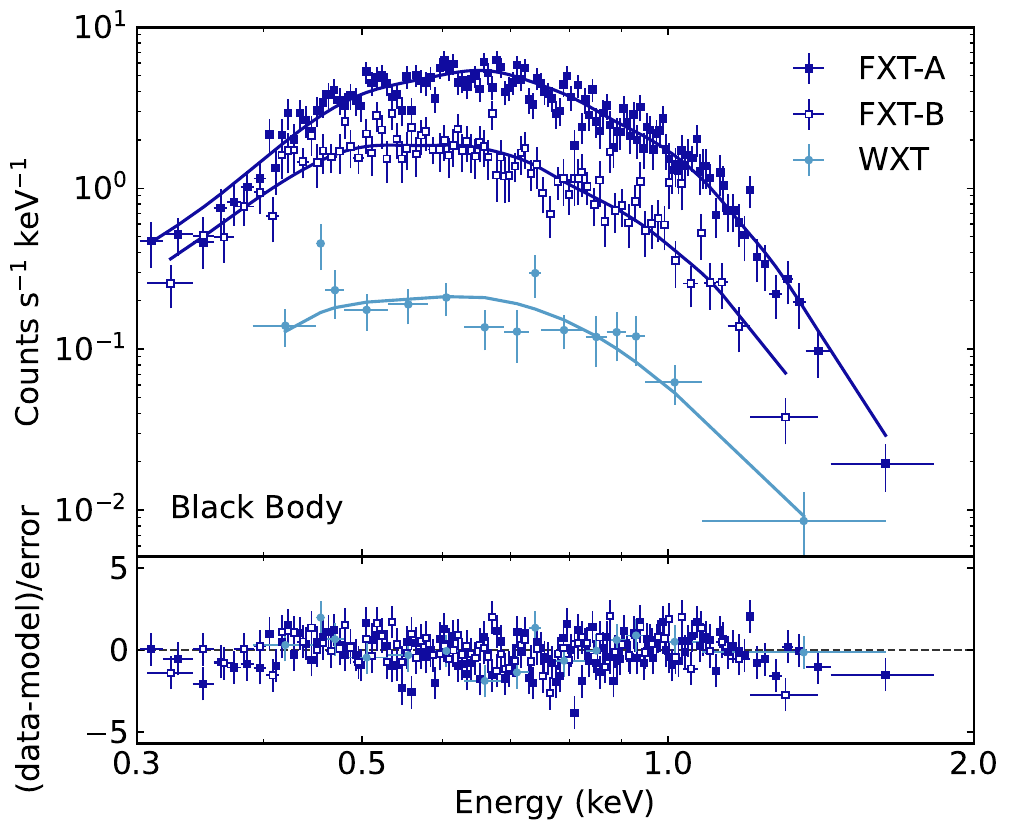}\put(0,83){\textbf{(A)}} 
\end{overpic}
\begin{overpic}[width=0.45\textwidth]{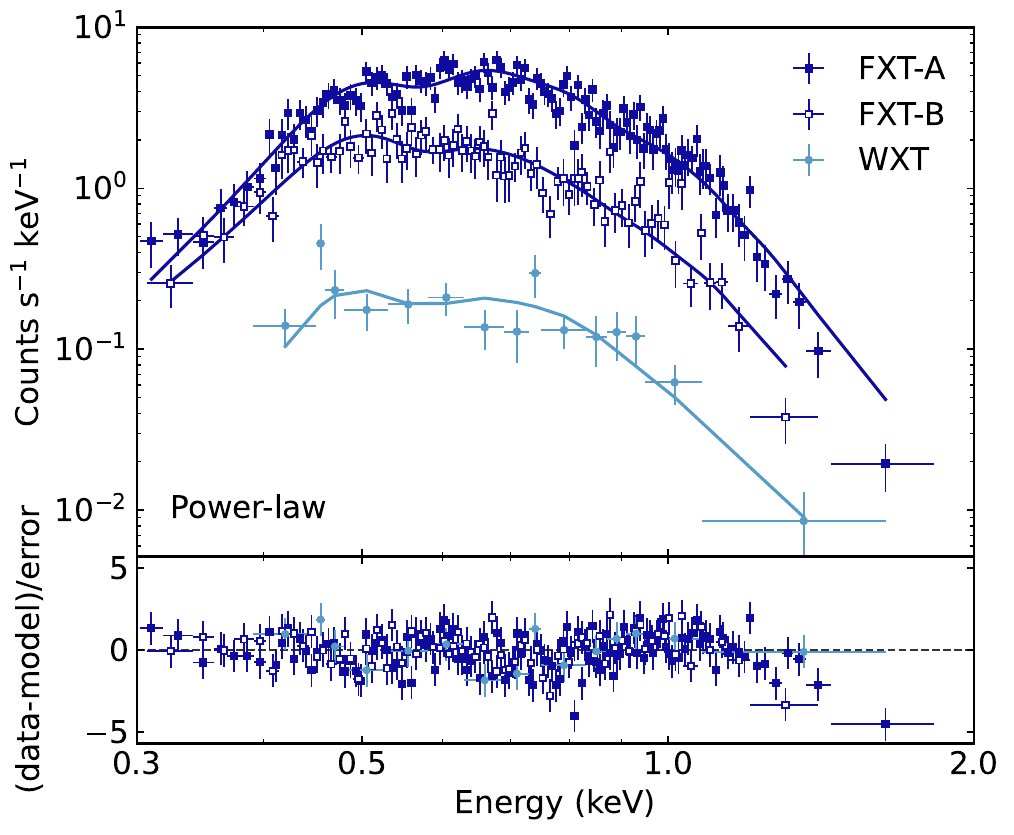}\put(0,83){\textbf{(B)}} 
\end{overpic}
\caption{{\bf X-ray spectra of EP260321a.} The WXT spectrum includes data from $T_0$ to $T_0+2200$ s; the FXT spectra include data from $T_0+1065$ s to $T_0+2359$ s. The count rate of the FXT-B spectrum is significantly lower than that of FXT-A, since the counts in the core of the point-spread function of FXT-B were excluded in the spectrum to mitigate pile-up effect (Methods). The data points are shown with $1\sigma$ uncertainty and the lines represent the best-fit absorbed blackbody {{(A)}} and power-law {{(B)}} models.}
\label{fig:Xrayspectra}
\end{figure}

\begin{figure}[h]
\centering
\begin{overpic}[width=0.5\textwidth]{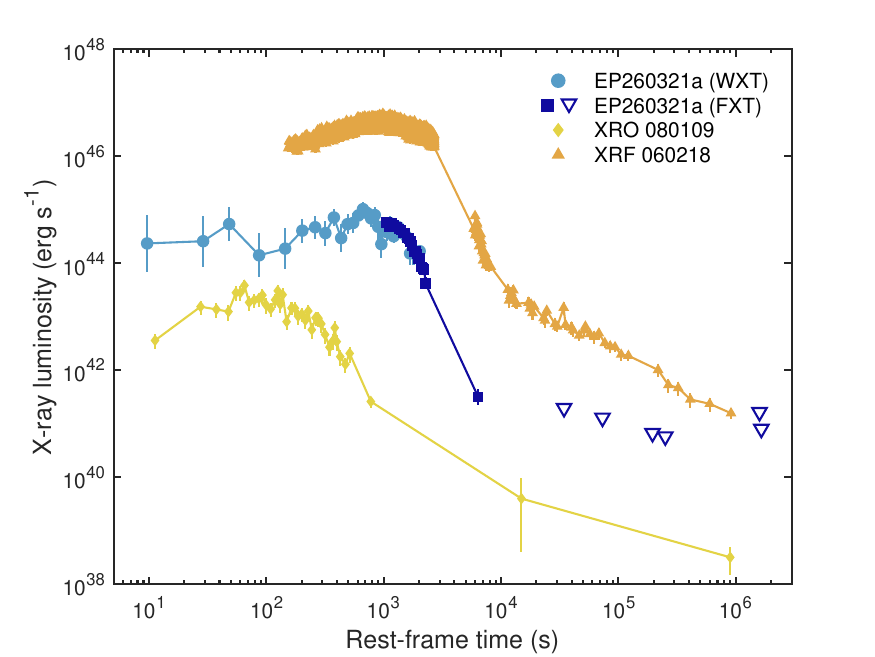}\put(0,70){\textbf{(A)}} 
\end{overpic}
\begin{overpic}[width=0.44\textwidth]{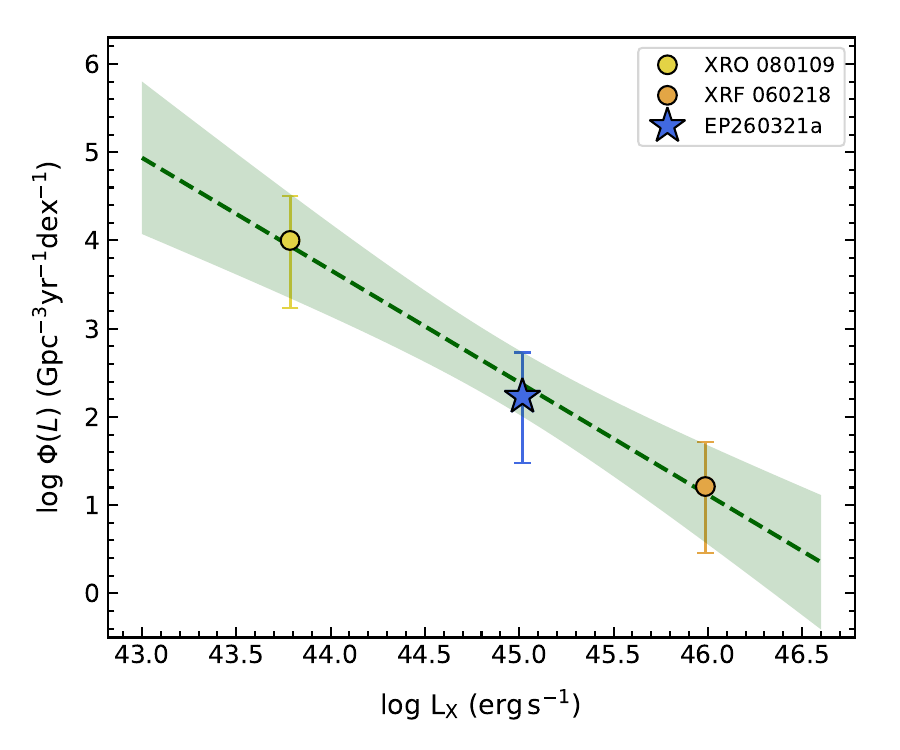}\put(0,80){\textbf{(B)}} 
\end{overpic}
\caption{\textbf{X-ray light curves and luminosity function of SN SBOs.} {{(A)}} X-ray
luminosity light curves of SN SBO events, including EP260321a, XRF 060218\cite{Campana2006,Mazzali2006Natur.442.1018M}, and XRO 080109\cite{Mazzali2008,Soderberg2008Natur.453..469S}. 
The data points are shown with $1\sigma$ uncertainty. The open triangles represent the upper limits derived from the FXT monitoring observations of EP260321a. The data of XRF 060218 are obtained from the \textit{Swift} X-ray telescope light-curve repository\cite{Evans2007,Evans2009}. The data of XRO 080109 are obtained from Ref. \cite{Modjaz2009}. {{(B)}} X-ray luminosity function of SN SBOs. The best-fit result from a power-law model and its 1$\sigma$ uncertainty are shown as a dashed green line and the light-green shaded region, respectively. The peak luminosity of the thermal component of XRO 060218 is adopted\cite{Campana2006}. All luminosities are in the 0.3–10 keV band.}
\label{fig:ComXrayLC}
\end{figure}

\begin{figure}[h]
\centering
\includegraphics[width=0.8\textwidth]{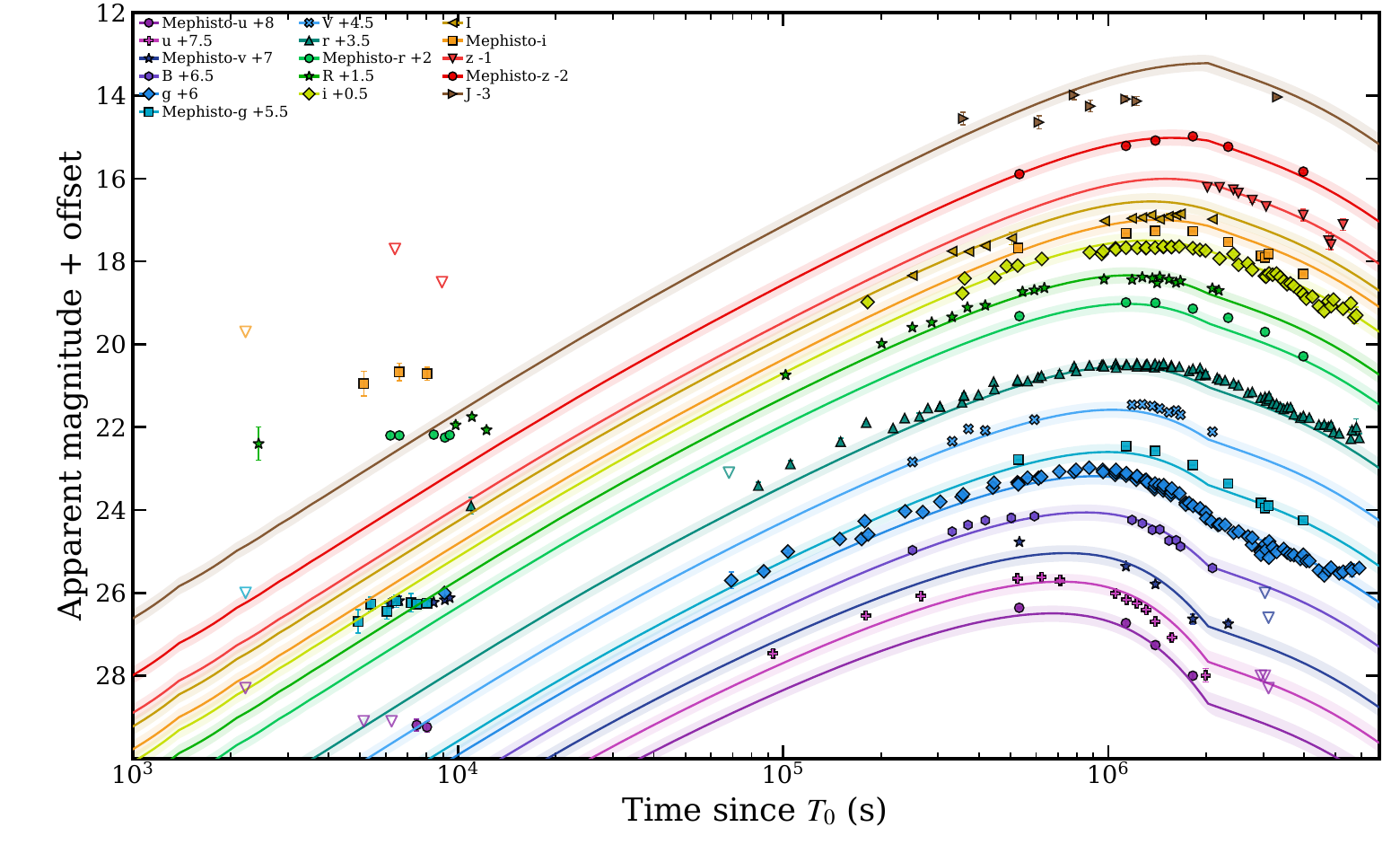}
\caption{
\textbf{Multiband optical and near-infrared evolution of SN~2026gzf.}
Apparent magnitudes are plotted as a function of time, with vertical offsets applied to individual filters for clarity. Open downward triangles mark upper limits. The solid lines give a theoretical fit of the light curves by invoking power due to the radioactive decay of $^{56}$Ni, with the shaded bands representing the $1\sigma$ model uncertainty.
}
\label{fig:SNlightcurves}
\end{figure}



\clearpage 

%
\bibliography{main,method} 

@ARTICLE{Huang2026GCN.44068....1H,
       author = {{Huang}, Q.~J. and {Zou}, Z.~C. and {Mao}, X. and {Li}, D.~Y. and {Pan}, H.~W. and {Einstein Probe Team}},
        title = "{EP260321a: Einstein Probe detection of an X-ray transient}",
      journal = {GRB Coordinates Network},
         year = 2026,
        month = mar,
       volume = {44068},
        pages = {1},
       adsurl = {https://ui.adsabs.harvard.edu/abs/2026GCN.44068....1H}
}

@ARTICLE{Pian2006Natur.442.1011P,
       author = {{Pian}, E. and {Mazzali}, P.~A. and {Masetti}, N. and {Ferrero}, P. and {Klose}, S. and {Palazzi}, E. and {Ramirez-Ruiz}, E. and {Woosley}, S.~E. and {Kouveliotou}, C. and {Deng}, J. and {Filippenko}, A.~V. and {Foley}, R.~J. and {Fynbo}, J.~P.~U. and {Kann}, D.~A. and {Li}, W. and {Hjorth}, J. and {Nomoto}, K. and {Patat}, F. and {Sauer}, D.~N. and {Sollerman}, J. and {Vreeswijk}, P.~M. and {Guenther}, E.~W. and {Levan}, A. and {O'Brien}, P. and {Tanvir}, N.~R. and {Wijers}, R.~A.~M.~J. and {Dumas}, C. and {Hainaut}, O. and {Wong}, D.~S. and {Baade}, D. and {Wang}, L. and {Amati}, L. and {Cappellaro}, E. and {Castro-Tirado}, A.~J. and {Ellison}, S. and {Frontera}, F. and {Fruchter}, A.~S. and {Greiner}, J. and {Kawabata}, K. and {Ledoux}, C. and {Maeda}, K. and {M{\o}ller}, P. and {Nicastro}, L. and {Rol}, E. and {Starling}, R.},
        title = "{An optical supernova associated with the X-ray flash XRF 060218}",
      journal = {\nat},
         year = 2006,
        month = aug,
       volume = {442},
       number = {7106},
        pages = {1011-1013},
          doi = {10.1038/nature05082},
archivePrefix = {arXiv},
       eprint = {astro-ph/0603530},
 primaryClass = {astro-ph},
       adsurl = {https://ui.adsabs.harvard.edu/abs/2006Natur.442.1011P}
}

@ARTICLE{Mazzali2006Natur.442.1018M,
       author = {{Mazzali}, Paolo A. and {Deng}, Jinsong and {Nomoto}, Ken'ichi and {Sauer}, Daniel N. and {Pian}, Elena and {Tominaga}, Nozomu and {Tanaka}, Masaomi and {Maeda}, Keiichi and {Filippenko}, Alexei V.},
        title = "{A neutron-star-driven X-ray flash associated with supernova SN 2006aj}",
      journal = {\nat},
         year = 2006,
        month = aug,
       volume = {442},
       number = {7106},
        pages = {1018-1020},
          doi = {10.1038/nature05081},
archivePrefix = {arXiv},
       eprint = {astro-ph/0603567},
 primaryClass = {astro-ph},
       adsurl = {https://ui.adsabs.harvard.edu/abs/2006Natur.442.1018M}
}

@ARTICLE{Kotake2006RPPh...69..971K,
       author = {{Kotake}, Kei and {Sato}, Katsuhiko and {Takahashi}, Keitaro},
        title = "{Explosion mechanism, neutrino burst and gravitational wave in core-collapse supernovae}",
      journal = {Reports on Progress in Physics},
         year = 2006,
        month = apr,
       volume = {69},
       number = {4},
        pages = {971-1143},
          doi = {10.1088/0034-4885/69/4/R03},
archivePrefix = {arXiv},
       eprint = {astro-ph/0509456},
 primaryClass = {astro-ph},
       adsurl = {https://ui.adsabs.harvard.edu/abs/2006RPPh...69..971K}
}

@ARTICLE{MartinCarrillo2026,
       author = {{Martin-Carrillo}, Antonio and {Th{\"o}ne}, Christina C. and {Leung}, James K. and {Corcoran}, Gregory and {de Ugarte Postigo}, Antonio and {Jonker}, Peter G. and {Izzo}, Luca and {Levan}, Andrew J. and {Gompertz}, Benjamin P. and {Basa}, St{\'e}phane and {Sarin}, Nikhil and {Quirola-V{\'a}squez}, Jonathan and {Eyles-Ferris}, Rob A.~J. and {Brivio}, Riccardo and {Watson}, Alan M. and {Cotter}, Laura and {Chac{\'o}n}, Jennifer Alexandra and {Rossi}, Andrea and {Melandri}, Andrea and {Kumnurdmanee}, Piramon and {Tanvir}, Nial R. and {Gupta}, Anshika and {Bauer}, Franz E. and {Ducoin}, Jean-Gr{\'e}goire and {Reguitti}, Andrea and {Misra}, Kuntal and {Xu}, Dong and {Vergani}, Susanna D. and {Fong}, Wen-fai and {Ackley}, Kendall and {Aguilar-Ruiz}, Edilberto and {Akl}, Dalya and {Aloy}, Miguel {\'A}ngel and {An}, Jie and {Angulo-Valdez}, Camila and {Antier}, Sarah and {Atteia}, Jean-Luc and {Becerra}, Rosa L. and {Breton}, Rene P. and {Butler}, Nathaniel R. and {Campana}, Sergio and {Carotenuto}, Francesco and {Casares Vel{\'a}zquez}, Jorge and {Chrimes}, Ashley A. and {D'Elia}, Valerio and {van Dalen}, Joyce N.~D. and {De Colle}, Fabio and {De Pasquale}, Massimiliano and {Dhillon}, Vik S. and {Dornic}, Damien and {Dyer}, Martin J. and {Ferro}, Matteo and {Fraser}, Morgan and {Fruchter}, Andrew S. and {Fortin}, Francis and {Galloway}, Duncan K. and {Garc{\'\i}a-Garc{\'\i}a}, Leonardo and {Geier}, Stefan and {Gill}, Ramandeep and {Globus}, No{\'e}mie and {Gualandi}, Roberto and {Guelfand}, Marion and {Guidolin}, Francesco and {Hartmann}, Dieter H. and {van Hoof}, Agnes P.~C. and {Jakobsson}, Pall and {Janghel}, Divyanshu and {Killestein}, Tom L. and {Klose}, Sylvio and {Kobayashi}, Shiho and {Kotak}, Rubina and {Kumar}, Amit and {Kuwata}, Asuka and {Laskar}, Tanmoy and {Lee}, William H. and {Lincetto}, Massimiliano and {Lombardi}, Gianluca and {L{\'o}pez-C{\'a}mara}, Diego and {Lyman}, Joseph D. and {Maiorano}, Elisabetta and {Maeda}, Keiichi and {Mandarakas}, Nikos and {Magnani}, Francesco and {Mao}, Jirong and {Moreno M{\'e}ndez}, Enrique and {Mar{\'\i}a Nicuesa Guelbenzu}, Ana and {Noysena}, Kanthanakorn and {Nuttall}, Laura K. and {O'Brien}, Paul T. and {O'Neill}, David and {Ochner}, Paolo and {Pereyra}, Margarita and {Pugliese}, Giovanna and {Ramsay}, Gavin and {Rhodes}, Lauren and {Saccardi}, Andrea and {Salvaterra}, Ruben and {S{\'a}nchez {\'A}lvarez}, Fredd and {Schneider}, Benjamin and {Schulze}, Steve and {Starling}, Rhaana L.~C. and {Steeghs}, Danny and {Ulaczyk}, Kzrysztof and {Ventura}, Chiara and {Zafar}, Tayyaba and {Zhu}, Zi-Pei},
        title = "{Failed jet breakout in the metal-poor broad-lined type Ic supernova 2026gzf}",
      journal = {arXiv e-prints},
         year = 2026,
        month = jun,
          eid = {arXiv:2606.10002},
        pages = {arXiv:2606.10002},
          doi = {10.48550/arXiv.2606.10002},
archivePrefix = {arXiv},
       eprint = {2606.10002},
 primaryClass = {astro-ph.HE},
       adsurl = {https://ui.adsabs.harvard.edu/abs/2026arXiv260610002M}
}

@ARTICLE{Chen2026,
       author = {{Chen}, Ting-Wan and {Aryan}, Amar and {Yang}, Sheng and {Smartt}, Stephen J. and {Moriya}, Takashi J. and {Brennan}, Se{\'a}n J. and {Stritzinger}, Maximilian D. and {Martin}, Bailey and {Nicholl}, Matt and {Kong}, Albert K.~H. and {Gillanders}, James H. and {Dutta}, Anirban and {Schmidt}, Brian P. and {Cheng}, Yu-Chi and {Huber}, Mark E. and {Lai}, Cheng-Han and {Lee}, Chien-Hsiu and {Lee}, Yu-Hsing and {Ngeow}, Chow-Choong and {Smith}, Ken W. and {Ashall}, Christopher and {Auchettl}, Katie and {Burns}, Chris R. and {Chambers}, Kenneth C. and {Chen}, Zhi-Yue and {de Boer}, Thomas and {Hsiao}, Eric Y. and {Ngo Thanh Ho}, Khoa and {Hoogendam}, Willem B. and {Jones}, David O. and {Kankare}, Erkki and {Killestein}, Tom L. and {Kuncarayakti}, Hanindyo and {Lee}, Meng-Han and {Li}, Chuan-Jui and {Lin}, Chien-Cheng and {Lidman}, Christopher and {Lowe}, Thomas B. and {Magnier}, Eugene A. and {Medler}, Kyle and {M{\"o}ller}, Anais and {Moore}, Thomas and {Morrell}, Nidia and {Paek}, Gregory S.~H. and {Pfeffer}, Cameron M. and {Qiang}, Da-Chun and {Rauf}, Liana and {Reynolds}, Thomas M. and {Sankar. K}, Aiswarya and {Srivastav}, Shubham and {Tweddle}, Jack and {Wainscoat}, Richard and {Wang}, Ze-Ning and {Xiao}, Huangfei and {Zhu}, Zonghong},
        title = "{Decadal pre-explosion activity and circumstellar interaction in a supernova}",
      journal = {arXiv e-prints},
         year = 2026,
        month = jun,
          eid = {arXiv:2606.10009},
        pages = {arXiv:2606.10009},
          doi = {10.48550/arXiv.2606.10009},
archivePrefix = {arXiv},
       eprint = {2606.10009},
 primaryClass = {astro-ph.HE},
       adsurl = {https://ui.adsabs.harvard.edu/abs/2026arXiv260610009C}
}

@ARTICLE{Rastinejad2026,
       author = {{Rastinejad}, Jillian C. and {Srinivasaragavan}, Gokul and {Sarin}, Nikhil and {O'Dwyer}, Tanner and {Cenko}, S. Bradley and {Leung}, James K. and {Nugent}, Anya E. and {Perley}, Daniel A. and {Schroeder}, Genevieve and {Anand}, Shreya and {Ahumada}, Tomas and {Andreoni}, Igor and {Bochenek}, Aleksandra and {Corsi}, Alessandra and {Fremling}, Christoffer and {Ho}, Anna Y.~Q. and {Kasliwal}, Mansi M. and {Mo}, Geoffrey and {Salgundi}, Anirudh and {Sippy}, Kendall I. and {Sollerman}, J. and {Bellm}, Eric C. and {Chen}, Tracy X. and {Coughlin}, Michael W. and {Davis}, Michael C. and {De Colle}, Fabio and {Frostig}, Danielle and {Fryer}, Christopher L. and {Graham}, Michael J. and {Hall}, Xander J. and {Hinds}, K.-R. and {Izzo}, Luca and {Jacobson-Galan}, Wynn and {Lourie}, Nathan P. and {Maeda}, Keiichi and {Purdum}, Josiah and {Rusholme}, Ben and {Singh}, Avinash and {Stein}, Robert},
        title = "{A Multi-Wavelength View of the First Type Ic-BL Supernova with an Einstein Probe X-ray Shock Breakout}",
      journal = {arXiv e-prints},
         year = 2026,
        month = jun,
          eid = {arXiv:2606.10011},
        pages = {arXiv:2606.10011},
          doi = {10.48550/arXiv.2606.10011},
archivePrefix = {arXiv},
       eprint = {2606.10011},
 primaryClass = {astro-ph.HE},
       adsurl = {https://ui.adsabs.harvard.edu/abs/2026arXiv260610011R}
}

@ARTICLE{OConnor2026,
       author = {{O'Connor}, Brendan and {Hall}, Xander J. and {Busmann}, Malte and {Gruen}, Daniel and {Floris}, Alberto and {Cabrera}, Tomas and {Zhu}, Ziyuan and {Palmese}, Antonella and {Green}, Dylan and {Banovetz}, John and {Gassert}, Julius and {Fryer}, Christopher L. and {Ricci}, Roberto and {Troja}, Eleonora and {Shivaprasad}, Surya and {Zeimann}, Gregory R. and {Amsellem}, Ariel J. and {Bailey}, Stephen and {BenZvi}, Segev and {Dichiara}, Simone and {van Eerten}, Hendrik and {Hare}, Jeremy and {Hu}, Lei and {Irwin}, Christopher M. and {Kunnumkai}, Keerthi and {Malanchev}, Konstantin and {Maleki}, Mitra and {Moss}, Michael J. and {Myers}, Adam D. and {Pasham}, Dheeraj and {Ries}, Christoph and {Ryan}, Geoffrey and {Schlegel}, David and {Schmidt}, Michael and {Wilke}, Silona and {Yang}, Yu-Han},
        title = "{EP260321a/SN 2026gzf: The Faintest Shock Breakout Associated with a Broad-Lined Supernova}",
      journal = {arXiv e-prints},
         year = 2026,
        month = jun,
          eid = {arXiv:2606.09992},
        pages = {arXiv:2606.09992},
          doi = {10.48550/arXiv.2606.09992},
archivePrefix = {arXiv},
       eprint = {2606.09992},
 primaryClass = {astro-ph.HE},
       adsurl = {https://ui.adsabs.harvard.edu/abs/2026arXiv260609992O}
}

@ARTICLE{Chandra2012,
       author = {{Chandra}, Poonam and {Frail}, Dale A.},
        title = "{A Radio-selected Sample of Gamma-Ray Burst Afterglows}",
      journal = {\apj},
         year = 2012,
        month = feb,
       volume = {746},
       number = {2},
          eid = {156},
        pages = {156},
          doi = {10.1088/0004-637X/746/2/156},
archivePrefix = {arXiv},
       eprint = {1110.4124},
 primaryClass = {astro-ph.CO},
       adsurl = {https://ui.adsabs.harvard.edu/abs/2012ApJ...746..156C}
}

@ARTICLE{Bersten2018,
       author = {{Bersten}, M.~C. and {Folatelli}, G. and {Garc{\'\i}a}, F. and {van Dyk}, S.~D. and {Benvenuto}, O.~G. and {Orellana}, M. and {Buso}, V. and {S{\'a}nchez}, J.~L. and {Tanaka}, M. and {Maeda}, K. and {Filippenko}, A.~V. and {Zheng}, W. and {Brink}, T.~G. and {Cenko}, S.~B. and {de Jaeger}, T. and {Kumar}, S. and {Moriya}, T.~J. and {Nomoto}, K. and {Perley}, D.~A. and {Shivvers}, I. and {Smith}, N.},
        title = "{A surge of light at the birth of a supernova}",
      journal = {\nat},
         year = 2018,
        month = feb,
       volume = {554},
       number = {7693},
        pages = {497-499},
          doi = {10.1038/nature25151},
archivePrefix = {arXiv},
       eprint = {1802.09360},
 primaryClass = {astro-ph.HE},
       adsurl = {https://ui.adsabs.harvard.edu/abs/2018Natur.554..497B}
}

@ARTICLE{Garnavich2016,
  author        = {{Garnavich}, P.~M. and {Tucker}, B.~E. and {Rest}, A. and
                   {Shaya}, E.~J. and {Olling}, R.~P. and {Kasen}, D. and
                   {Villar}, A.},
  title         = {Shock Breakout and Early Light Curves of Type {II-P} Supernovae Observed with {Kepler}},
  journal       = {\apj},
  year          = {2016},
  month         = mar,
  volume        = {820},
  number        = {1},
  eid           = {23},
  pages         = {23},
  doi           = {10.3847/0004-637X/820/1/23},
  archivePrefix = {arXiv},
  eprint        = {1603.05657},
  primaryClass  = {astro-ph.SR}
}

@ARTICLE{Li2024,
       author = {{Li}, Gaici and {Hu}, Maokai and {Li}, Wenxiong and {Yang}, Yi and {Wang}, Xiaofeng and {Yan}, Shengyu and {Hu}, Lei and {Zhang}, Jujia and {Mao}, Yiming and {Riise}, Henrik and {Gao}, Xing and {Sun}, Tianrui and {Liu}, Jialian and {Xiong}, Dingrong and {Wang}, Lifan and {Mo}, Jun and {Iskandar}, Abdusamatjan and {Xi}, Gaobo and {Xiang}, Danfeng and {Wang}, Lingzhi and {Sun}, Guoyou and {Zhang}, Keming and {Chen}, Jian and {Lin}, Weili and {Guo}, Fangzhou and {Liu}, Qichun and {Cai}, Guangyao and {Zhou}, Wenjie and {Zhao}, Jingyuan and {Chen}, Jin and {Zheng}, Xin and {Li}, Keying and {Zhang}, Mi and {Xu}, Shijun and {Lyu}, Xiaodong and {Castro-Tirado}, Alberto J. and {Chufarin}, Vasilii and {Potapov}, Nikolay and {Ionov}, Ivan and {Korotkiy}, Stanislav and {Nazarov}, Sergey and {Sokolovsky}, Kirill and {Hamann}, Norman and {Herman}, Eliot},
        title = "{A shock flash breaking out of a dusty red supergiant}",
      journal = {\nat},
         year = 2024,
        month = mar,
       volume = {627},
       number = {8005},
        pages = {754-758},
          doi = {10.1038/s41586-023-06843-6},
archivePrefix = {arXiv},
       eprint = {2311.14409},
 primaryClass = {astro-ph.HE},
       adsurl = {https://ui.adsabs.harvard.edu/abs/2024Natur.627..754L}
}

@ARTICLE{AguileraDena2018,
       author = {{Aguilera-Dena}, David R. and {Langer}, Norbert and {Moriya}, Takashi J. and {Schootemeijer}, Abel},
        title = "{Related Progenitor Models for Long-duration Gamma-Ray Bursts and Type Ic Superluminous Supernovae}",
      journal = {\apj},
         year = 2018,
        month = may,
       volume = {858},
       number = {2},
          eid = {115},
        pages = {115},
          doi = {10.3847/1538-4357/aabfc1},
archivePrefix = {arXiv},
       eprint = {1804.07317},
 primaryClass = {astro-ph.SR},
       adsurl = {https://ui.adsabs.harvard.edu/abs/2018ApJ...858..115A}
}

@ARTICLE{Taddia2018,
       author = {{Taddia}, F. and {Stritzinger}, M.~D. and {Bersten}, M. and {Baron}, E. and {Burns}, C. and {Contreras}, C. and {Holmbo}, S. and {Hsiao}, E.~Y. and {Morrell}, N. and {Phillips}, M.~M. and {Sollerman}, J. and {Suntzeff}, N.~B.},
        title = "{The Carnegie Supernova Project I. Analysis of stripped-envelope supernova light curves}",
      journal = {\aap},
         year = 2018,
        month = feb,
       volume = {609},
          eid = {A136},
        pages = {A136},
          doi = {10.1051/0004-6361/201730844},
archivePrefix = {arXiv},
       eprint = {1707.07614},
 primaryClass = {astro-ph.HE},
       adsurl = {https://ui.adsabs.harvard.edu/abs/2018A&A...609A.136T}
}

@ARTICLE{Lyman2016,
       author = {{Lyman}, J.~D. and {Bersier}, D. and {James}, P.~A. and {Mazzali}, P.~A. and {Eldridge}, J.~J. and {Fraser}, M. and {Pian}, E.},
        title = "{Bolometric light curves and explosion parameters of 38 stripped-envelope core-collapse supernovae}",
      journal = {\mnras},
         year = 2016,
        month = mar,
       volume = {457},
       number = {1},
        pages = {328-350},
          doi = {10.1093/mnras/stv2983},
archivePrefix = {arXiv},
       eprint = {1406.3667},
 primaryClass = {astro-ph.SR},
       adsurl = {https://ui.adsabs.harvard.edu/abs/2016MNRAS.457..328L}
}

@ARTICLE{Wongwathanarat2015,
       author = {{Wongwathanarat}, A. and {M{\"u}ller}, E. and {Janka}, H.-Th.},
        title = "{Three-dimensional simulations of core-collapse supernovae: from shock revival to shock breakout}",
      journal = {\aap},
         year = 2015,
        month = may,
       volume = {577},
          eid = {A48},
        pages = {A48},
          doi = {10.1051/0004-6361/201425025},
archivePrefix = {arXiv},
       eprint = {1409.5431},
 primaryClass = {astro-ph.HE},
       adsurl = {https://ui.adsabs.harvard.edu/abs/2015A&A...577A..48W}
}

@ARTICLE{Hammer2010,
       author = {{Hammer}, N.~J. and {Janka}, H.-Th. and {M{\"u}ller}, E.},
        title = "{Three-dimensional Simulations of Mixing Instabilities in Supernova Explosions}",
      journal = {\apj},
         year = 2010,
        month = may,
       volume = {714},
       number = {2},
        pages = {1371-1385},
          doi = {10.1088/0004-637X/714/2/1371},
archivePrefix = {arXiv},
       eprint = {0908.3474},
 primaryClass = {astro-ph.SR},
       adsurl = {https://ui.adsabs.harvard.edu/abs/2010ApJ...714.1371H}
}

@ARTICLE{Evans2007,
       author = {{Evans}, P.~A. and {Beardmore}, A.~P. and {Page}, K.~L. and {Tyler}, L.~G. and {Osborne}, J.~P. and {Goad}, M.~R. and {O'Brien}, P.~T. and {Vetere}, L. and {Racusin}, J. and {Morris}, D. and {Burrows}, D.~N. and {Capalbi}, M. and {Perri}, M. and {Gehrels}, N. and {Romano}, P.},
        title = "{An online repository of Swift/XRT light curves of {\ensuremath{\gamma}}-ray bursts}",
      journal = {\aap},
         year = 2007,
        month = jul,
       volume = {469},
       number = {1},
        pages = {379-385},
          doi = {10.1051/0004-6361:20077530},
archivePrefix = {arXiv},
       eprint = {0704.0128},
 primaryClass = {astro-ph},
       adsurl = {https://ui.adsabs.harvard.edu/abs/2007A&A...469..379E}
}

@ARTICLE{Evans2009,
       author = {{Evans}, P.~A. and {Beardmore}, A.~P. and {Page}, K.~L. and {Osborne}, J.~P. and {O'Brien}, P.~T. and {Willingale}, R. and {Starling}, R.~L.~C. and {Burrows}, D.~N. and {Godet}, O. and {Vetere}, L. and {Racusin}, J. and {Goad}, M.~R. and {Wiersema}, K. and {Angelini}, L. and {Capalbi}, M. and {Chincarini}, G. and {Gehrels}, N. and {Kennea}, J.~A. and {Margutti}, R. and {Morris}, D.~C. and {Mountford}, C.~J. and {Pagani}, C. and {Perri}, M. and {Romano}, P. and {Tanvir}, N.},
        title = "{Methods and results of an automatic analysis of a complete sample of Swift-XRT observations of GRBs}",
      journal = {\mnras},
         year = 2009,
        month = aug,
       volume = {397},
       number = {3},
        pages = {1177-1201},
          doi = {10.1111/j.1365-2966.2009.14913.x},
archivePrefix = {arXiv},
       eprint = {0812.3662},
 primaryClass = {astro-ph},
       adsurl = {https://ui.adsabs.harvard.edu/abs/2009MNRAS.397.1177E}
}

@ARTICLE{Modjaz2009,
       author = {{Modjaz}, M. and {Li}, W. and {Butler}, N. and {Chornock}, R. and {Perley}, D. and {Blondin}, S. and {Bloom}, J.~S. and {Filippenko}, A.~V. and {Kirshner}, R.~P. and {Kocevski}, D. and {Poznanski}, D. and {Hicken}, M. and {Foley}, R.~J. and {Stringfellow}, G.~S. and {Berlind}, P. and {Barrado y Navascues}, D. and {Blake}, C.~H. and {Bouy}, H. and {Brown}, W.~R. and {Challis}, P. and {Chen}, H. and {de Vries}, W.~H. and {Dufour}, P. and {Falco}, E. and {Friedman}, A. and {Ganeshalingam}, M. and {Garnavich}, P. and {Holden}, B. and {Illingworth}, G. and {Lee}, N. and {Liebert}, J. and {Marion}, G.~H. and {Olivier}, S.~S. and {Prochaska}, J.~X. and {Silverman}, J.~M. and {Smith}, N. and {Starr}, D. and {Steele}, T.~N. and {Stockton}, A. and {Williams}, G.~G. and {Wood-Vasey}, W.~M.},
        title = "{From Shock Breakout to Peak and Beyond: Extensive Panchromatic Observations of the Type Ib Supernova 2008D Associated with Swift X-ray Transient 080109}",
      journal = {\apj},
         year = 2009,
        month = sep,
       volume = {702},
       number = {1},
        pages = {226-248},
          doi = {10.1088/0004-637X/702/1/226},
archivePrefix = {arXiv},
       eprint = {0805.2201},
 primaryClass = {astro-ph},
       adsurl = {https://ui.adsabs.harvard.edu/abs/2009ApJ...702..226M}
}

@ARTICLE{Nakar2015,
       author = {{Nakar}, Ehud},
        title = "{A Unified Picture for Low-luminosity and Long Gamma-Ray Bursts Based on the Extended Progenitor of llGRB 060218/SN 2006aj}",
      journal = {\apj},
         year = 2015,
        month = jul,
       volume = {807},
       number = {2},
          eid = {172},
        pages = {172},
          doi = {10.1088/0004-637X/807/2/172},
archivePrefix = {arXiv},
       eprint = {1503.00441},
 primaryClass = {astro-ph.HE},
       adsurl = {https://ui.adsabs.harvard.edu/abs/2015ApJ...807..172N}
}

@ARTICLE{Zheng2026,
       author = {{Zheng}, Jian-He and {Lu}, Wenbin},
        title = "{Fast X-Ray Transients Produced by Off-axis Jet Cocoons from Long Gamma-Ray Bursts}",
      journal = {\apjl},
         year = 2026,
        month = may,
       volume = {1003},
       number = {1},
          eid = {L19},
        pages = {L19},
          doi = {10.3847/2041-8213/ae67f2},
archivePrefix = {arXiv},
       eprint = {2603.09674},
 primaryClass = {astro-ph.HE},
       adsurl = {https://ui.adsabs.harvard.edu/abs/2026ApJ..1003L..19Z}
}

@ARTICLE{Li2025,
       author = {{Li}, W.-X. and {Zhu}, Z.-P. and {Zou}, X.-Z. and {Geng}, J.-J. and {Liu}, L.-D. and {Wang}, Y.-H. and {Li}, R.-Z. and {Xu}, D. and {Sun}, H. and {Wang}, X.-F. and {Yu}, Y.-W. and {Zhang}, B. and {Wu}, X.-F. and {Yang}, Y. and {Filippenko}, A.~V. and {Liu}, X.-W. and {Yuan}, W.-M. and {Aguado}, D. and {An}, J. and {An}, T. and {Buckley}, D.~A.~H. and {Castro-Tirado}, A.~J. and {Fu}, S.-Y. and {Fynbo}, J.~P.~U. and {Howell}, D.~A. and {Hu}, J.-W. and {Jiang}, S.-Q. and {Kumar}, A. and {Mao}, J.-R. and {Maund}, J.~R. and {Liu}, X. and {Mockler}, B. and {Moskvitin}, A. and {Andrews}, M. and {Bom}, C.~R. and {Brink}, T.~G. and {Chatterjee}, K. and {Chen}, Y. and {Cheng}, H.-Q. and {Cooke}, J. and {Dai}, J.~L. and {Du}, G.-W. and {Erasmus}, N. and {Fang}, Y. and {Farah}, J. and {Goranskij}, V. and {Gritsevich}, M. and {Gu}, M. and {Guo}, Z. and {Hsiao}, E. and {Hu}, Y.-D. and {Hua}, Y.-L. and {Jacobson-Gal{\'a}n}, W. and {Jia}, S.-M. and {Jin}, C.-C. and {Kasliwal}, M.~M. and {Kilpatrick}, C.~D. and {Kumar}, B. and {Lei}, W.-H. and {Li}, C.-K. and {Li}, D.-Y. and {Li}, L.-P. and {Ling}, Z.-X. and {Liu}, Q.-C. and {Liu}, Y. and {Liu}, Y.-Q. and {L{\'o}pez-Oramas}, A. and {Maslennikova}, O. and {McCully}, C. and {Monageng}, I. and {Newsone}, M. and {Padilla Gonzalez}, E. and {Pan}, H.-W. and {Peng}, H.-W. and {Pignata}, G. and {Poidevin}, F. and {Potter}, S.~B. and {P{\'e}rez-Fournon}, I. and {Santana-Silva}, L. and {Santos}, A. and {Song}, C.-Y. and {Song}, F.-F. and {Spiridonova}, O. and {Sun}, N.-C. and {Sun}, X.-J. and {Terreran}, G. and {Wang}, L.-Z. and {Wang}, L.-F. and {Wang}, B.-T. and {Wang}, Z.-Y. and {Wu}, G.-L. and {Xiang}, D.-F. and {Xiao}, H.-F. and {Xu}, Y.-F. and {Xue}, S.-J. and {Yan}, S.-Y. and {Yang}, Y.-P. and {Yu}, L.-X. and {Zhang}, Y.-H. and {Zhang}, Y.-H. and {Zhang}, C. and {Zhang}, J.-H. and {Zhang}, J.-J. and {Zheng}, W. and {Zou}, H.},
        title = "{An extremely soft and weak fast X-ray transient associated with a luminous supernova}",
      journal = {arXiv e-prints},
         year = 2025,
        month = apr,
          eid = {arXiv:2504.17034},
        pages = {arXiv:2504.17034},
          doi = {10.48550/arXiv.2504.17034},
archivePrefix = {arXiv},
       eprint = {2504.17034},
 primaryClass = {astro-ph.HE},
       adsurl = {https://ui.adsabs.harvard.edu/abs/2025arXiv250417034L}
}

@ARTICLE{Sun2025,
       author = {{Sun}, H. and {Li}, W.-X. and {Liu}, L.-D. and {Gao}, H. and {Wang}, X.-F. and {Yuan}, W. and {Zhang}, B. and {Filippenko}, A.~V. and {Xu}, D. and {An}, T. and {Ai}, S. and {Brink}, T.~G. and {Liu}, Y. and {Liu}, Y.-Q. and {Wang}, C.-Y. and {Wu}, Q.-Y. and {Wu}, X.-F. and {Yang}, Y. and {Zhang}, B.-B. and {Zheng}, W.-K. and {Ahumada}, T. and {Dai}, Z.-G. and {Delaunay}, J. and {Elias-Rosa}, N. and {Benetti}, S. and {Fu}, S.-Y. and {Howell}, D.~A. and {Huang}, Y.-F. and {Kasliwal}, M.~M. and {Karambelkar}, V. and {Stein}, R. and {Lei}, W.-H. and {Lian}, T.-Y. and {Peng}, Z.-K. and {Frederiks}, D.~D. and {Ridnaia}, A.~V. and {Svinkin}, D.~S. and {Wang}, X.-Y. and {Wang}, A.-L. and {Wei}, D.-M. and {An}, J. and {Andrews}, M. and {Bai}, J.-M. and {Dai}, C.-Y. and {Ehgamberdiev}, S.~A. and {Fan}, Z. and {Farah}, J. and {Feng}, H.-C. and {Fynbo}, J.~P.~U. and {Guo}, W.-J. and {Guo}, Z. and {Hu}, M.-K. and {Hu}, J.-W. and {Jiang}, S.-Q. and {Jin}, J.-J. and {Li}, A. and {Li}, J.-D. and {Li}, R.-Z. and {Liang}, Y.-F. and {Ling}, Z.-X. and {Liu}, X. and {Mao}, J.-R. and {McCully}, C. and {Mirzaqulov}, D. and {Newsome}, M. and {Padilla Gonzalez}, E. and {Pan}, X. and {Terreran}, G. and {Tinyanont}, S. and {Wang}, B.-T. and {Wang}, L.-Z. and {Wen}, X.-D. and {Xiang}, D.-F. and {Xue}, S.-J. and {Yang}, J. and {Zhu}, Z.-P. and {Cai}, Z.-M. and {Castro-Tirado}, A.~J. and {Chen}, F.-S. and {Chen}, H.-L. and {Chen}, T.-X. and {Chen}, W. and {Chen}, Y.-H. and {Chen}, Y.-F. and {Chen}, Y. and {Cheng}, H.-Q. and {Cordier}, B. and {Cui}, C.-Z. and {Cui}, W.-W. and {Dai}, Y.-F. and {Fan}, D.-W. and {Feng}, H. and {Guan}, J. and {Han}, D.-W. and {Hou}, D.-J. and {Hu}, H.-B. and {Huang}, M.-H. and {Huo}, J. and {Jia}, S.-M. and {Jia}, Z.-Q. and {Jiang}, B.-W. and {Jin}, C.-C. and {Jin}, G. and {Kuulkers}, E. and {Li}, C.-K. and {Li}, D.-Y. and {Li}, J.-F. and {Li}, L.-H. and {Li}, M.-S. and {Li}, W. and {Li}, Z.-D. and {Liu}, C.-Z. and {Liu}, H.-Y. and {Liu}, H.-Q. and {Liu}, M.-J. and {Lu}, F.-J. and {Luo}, L.-D. and {Ma}, J. and {Mao}, X. and {Nandra}, K. and {O'Brien}, P. and {Pan}, H.-W. and {Rau}, A. and {Rea}, N. and {Sanders}, J. and {Song}, L.-M. and {Sun}, S.-L. and {Sun}, X.-J. and {Tan}, Y.-Y. and {Tang}, Q.-J. and {Tao}, Y.-H. and {Wang}, H. and {Wang}, J. and {Wang}, L. and {Wang}, W.-X. and {Wang}, Y.-L. and {Wang}, Y.-S. and {Xiong}, D.-R. and {Xu}, H.-T. and {Xu}, J.-J. and {Xu}, X.-P. and {Xu}, Y.-F. and {Xu}, Z. and {Xue}, C.-B. and {Xue}, Y.-L. and {Yan}, A.-L. and {Yang}, H.-N. and {Yang}, X.-T. and {Yang}, Y.-J. and {Zhang}, C. and {Zhang}, J. and {Zhang}, M. and {Zhang}, S.-N. and {Zhang}, W.-D. and {Zhang}, W.-J. and {Zhang}, Y.-H. and {Zhang}, Z. and {Zhang}, Z. and {Zhang}, Z.-L. and {Zhao}, D.-H. and {Zhao}, H.-S. and {Zhao}, X.-F. and {Zhao}, Z.-J. and {Zhou}, Y.-L. and {Zhu}, Y.-X. and {Zhu}, Z.-C. and {Zou}, H.},
        title = "{A fast X-ray transient from a weak relativistic jet associated with a type Ic-BL supernova}",
      journal = {Nature Astronomy},
         year = 2025,
        month = jul,
       volume = {9},
        pages = {1073-1085},
          doi = {10.1038/s41550-025-02571-1},
archivePrefix = {arXiv},
       eprint = {2410.02315},
 primaryClass = {astro-ph.HE},
       adsurl = {https://ui.adsabs.harvard.edu/abs/2025NatAs...9.1073S}
}

@ARTICLE{Crowther2007,
       author = {{Crowther}, Paul A.},
        title = "{Physical Properties of Wolf-Rayet Stars}",
      journal = {\araa},
         year = 2007,
        month = sep,
       volume = {45},
       number = {1},
        pages = {177-219},
          doi = {10.1146/annurev.astro.45.051806.110615},
archivePrefix = {arXiv},
       eprint = {astro-ph/0610356},
 primaryClass = {astro-ph},
       adsurl = {https://ui.adsabs.harvard.edu/abs/2007ARA&A..45..177C}
}

@ARTICLE{Yu2015ApJ...806L...6Y,
       author = {{Yu}, Yun-Wei and {Li}, Shao-Ze and {Dai}, Zi-Gao},
        title = "{Rapidly Evolving and Luminous Transients Driven by Newly Born Neutron Stars}",
      journal = {\apjl},
         year = 2015,
        month = jun,
       volume = {806},
       number = {1},
          eid = {L6},
        pages = {L6},
          doi = {10.1088/2041-8205/806/1/L6},
archivePrefix = {arXiv},
       eprint = {1505.03251},
 primaryClass = {astro-ph.HE},
       adsurl = {https://ui.adsabs.harvard.edu/abs/2015ApJ...806L...6Y}
}

@ARTICLE{Chevalier2006,
       author = {{Chevalier}, Roger A. and {Fransson}, Claes},
        title = "{Circumstellar Emission from Type Ib and Ic Supernovae}",
      journal = {\apj},
         year = 2006,
        month = nov,
       volume = {651},
       number = {1},
        pages = {381-391},
          doi = {10.1086/507606},
archivePrefix = {arXiv},
       eprint = {astro-ph/0607196},
 primaryClass = {astro-ph},
       adsurl = {https://ui.adsabs.harvard.edu/abs/2006ApJ...651..381C}
}

@ARTICLE{Santana2014,
       author = {{Santana}, Rodolfo and {Barniol Duran}, Rodolfo and {Kumar}, Pawan},
        title = "{Magnetic Fields in Relativistic Collisionless Shocks}",
      journal = {\apj},
         year = 2014,
        month = apr,
       volume = {785},
       number = {1},
          eid = {29},
        pages = {29},
          doi = {10.1088/0004-637X/785/1/29},
archivePrefix = {arXiv},
       eprint = {1309.3277},
 primaryClass = {astro-ph.HE},
       adsurl = {https://ui.adsabs.harvard.edu/abs/2014ApJ...785...29S}
}

@ARTICLE{Chen2024ApJ...971L...2C,
       author = {{Chen}, Xinlei and {Kumar}, Brajesh and {Er}, Xinzhong and {Guo}, Helong and {Yang}, Yuan-Pei and {Lin}, Weikang and {Fang}, Yuan and {Du}, Guowang and {Liu}, Chenxu and {Zhao}, Jiewei and {Zhang}, Tianyu and {Bao}, Yuxi and {Zou}, Xingzhu and {Pan}, Yu and {Wang}, Yu and {Zhu}, Xufeng and {Chatterjee}, Kaushik and {Liu}, Xiangkun and {Liu}, Dezi and {Lagioia}, Edoardo P. and {Rangwal}, Geeta and {Zhong}, Shiyan and {Zhang}, Jinghua and {Lian}, Jianhui and {Cai}, Yongzhi and {Zhang}, Yangwei and {Liu}, Xiaowei},
        title = "{Early-phase Simultaneous Multiband Observations of the Type II Supernova SN 2024ggi with Mephisto}",
      journal = {\apjl},
         year = 2024,
        month = aug,
       volume = {971},
       number = {1},
          eid = {L2},
        pages = {L2},
          doi = {10.3847/2041-8213/ad62f7},
archivePrefix = {arXiv},
       eprint = {2405.07964},
 primaryClass = {astro-ph.HE},
       adsurl = {https://ui.adsabs.harvard.edu/abs/2024ApJ...971L...2C}
}

@ARTICLE{Yang2024ApJ...969..126Y,
       author = {{Yang}, Yuan-Pei and {Liu}, Xiangkun and {Pan}, Yu and {Er}, Xinzhong and {Liu}, Dezi and {Fang}, Yuan and {Du}, Guowang and {Cai}, Yongzhi and {Xu}, Xian and {Chen}, Xinlei and {Zou}, Xingzhu and {Guo}, Helong and {Liu}, Chenxu and {Cheng}, Yehao and {Kumar}, Brajesh and {Liu}, Xiaowei},
        title = "{Multiband Simultaneous Photometry of Type II SN 2023ixf with Mephisto and the Twin 50 cm Telescopes}",
      journal = {\apj},
         year = 2024,
        month = jul,
       volume = {969},
       number = {2},
          eid = {126},
        pages = {126},
          doi = {10.3847/1538-4357/ad4be3},
archivePrefix = {arXiv},
       eprint = {2405.08327},
 primaryClass = {astro-ph.HE},
       adsurl = {https://ui.adsabs.harvard.edu/abs/2024ApJ...969..126Y}
}

@ARTICLE{Srinivasaragavan2025,
       author = {{Srinivasaragavan}, Gokul P. and {Li}, Dongyue and {Hall}, Xander J. and {Gottlieb}, Ore and {Schroeder}, Genevieve and {Liu}, Heyang and {O'Connor}, Brendan and {Jin}, Chichuan and {Kasliwal}, Mansi and {Ahumada}, Tom{\'a}s and {Wu}, Qinyu and {Fryer}, Christopher L. and {Niblett}, Annabelle E. and {Xu}, Dong and {Edvige Ravasio}, Maria and {Daja}, Grace and {Li}, Wenxiong and {Anand}, Shreya and {Ho}, Anna Y.~Q. and {Sun}, Hui and {Perley}, Daniel A. and {Yan}, Lin and {Burns}, Eric and {Cenko}, S. Bradley and {Sollerman}, Jesper and {Sarin}, Nikhil and {Piro}, Anthony L. and {Aryan}, Amar and {Miller}, M. Coleman and {An}, Jie and {An}, Tao and {Andrews}, Moira and {Augustin}, Jule and {Bellm}, Eric C. and {Bochenek}, Aleksandra and {Busmann}, Malte and {Chanchaiworawit}, Krittapas and {Chen}, Huaqing and {Caballero-Garc{\'\i}a}, Maria D. and {Castro-Tirado}, Alberto J. and {Esamdin}, Ali and {Faba-Moreno}, Jennifer and {Farah}, Joseph and {Fern{\'a}ndez-Garc{\'\i}a}, Emilio and {Fu}, Shaoyu and {Fynbo}, Johan P.~U. and {Gassert}, Julius and {Padilla Gonzalez}, Estefania and {P{\'e}rez-Garc{\'\i}a}, Ignacio and {Graham}, Matthew and {Gritsevich}, Maria and {Gruen}, Daniel and {Guziy}, Sergiy and {Howell}, D. Andrew and {He}, Linbo and {Hu}, Jingwei and {Hu}, You-Dong and {Iskandar}, Abdusamatjan and {Castaneda Jaims}, Joahan and {Jiang}, Ji-An and {Jiang}, Ning and {Jiang}, Shuaijiao and {Liang}, Runduo and {Ling}, Zhixing and {Liu}, Jialian and {Liu}, Xing and {Liu}, Yuan and {Masci}, Frank J. and {McCully}, Curtis and {Newsome}, Megan and {Noysena}, Kanthanakorn and {Pandey}, Shashi B. and {Ni}, Kangrui and {Palmese}, Antonella and {Peng}, Han-Long and {Purdum}, Josiah and {Qin}, Yu-Jing and {Rose}, Sam and {Rusholme}, Ben and {S{\'a}nchez-Ram{\'\i}rez}, Rub{\'e}n and {Sevilla}, Cassie and {Smith}, Roger and {Song}, Yujia and {Sravan}, Niharika and {Stein}, Robert and {Tabor}, Constantin and {Terreran}, Giacomo and {Tinyanont}, Samaporn and {Vega}, Pablo and {Wang}, Letian and {Wang}, Tinggu and {Wang}, Xiaofeng and {Wu}, Siyu and {Wu}, Xuefeng and {Wynn}, Kathryn and {Xu}, Yunfei and {Yan}, Shengyu and {Yuan}, Weimin and {Zhang}, Binbin and {Zhang}, Chen and {Zhu}, Zipei and {Zuo}, Xiaoxiong and {Bhullar}, Gursimran},
        title = "{EP250827b/SN 2025wkm: An X-ray Flash-Supernova Powered by a Central Engine and Circumstellar Interaction}",
      journal = {arXiv e-prints},
         year = 2025,
        month = dec,
          eid = {arXiv:2512.10239},
        pages = {arXiv:2512.10239},
          doi = {10.48550/arXiv.2512.10239},
archivePrefix = {arXiv},
       eprint = {2512.10239},
 primaryClass = {astro-ph.HE},
       adsurl = {https://ui.adsabs.harvard.edu/abs/2025arXiv251210239S}
}

@ARTICLE{Zhu2025MNRAS.544L.139Z,
       author = {{Zhu}, Jin-Ping and {Zheng}, Jian-He and {Zhang}, Bing},
        title = "{EP 250108a/SN 2025kg: a magnetar-powered gamma-ray burst supernova originating from a close helium-star binary via isolated binary evolution}",
      journal = {\mnras},
         year = 2025,
        month = nov,
       volume = {544},
       number = {1},
        pages = {L139-L149},
          doi = {10.1093/mnrasl/slaf114},
archivePrefix = {arXiv},
       eprint = {2507.18544},
 primaryClass = {astro-ph.HE},
       adsurl = {https://ui.adsabs.harvard.edu/abs/2025MNRAS.544L.139Z}
}

@ARTICLE{Zhu2026,
       author = {{Zhu}, Jin-Ping and {Zhang}, Bing},
        title = "{Magnetar Engines in Broad-lined Type Ic Supernovae and a Unified Picture for Magnetar-powered Stripped-envelope Supernovae}",
      journal = {arXiv e-prints},
         year = 2026,
        month = apr,
          eid = {arXiv:2604.21759},
        pages = {arXiv:2604.21759},
          doi = {10.48550/arXiv.2604.21759},
archivePrefix = {arXiv},
       eprint = {2604.21759},
 primaryClass = {astro-ph.HE},
       adsurl = {https://ui.adsabs.harvard.edu/abs/2026arXiv260421759Z}
}

@INCOLLECTION{Waxman2017hsn..book..967W,
       author = {{Waxman}, Eli and {Katz}, Boaz},
        title = "{Shock Breakout Theory}",
    booktitle = {Handbook of Supernovae},
         year = 2017,
       editor = {{Alsabti}, Athem W. and {Murdin}, Paul},
        pages = {967},
          doi = {10.1007/978-3-319-21846-5_33},
       adsurl = {https://ui.adsabs.harvard.edu/abs/2017hsn..book..967W}
}

@ARTICLE{Chevalier2011ApJ...729L...6C,
       author = {{Chevalier}, Roger A. and {Irwin}, Christopher M.},
        title = "{Shock Breakout in Dense Mass Loss: Luminous Supernovae}",
      journal = {\apjl},
         year = 2011,
        month = mar,
       volume = {729},
       number = {1},
          eid = {L6},
        pages = {L6},
          doi = {10.1088/2041-8205/729/1/L6},
archivePrefix = {arXiv},
       eprint = {1101.1111},
 primaryClass = {astro-ph.HE},
       adsurl = {https://ui.adsabs.harvard.edu/abs/2011ApJ...729L...6C}
}

@ARTICLE{Fuller2018,
       author = {{Fuller}, Jim and {Ro}, Stephen},
        title = "{Pre-supernova outbursts via wave heating in massive stars - II. Hydrogen-poor stars}",
      journal = {\mnras},
         year = 2018,
        month = may,
       volume = {476},
       number = {2},
        pages = {1853-1868},
          doi = {10.1093/mnras/sty369},
archivePrefix = {arXiv},
       eprint = {1710.04251},
 primaryClass = {astro-ph.SR},
       adsurl = {https://ui.adsabs.harvard.edu/abs/2018MNRAS.476.1853F}
}

@ARTICLE{Wu2021,
       author = {{Wu}, Samantha and {Fuller}, Jim},
        title = "{A Diversity of Wave-driven Presupernova Outbursts}",
      journal = {\apj},
         year = 2021,
        month = jan,
       volume = {906},
       number = {1},
          eid = {3},
        pages = {3},
          doi = {10.3847/1538-4357/abc87c},
archivePrefix = {arXiv},
       eprint = {2011.05453},
 primaryClass = {astro-ph.HE},
       adsurl = {https://ui.adsabs.harvard.edu/abs/2021ApJ...906....3W}
}

@ARTICLE{Wu2022,
       author = {{Wu}, Samantha C. and {Fuller}, Jim},
        title = "{Wave-driven Outbursts and Variability of Low-mass Supernova Progenitors}",
      journal = {\apj},
         year = 2022,
        month = may,
       volume = {930},
       number = {2},
          eid = {119},
        pages = {119},
          doi = {10.3847/1538-4357/ac660c},
archivePrefix = {arXiv},
       eprint = {2205.03319},
 primaryClass = {astro-ph.HE},
       adsurl = {https://ui.adsabs.harvard.edu/abs/2022ApJ...930..119W}
}

@ARTICLE{Leung2021,
       author = {{Leung}, Shing-Chi and {Wu}, Samantha and {Fuller}, Jim},
        title = "{Wave-driven Mass Loss of Stripped Envelope Massive Stars: Progenitor-dependence, Mass Ejection, and Supernovae}",
      journal = {\apj},
         year = 2021,
        month = dec,
       volume = {923},
       number = {1},
          eid = {41},
        pages = {41},
          doi = {10.3847/1538-4357/ac2c63},
archivePrefix = {arXiv},
       eprint = {2110.01565},
 primaryClass = {astro-ph.SR},
       adsurl = {https://ui.adsabs.harvard.edu/abs/2021ApJ...923...41L}
}

@ARTICLE{Quataert2012,
       author = {{Quataert}, E. and {Shiode}, J.},
        title = "{Wave-driven mass loss in the last year of stellar evolution: setting the stage for the most luminous core-collapse supernovae}",
      journal = {\mnras},
         year = 2012,
        month = jun,
       volume = {423},
       number = {1},
        pages = {L92-L96},
          doi = {10.1111/j.1745-3933.2012.01264.x},
archivePrefix = {arXiv},
       eprint = {1202.5036},
 primaryClass = {astro-ph.SR},
       adsurl = {https://ui.adsabs.harvard.edu/abs/2012MNRAS.423L..92Q}
}

@ARTICLE{Colgate1974ApJ...187..333C,
       author = {{Colgate}, Stirling A.},
        title = "{Early Gamma Rays from Supernovae}",
      journal = {\apj},
         year = 1974,
        month = jan,
       volume = {187},
        pages = {333-336},
          doi = {10.1086/152632},
       adsurl = {https://ui.adsabs.harvard.edu/abs/1974ApJ...187..333C}
}

@ARTICLE{Soderberg2008Natur.453..469S,
       author = {{Soderberg}, A.~M. and {Berger}, E. and {Page}, K.~L. and {Schady}, P. and {Parrent}, J. and {Pooley}, D. and {Wang}, X.-Y. and {Ofek}, E.~O. and {Cucchiara}, A. and {Rau}, A. and {Waxman}, E. and {Simon}, J.~D. and {Bock}, D.~C.-J. and {Milne}, P.~A. and {Page}, M.~J. and {Barentine}, J.~C. and {Barthelmy}, S.~D. and {Beardmore}, A.~P. and {Bietenholz}, M.~F. and {Brown}, P. and {Burrows}, A. and {Burrows}, D.~N. and {Byrngelson}, G. and {Cenko}, S.~B. and {Chandra}, P. and {Cummings}, J.~R. and {Fox}, D.~B. and {Gal-Yam}, A. and {Gehrels}, N. and {Immler}, S. and {Kasliwal}, M. and {Kong}, A.~K.~H. and {Krimm}, H.~A. and {Kulkarni}, S.~R. and {Maccarone}, T.~J. and {M{\'e}sz{\'a}ros}, P. and {Nakar}, E. and {O'Brien}, P.~T. and {Overzier}, R.~A. and {de Pasquale}, M. and {Racusin}, J. and {Rea}, N. and {York}, D.~G.},
        title = "{An extremely luminous X-ray outburst at the birth of a supernova}",
      journal = {\nat},
         year = 2008,
        month = may,
       volume = {453},
       number = {7194},
        pages = {469-474},
          doi = {10.1038/nature06997},
archivePrefix = {arXiv},
       eprint = {0802.1712},
 primaryClass = {astro-ph},
       adsurl = {https://ui.adsabs.harvard.edu/abs/2008Natur.453..469S}
}

@ARTICLE{EPGCN2,
       author = {{Huang}, Q.~J. and {Zou}, Z.-C. and {Li}, D.~Y. and {Mao}, X. and {Pan}, H.~W. and {Einstein Probe Team}},
        title = "{EP260321a: refined analysis of the EP-WXT and EP-FXT observations, implying a possible supernova shock breakout candidate}",
      journal = {GRB Coordinates Network},
         year = 2026,
        month = mar,
       volume = {44075},
        pages = {1},
       adsurl = {https://ui.adsabs.harvard.edu/abs/2026GCN.44075....1H}
}

@ARTICLE{EPGCN1,
       author = {{Huang}, Q.~J. and {Zou}, Z.~C. and {Mao}, X. and {Li}, D.~Y. and {Pan}, H.~W. and {Einstein Probe Team}},
        title = "{EP260321a: Einstein Probe detection of an X-ray transient}",
      journal = {GRB Coordinates Network},
         year = 2026,
        month = mar,
       volume = {44068},
        pages = {1},
       adsurl = {https://ui.adsabs.harvard.edu/abs/2026GCN.44068....1H}
}

@ARTICLE{OConnor2026GCN.44250....1O,
       author = {{O'Connor}, Brendan and {Troja}, Eleonora},
        title = "{EP260321a: Chandra X-ray Non-detection}",
      journal = {GRB Coordinates Network},
         year = 2026,
        month = apr,
       volume = {44250},
        pages = {1},
       adsurl = {https://ui.adsabs.harvard.edu/abs/2026GCN.44250....1O}
}

@ARTICLE{Lee2026GCN.44070....1L,
       author = {{Lee}, M.-H. and {Aryan}, A. and {Chen}, T.-W. and {Hou}, W.-J. and {Yang}, S. and {Smartt}, S.~J. and {Gillanders}, J. and {Kong}, A.~K.~H. and {Yang}, Y.~J. and {Lee}, Y.-H. and {Sankar. K}, A. and {Pan}, Y.-C. and {Ngeow}, C.-C. and {Lai}, C.-H. and {Lin}, C.-S. Lin H.-C. and {Hsiao}, H.-Y. and {Guo}, J.-K. and {Wang}, Z.~N. and {Qiang}, D.~C. and {Fan}, L.~L. and {Lin}, H.-W. and {Stevance}, H.~F. and {Srivastav}, S. and {Rhodes}, L. and {Nicholl}, M. and {Fulton}, M. and {Moore}, T. and {Smith}, K.~W. and {Angus}, C. and {Aamer}, A. and {Schultz}, A. and {Huber}, M.},
        title = "{EP260321a: Kinder observations detect a blue variable star and set limits on a source from the z =0.034 galaxy within the error circle}",
      journal = {GRB Coordinates Network},
         year = 2026,
        month = mar,
       volume = {44070},
        pages = {1},
       adsurl = {https://ui.adsabs.harvard.edu/abs/2026GCN.44070....1L}
}

@INCOLLECTION{Yuan2022,
       author = {{Yuan}, Weimin and {Zhang}, Chen and {Chen}, Yong and {Ling}, Zhixing},
        title = "{The Einstein Probe Mission}",
    booktitle = {Handbook of X-ray and Gamma-ray Astrophysics},
         year = 2022,
       editor = {{Bambi}, Cosimo and {Sangangelo}, Andrea},
          eid = {86},
        pages = {86},
          doi = {10.1007/978-981-16-4544-0_151-1},
       adsurl = {https://ui.adsabs.harvard.edu/abs/2022hxga.book...86Y}
}

@ARTICLE{Yuan2025,
       author = {{Yuan}, Weimin and {Dai}, Lixin and {Feng}, Hua and {Jin}, Chichuan and {Jonker}, Peter and {Kuulkers}, Erik and {Liu}, Yuan and {Nandra}, Kirpal and {O'Brien}, Paul and {Piro}, Luigi and et al.},
        title = "{Science objectives of the Einstein Probe mission}",
      journal = {Science China Physics, Mechanics, and Astronomy},
         year = 2025,
        month = mar,
       volume = {68},
       number = {3},
          eid = {239501},
        pages = {239501},
          doi = {10.1007/s11433-024-2600-3},
archivePrefix = {arXiv},
       eprint = {2501.07362},
 primaryClass = {astro-ph.HE},
       adsurl = {https://ui.adsabs.harvard.edu/abs/2025SCPMA..6839501Y}
}

@ARTICLE{Mazzali2008,
       author = {{Mazzali}, Paolo A. and {Valenti}, Stefano and {Della Valle}, Massimo and {Chincarini}, Guido and {Sauer}, Daniel N. and {Benetti}, Stefano and {Pian}, Elena and {Piran}, Tsvi and {D'Elia}, Valerio and {Elias-Rosa}, Nancy and {Margutti}, Raffaella and {Pasotti}, Francesco and {Antonelli}, L. Angelo and {Bufano}, Filomena and {Campana}, Sergio and {Cappellaro}, Enrico and {Covino}, Stefano and {D'Avanzo}, Paolo and {Fiore}, Fabrizio and {Fugazza}, Dino and {Gilmozzi}, Roberto and {Hunter}, Deborah and {Maguire}, Kate and {Maiorano}, Elisabetta and {Marziani}, Paola and {Masetti}, Nicola and {Mirabel}, Felix and {Navasardyan}, Hripsime and {Nomoto}, Ken'ichi and {Palazzi}, Eliana and {Pastorello}, Andrea and {Panagia}, Nino and {Pellizza}, L.~J. and {Sari}, Re'em and {Smartt}, Stephen and {Tagliaferri}, Gianpiero and {Tanaka}, Masaomi and {Taubenberger}, Stefan and {Tominaga}, Nozomu and {Trundle}, Carrie and {Turatto}, Massimo},
        title = "{The Metamorphosis of Supernova SN 2008D/XRF 080109: A Link Between Supernovae and GRBs/Hypernovae}",
      journal = {Science},
         year = 2008,
        month = aug,
       volume = {321},
       number = {5893},
        pages = {1185},
          doi = {10.1126/science.1158088},
archivePrefix = {arXiv},
       eprint = {0807.1695},
 primaryClass = {astro-ph},
       adsurl = {https://ui.adsabs.harvard.edu/abs/2008Sci...321.1185M}
}

@ARTICLE{Campana2006,
       author = {{Campana}, S. and {Mangano}, V. and {Blustin}, A.~J. and {Brown}, P. and {Burrows}, D.~N. and {Chincarini}, G. and {Cummings}, J.~R. and {Cusumano}, G. and {Della Valle}, M. and {Malesani}, D. and {M{\'e}sz{\'a}ros}, P. and {Nousek}, J.~A. and {Page}, M. and {Sakamoto}, T. and {Waxman}, E. and {Zhang}, B. and {Dai}, Z.~G. and {Gehrels}, N. and {Immler}, S. and {Marshall}, F.~E. and {Mason}, K.~O. and {Moretti}, A. and {O'Brien}, P.~T. and {Osborne}, J.~P. and {Page}, K.~L. and {Romano}, P. and {Roming}, P.~W.~A. and {Tagliaferri}, G. and {Cominsky}, L.~R. and {Giommi}, P. and {Godet}, O. and {Kennea}, J.~A. and {Krimm}, H. and {Angelini}, L. and {Barthelmy}, S.~D. and {Boyd}, P.~T. and {Palmer}, D.~M. and {Wells}, A.~A. and {White}, N.~E.},
        title = "{The association of GRB 060218 with a supernova and the evolution of the shock wave}",
      journal = {\nat},
         year = 2006,
        month = aug,
       volume = {442},
       number = {7106},
        pages = {1008-1010},
          doi = {10.1038/nature04892},
archivePrefix = {arXiv},
       eprint = {astro-ph/0603279},
 primaryClass = {astro-ph},
       adsurl = {https://ui.adsabs.harvard.edu/abs/2006Natur.442.1008C}
}

@ARTICLE{Katz2010,
       author = {{Katz}, Boaz and {Budnik}, Ran and {Waxman}, Eli},
        title = "{Fast Radiation Mediated Shocks and Supernova Shock Breakouts}",
      journal = {\apj},
         year = 2010,
        month = jun,
       volume = {716},
       number = {1},
        pages = {781-791},
          doi = {10.1088/0004-637X/716/1/781},
archivePrefix = {arXiv},
       eprint = {0902.4708},
 primaryClass = {astro-ph.HE},
       adsurl = {https://ui.adsabs.harvard.edu/abs/2010ApJ...716..781K}
}

@ARTICLE{Alp2020Blast,
       author = {{Alp}, Dennis and {Larsson}, Josefin},
        title = "{Blasts from the Past: Supernova Shock Breakouts among X-Ray Transients in the XMM-Newton Archive}",
      journal = {\apj},
         year = 2020,
        month = jun,
       volume = {896},
       number = {1},
          eid = {39},
        pages = {39},
          doi = {10.3847/1538-4357/ab91ba},
archivePrefix = {arXiv},
       eprint = {2004.09519},
 primaryClass = {astro-ph.HE},
       adsurl = {https://ui.adsabs.harvard.edu/abs/2020ApJ...896...39A}
}

@ARTICLE{Eappachen2024,
       author = {{Eappachen}, D. and {Jonker}, P.~G. and {Quirola-V{\'a}squez}, J. and {Mata S{\'a}nchez}, D. and {Inkenhaag}, A. and {Levan}, A.~J. and {Fraser}, M. and {Torres}, M.~A.~P. and {Bauer}, F.~E. and {Chrimes}, A.~A. and {Stern}, D. and {Graham}, M.~J. and {Smartt}, S.~J. and {Smith}, K.~W. and {Ravasio}, M.~E. and {Zabludoff}, A.~I. and {Yue}, M. and {Stoppa}, F. and {Malesani}, D.~B. and {Stone}, N.~C. and {Wen}, S.},
        title = "{XMM-Newton-discovered Fast X-ray Transients: host galaxies and limits on contemporaneous detections of optical counterparts}",
      journal = {\mnras},
         year = 2024,
        month = feb,
       volume = {527},
       number = {4},
        pages = {11823-11839},
          doi = {10.1093/mnras/stad3924},
archivePrefix = {arXiv},
       eprint = {2312.10786},
 primaryClass = {astro-ph.HE},
       adsurl = {https://ui.adsabs.harvard.edu/abs/2024MNRAS.52711823E}
}

@ARTICLE{Chevalier1989,
       author = {{Chevalier}, Roger A. and {Soker}, Noam},
        title = "{Asymmetric Envelope Expansion of Supernova 1987A}",
      journal = {\apj},
         year = 1989,
        month = jun,
       volume = {341},
        pages = {867},
          doi = {10.1086/167545},
       adsurl = {https://ui.adsabs.harvard.edu/abs/1989ApJ...341..867C}
}

@ARTICLE{Piran2013MNRAS.430.2121P,
       author = {{Piran}, Tsvi and {Nakar}, Ehud and {Rosswog}, Stephan},
        title = "{The electromagnetic signals of compact binary mergers}",
      journal = {\mnras},
         year = 2013,
        month = apr,
       volume = {430},
       number = {3},
        pages = {2121-2136},
          doi = {10.1093/mnras/stt037},
archivePrefix = {arXiv},
       eprint = {1204.6242},
 primaryClass = {astro-ph.HE},
       adsurl = {https://ui.adsabs.harvard.edu/abs/2013MNRAS.430.2121P}
}

@ARTICLE{Yan2018ApJ...858...91Y,
       author = {{Yan}, Lin and {Perley}, D.~A. and {De Cia}, A. and {Quimby}, R. and {Lunnan}, R. and {Rubin}, Kate H.~R. and {Brown}, P.~J.},
        title = "{Far-UV HST  Spectroscopy of an Unusual Hydrogen-poor Superluminous Supernova: SN2017egm}",
      journal = {\apj},
         year = 2018,
        month = may,
       volume = {858},
       number = {2},
          eid = {91},
        pages = {91},
          doi = {10.3847/1538-4357/aabad5},
archivePrefix = {arXiv},
       eprint = {1711.01534},
 primaryClass = {astro-ph.HE},
       adsurl = {https://ui.adsabs.harvard.edu/abs/2018ApJ...858...91Y}
}

@ARTICLE{Li2016ApJ...819..120L,
       author = {{Li}, Shao-Ze and {Yu}, Yun-Wei},
        title = "{Shock Breakout Driven by the Remnant of a Neutron Star Binary Merger: An X-Ray Precursor of Mergernova Emission}",
      journal = {\apj},
         year = 2016,
        month = mar,
       volume = {819},
       number = {2},
          eid = {120},
        pages = {120},
          doi = {10.3847/0004-637X/819/2/120},
archivePrefix = {arXiv},
       eprint = {1511.01229},
 primaryClass = {astro-ph.HE},
       adsurl = {https://ui.adsabs.harvard.edu/abs/2016ApJ...819..120L}
}

@ARTICLE{Kasen2010ApJ...717..245K,
       author = {{Kasen}, Daniel and {Bildsten}, Lars},
        title = "{Supernova Light Curves Powered by Young Magnetars}",
      journal = {\apj},
         year = 2010,
        month = jul,
       volume = {717},
       number = {1},
        pages = {245-249},
          doi = {10.1088/0004-637X/717/1/245},
archivePrefix = {arXiv},
       eprint = {0911.0680},
 primaryClass = {astro-ph.HE},
       adsurl = {https://ui.adsabs.harvard.edu/abs/2010ApJ...717..245K}
}

@ARTICLE{Nakar2010,
       author = {{Nakar}, Ehud and {Sari}, Re'em},
        title = "{Early Supernovae Light Curves Following the Shock Breakout}",
      journal = {\apj},
         year = 2010,
        month = dec,
       volume = {725},
       number = {1},
        pages = {904-921},
          doi = {10.1088/0004-637X/725/1/904},
archivePrefix = {arXiv},
       eprint = {1004.2496},
 primaryClass = {astro-ph.HE},
       adsurl = {https://ui.adsabs.harvard.edu/abs/2010ApJ...725..904N}
}

@ARTICLE{Wang2008,
       author = {{Wang}, Xiaofeng and {Li}, Weidong and {Filippenko}, Alexei V. and {Krisciunas}, Kevin and {Suntzeff}, Nicholas B. and {Li}, Junzheng and {Zhang}, Tianmeng and {Deng}, Jingsong and {Foley}, Ryan J. and {Ganeshalingam}, Mohan and {Li}, Tipei and {Lou}, YuQing and {Qiu}, Yulei and {Shang}, Rencheng and {Silverman}, Jeffrey M. and {Zhang}, Shuangnan and {Zhang}, Youhong},
        title = "{Optical and Near-Infrared Observations of the Highly Reddened, Rapidly Expanding Type Ia Supernova SN 2006X in M100}",
      journal = {\apj},
         year = 2008,
        month = mar,
       volume = {675},
       number = {1},
        pages = {626-643},
          doi = {10.1086/526413},
archivePrefix = {arXiv},
       eprint = {0708.0140},
 primaryClass = {astro-ph},
       adsurl = {https://ui.adsabs.harvard.edu/abs/2008ApJ...675..626W}
}

@ARTICLE{Schwarz1978,
       author = {{Schwarz}, Gideon},
        title = "{Estimating the Dimension of a Model}",
      journal = {Annals of Statistics},
         year = 1978,
        month = jul,
       volume = {6},
       number = {2},
        pages = {461-464},
       adsurl = {https://ui.adsabs.harvard.edu/abs/1978AnSta...6..461S}
}

@ARTICLE{Cash1979,
       author = {{Cash}, W.},
        title = "{Parameter estimation in astronomy through application of the likelihood ratio.}",
      journal = {\apj},
         year = 1979,
        month = mar,
       volume = {228},
        pages = {939-947},
          doi = {10.1086/156922},
       adsurl = {https://ui.adsabs.harvard.edu/abs/1979ApJ...228..939C}
}

@ARTICLE{Willingale2013,
       author = {{Willingale}, R. and {Starling}, R.~L.~C. and {Beardmore}, A.~P. and {Tanvir}, N.~R. and {O'Brien}, P.~T.},
        title = "{Calibration of X-ray absorption in our Galaxy}",
      journal = {\mnras},
         year = 2013,
        month = may,
       volume = {431},
       number = {1},
        pages = {394-404},
          doi = {10.1093/mnras/stt175},
       adsurl = {https://ui.adsabs.harvard.edu/abs/2013MNRAS.431..394W}
}

@ARTICLE{FXT2025,
       author = {{Chen}, Yong and {Wang}, Juan and {Yang}, Yanji and {Zhao}, Xiaofan and {Ma}, Jia and {Wang}, Yusa and {Han}, Dawei and {Cui}, Weiwei and {Ban}, Hanyu and {Bi}, Xiyan and {Bianucci}, Giovanni and {Burwitz}, Vadim and {Cai}, Hongbo and {Cai}, Zhiming and {Chen}, Houlei and {Chen}, Tianxiang and {Chen}, Yehai and {Cheng}, Junsheng and {Ding}, Fei and {Eder}, Josef and {Feng}, Hua and {Feng}, Jianchao and {Friedrich}, Peter and {Geng}, Haoyang and {Gao}, Min and {Gao}, Na and {Guan}, Ju and {Hou}, Dongjie and {Huo}, Jia and {Jia}, Shumei and {Keereman}, Arnoud and {Li}, Chengkui and {Li}, Fei and {Li}, Gang and {Li}, Maoshun and {Li}, Shuo and {Li}, Wei and {Li}, Xin and {Liao}, Qiuyan and {Liu}, Congzhan and {Liu}, Huaqiu and {Liu}, Yao and {Lu}, Fangjun and {Luo}, Laidan and {Meidinger}, Norbert and {Qiang}, Pengfei and {Santovincenzo}, Andrea and {Sheng}, Lizhi and {Sun}, Wenxin and {Tang}, Qingjun and {Tian}, Jinxin and {Valsecchi}, Giuseppe and {Vernani}, Dervis and {Wang}, Bo and {Wang}, Dianlong and {Wang}, Hao and {Wang}, Langping and {Wang}, Lei and {Wang}, Zequn and {Xie}, Dong and {Xu}, Jingjing and {Xue}, Jiadai and {Yan}, Yongqing and {Yang}, Xiongtao and {Yu}, Ke and {Yuan}, Weimin and {Zhang}, Juan and {Zhang}, Qian and {Zhang}, Shuangnan and {Zhang}, Xiaofeng and {Zhang}, Yifan and {Zhang}, Yonghe and {Zhang}, Ziliang and {Zhao}, Haisheng and {Zhao}, Zijian and {Zhu}, Yuxuan},
        title = "{Design and development of the follow-up X-ray telescope onboard Einstein Probe in China: a review}",
      journal = {Radiation Detection Technology and Methods},
         year = 2025,
        month = jun,
       volume = {9},
       number = {2},
        pages = {198-207},
          doi = {10.1007/s41605-025-00558-0},
       adsurl = {https://ui.adsabs.harvard.edu/abs/2025RDTM....9..198C}
}

@ARTICLE{Cheng2025,
       author = {{Cheng}, Huaqing and {Zhang}, Chen and {Ling}, Zhixing and {Sun}, Xiaojin and {Sun}, Shengli and {Liu}, Yuan and {Dai}, Yanfeng and {Jia}, Zhenqing and {Pan}, Haiwu and {Wang}, Wenxin and {Zhao}, Donghua and {Chen}, Yifan and {Cheng}, Zhiwei and {Fu}, Wei and {Han}, Yixiao and {Li}, Junfei and {Li}, Zhengda and {Ma}, Xiaohao and {Xue}, Yulong and {Yan}, Ailiang and {Zhang}, Qiang and {Wang}, Yusa and {Yang}, Xiongtao and {Zhao}, Zijian and {Li}, Longhui and {Jin}, Ge and {Yuan}, Weimin},
        title = "{Ground calibration result of the wide-field X-ray telescope (WXT) onboard the Einstein probe}",
      journal = {Experimental Astronomy},
         year = 2025,
        month = oct,
       volume = {60},
       number = {2},
          eid = {15},
        pages = {15},
          doi = {10.1007/s10686-025-10025-9},
archivePrefix = {arXiv},
       eprint = {2505.18939},
 primaryClass = {astro-ph.IM},
       adsurl = {https://ui.adsabs.harvard.edu/abs/2025ExA....60...15C}
}

@ARTICLE{cheng2024ExA,
       author = {{Cheng}, Huaqing and {Ling}, Zhixing and {Zhang}, Chen and {Sun}, Xiaojin and {Sun}, Shengli and {Liu}, Yuan and {Dai}, Yanfeng and {Jia}, Zhenqing and {Pan}, Haiwu and {Wang}, Wenxin and {Zhao}, Donghua and {Chen}, Yifan and {Cheng}, Zhiwei and {Fu}, Wei and {Han}, Yixiao and {Li}, Junfei and {Li}, Zhengda and {Ma}, Xiaohao and {Xue}, Yulong and {Yan}, Ailiang and {Zhang}, Qiang and {Wang}, Yusa and {Yang}, Xiongtao and {Zhao}, Zijian and {Yuan}, Weimin},
        title = "{Ground calibration result of the Lobster Eye Imager for Astronomy}",
      journal = {Experimental Astronomy},
         year = 2024,
        month = mar,
       volume = {57},
       number = {2},
          eid = {10},
        pages = {10},
          doi = {10.1007/s10686-024-09932-0},
archivePrefix = {arXiv},
       eprint = {2312.06964},
 primaryClass = {astro-ph.IM},
       adsurl = {https://ui.adsabs.harvard.edu/abs/2024ExA....57...10C}
}

@ARTICLE{Granot2002,
       author = {{Granot}, Jonathan and {Sari}, Re'em},
        title = "{The Shape of Spectral Breaks in Gamma-Ray Burst Afterglows}",
      journal = {\apj},
         year = 2002,
        month = apr,
       volume = {568},
       number = {2},
        pages = {820-829},
          doi = {10.1086/338966},
archivePrefix = {arXiv},
       eprint = {astro-ph/0108027},
 primaryClass = {astro-ph},
       adsurl = {https://ui.adsabs.harvard.edu/abs/2002ApJ...568..820G}
}

@ARTICLE{Chevalier1998,
       author = {{Chevalier}, Roger A.},
        title = "{Synchrotron Self-Absorption in Radio Supernovae}",
      journal = {\apj},
         year = 1998,
        month = may,
       volume = {499},
       number = {2},
        pages = {810-819},
          doi = {10.1086/305676},
       adsurl = {https://ui.adsabs.harvard.edu/abs/1998ApJ...499..810C}
}

@ARTICLE{Gehrels1986ApJ,
       author = {{Gehrels}, N.},
        title = "{Confidence Limits for Small Numbers of Events in Astrophysical Data}",
      journal = {\apj},
         year = 1986,
        month = apr,
       volume = {303},
        pages = {336},
          doi = {10.1086/164079},
       adsurl = {https://ui.adsabs.harvard.edu/abs/1986ApJ...303..336G}
}

@ARTICLE{Yuksel2008ApJ,
       author = {{Y{\"u}ksel}, Hasan and {Kistler}, Matthew D. and {Beacom}, John F. and {Hopkins}, Andrew M.},
        title = "{Revealing the High-Redshift Star Formation Rate with Gamma-Ray Bursts}",
      journal = {\apjl},
         year = 2008,
        month = aug,
       volume = {683},
       number = {1},
        pages = {L5},
          doi = {10.1086/591449},
       adsurl = {https://ui.adsabs.harvard.edu/abs/2008ApJ...683L...5Y}
}

@ARTICLE{Nicholl2017,
       author = {{Nicholl}, Matt and {Guillochon}, James and {Berger}, Edo},
        title = "{The Magnetar Model for Type I Superluminous Supernovae. I. Bayesian Analysis of the Full Multicolor Light-curve Sample with MOSFiT}",
      journal = {\apj},
         year = 2017,
        month = nov,
       volume = {850},
       number = {1},
          eid = {55},
        pages = {55},
          doi = {10.3847/1538-4357/aa9334},
archivePrefix = {arXiv},
       eprint = {1706.00825},
 primaryClass = {astro-ph.HE},
       adsurl = {https://ui.adsabs.harvard.edu/abs/2017ApJ...850...55N}
}

@ARTICLE{Weiler2002,
       author = {{Weiler}, Kurt W. and {Panagia}, Nino and {Montes}, Marcos J. and {Sramek}, Richard A.},
        title = "{Radio Emission from Supernovae and Gamma-Ray Bursters}",
      journal = {\araa},
         year = 2002,
        month = jan,
       volume = {40},
        pages = {387-438},
          doi = {10.1146/annurev.astro.40.060401.093744},
       adsurl = {https://ui.adsabs.harvard.edu/abs/2002ARA&A..40..387W}
}

@ARTICLE{Piro2021,
       author = {{Piro}, Anthony L. and {Haynie}, Annastasia and {Yao}, Yuhan},
        title = "{Shock Cooling Emission from Extended Material Revisited}",
      journal = {\apj},
         year = 2021,
        month = mar,
       volume = {909},
       number = {2},
          eid = {209},
        pages = {209},
          doi = {10.3847/1538-4357/abe2b1},
archivePrefix = {arXiv},
       eprint = {2007.08543},
 primaryClass = {astro-ph.HE},
       adsurl = {https://ui.adsabs.harvard.edu/abs/2021ApJ...909..209P}
}

@ARTICLE{Zhu2025,
       author = {{Zhu}, Jin-Ping and {Zheng}, Jian-He and {Zhang}, Bing},
        title = "{EP 250108a/SN 2025kg: a magnetar-powered gamma-ray burst supernova originating from a close helium-star binary via isolated binary evolution}",
      journal = {\mnras},
         year = 2025,
        month = nov,
       volume = {544},
       number = {1},
        pages = {L139-L149},
          doi = {10.1093/mnrasl/slaf114},
archivePrefix = {arXiv},
       eprint = {2507.18544},
 primaryClass = {astro-ph.HE},
       adsurl = {https://ui.adsabs.harvard.edu/abs/2025MNRAS.544L.139Z}
}

@ARTICLE{Oke+1990,
       author = {{Oke}, J.~B.},
        title = "{Faint Spectrophotometric Standard Stars}",
      journal = {\aj},
         year = 1990,
        month = may,
       volume = {99},
        pages = {1621},
          doi = {10.1086/115444},
       adsurl = {https://ui.adsabs.harvard.edu/abs/1990AJ.....99.1621O}
}

@ARTICLE{Fan+2016,
       author = {{Fan}, Zhou and {Wang}, Huijuan and {Jiang}, Xiaojun and {Wu}, Hong and {Li}, Hongbin and {Huang}, Yang and {Xu}, Dawei and {Hu}, Zhongwen and {Zhu}, Yinan and {Wang}, Jianfeng and {Komossa}, Stefanie and {Zhang}, Xiaoming},
        title = "{The Xinglong 2.16-m Telescope: Current Instruments and Scientific Projects}",
      journal = {\pasp},
         year = 2016,
        month = nov,
       volume = {128},
       number = {969},
        pages = {115005},
          doi = {10.1088/1538-3873/128/969/115005},
archivePrefix = {arXiv},
       eprint = {1605.09166},
 primaryClass = {astro-ph.IM},
       adsurl = {https://ui.adsabs.harvard.edu/abs/2016PASP..128k5005F}
}

@ARTICLE{Appenzeller+1998,
       author = {{Appenzeller}, I. and {Fricke}, K. and {F{\"u}rtig}, W. and {G{\"a}ssler}, W. and {H{\"a}fner}, R. and {Harke}, R. and {Hess}, H.-J. and {Hummel}, W. and {J{\"u}rgens}, P. and {Kudritzki}, R.-P. and {Mantel}, K.-H. and {Meisl}, W. and {Muschielok}, B. and {Nicklas}, H. and {Rupprecht}, G. and {Seifert}, W. and {Stahl}, O. and {Szeifert}, T. and {Tarantik}, K.},
        title = "{Successful commissioning of FORS1 - the first optical instrument on the VLT.}",
      journal = {The Messenger},
         year = 1998,
        month = dec,
       volume = {94},
        pages = {1-6},
       adsurl = {https://ui.adsabs.harvard.edu/abs/1998Msngr..94....1A}
}

@ARTICLE{GCN44227,
       author = {{Gianfagna}, G. and {Balasubramanian}, A. and {Bruni}, G. and {Eappachen}, D. and {Piro}, L. and {Thakur}, A.~L. and {Anupama}, G.~C. and {Sahu}, D.~K. and {Beri}, Aru and {Bhalerao}, Varun and et al.},
        title = "{EP260321a: uGMRT observations}",
      journal = {GRB Coordinates Network},
         year = 2026,
        month = apr,
       volume = {44227},
        pages = {1},
       adsurl = {https://ui.adsabs.harvard.edu/abs/2026GCN.44227....1G}
}

@ARTICLE{GCN44229,
       author = {{Leung}, J.~K. and {Izzo}, L. and {de Colle}, F. and {Drout}, M.~R.},
        title = "{EP260321a: Upper limits from VLA radio observations}",
      journal = {GRB Coordinates Network},
         year = 2026,
        month = apr,
       volume = {44229},
        pages = {1},
       adsurl = {https://ui.adsabs.harvard.edu/abs/2026GCN.44229....1L}
}

@ARTICLE{GCN44239,
       author = {{O'Dwyer}, Tanner and {Corsi}, Alessandra and {Srinivasaragavan}, Gokul and {Cenko}, S. Bradley and {Anand}, Shreya and {Ho}, Anna and {Sollerman}, Jesper and {Zhou}, Bei and {Kamionkowski}, Marc and {Rastinejad}, Jillian},
        title = "{EP260321a: VLA radio detection}",
      journal = {GRB Coordinates Network},
         year = 2026,
        month = apr,
       volume = {44239},
        pages = {1},
       adsurl = {https://ui.adsabs.harvard.edu/abs/2026GCN.44239....1O}
}

@ARTICLE{GCN44357,
       author = {{O'Dwyer}, Tanner and {Corsi}, Alessandra and {Srinivasaragavan}, Gokul and {Cenko}, S. Bradley and {Anand}, Shreya and {Ho}, Anna and {Sollerman}, Jesper and {Zhou}, Bei and {Kamionkowski}, Marc and {Rastinejad}, Jillian},
        title = "{EP260321a: VLA radio follow-up}",
      journal = {GRB Coordinates Network},
         year = 2026,
        month = apr,
       volume = {44357},
        pages = {1},
       adsurl = {https://ui.adsabs.harvard.edu/abs/2026GCN.44357....1O}
}

@ARTICLE{GCN44403,
       author = {{Nayana}, A.J. and {Wiston}, Eli and {Margutti}, Raffaella and {Sfaradi}, Itai and {Chornock}, Ryan },
        title = "{EP260321a: Upper limits from ATCA radio observations}",
      journal = {GRB Coordinates Network},
         year = 2026,
        month = apr,
       volume = {44403},
        pages = {1},
       adsurl = {https://ui.adsabs.harvard.edu/abs/2026GCN.44403....1N}
}

@software{2020ascl.soft09003H,
       author = {{Heywood}, Ian},
        title = "{oxkat: Semi-automated imaging of MeerKAT observations}",
 howpublished = {Astrophysics Source Code Library, record ascl:2009.003},
         year = 2020,
        month = sep,
          eid = {ascl:2009.003},
archivePrefix = {ascl},
       eprint = {2009.003},
       adsurl = {https://ui.adsabs.harvard.edu/abs/2020ascl.soft09003H}
}

@ARTICLE{Yang2024,
       author = {{Yang}, Yuan-Pei and {Liu}, Xiangkun and {Pan}, Yu and {Er}, Xinzhong and {Liu}, Dezi and {Fang}, Yuan and {Du}, Guowang and {Cai}, Yongzhi and {Xu}, Xian and {Chen}, Xinlei and {Zou}, Xingzhu and {Guo}, Helong and {Liu}, Chenxu and {Cheng}, Yehao and {Kumar}, Brajesh and {Liu}, Xiaowei},
        title = "{Multiband Simultaneous Photometry of Type II SN 2023ixf with Mephisto and the Twin 50 cm Telescopes}",
      journal = {ApJL},
         year = 2024,
        month = jul,
       volume = {969},
       number = {2},
          eid = {126},
        pages = {126},
          doi = {10.3847/1538-4357/ad4be3},
       adsurl = {https://ui.adsabs.harvard.edu/abs/2024ApJ...969..126Y}
}

@ARTICLE{Chen2024,
       author = {{Chen}, Xinlei and {Kumar}, Brajesh and {Er}, Xinzhong and {Guo}, Helong and {Yang}, Yuan-Pei and {Lin}, Weikang and {Fang}, Yuan and {Du}, Guowang and {Liu}, Chenxu and {Zhao}, Jiewei and {Zhang}, Tianyu and {Bao}, Yuxi and {Zou}, Xingzhu and {Pan}, Yu and {Wang}, Yu and {Zhu}, Xufeng and {Chatterjee}, Kaushik and {Liu}, Xiangkun and {Liu}, Dezi and {Lagioia}, Edoardo P. and {Rangwal}, Geeta and {Zhong}, Shiyan and {Zhang}, Jinghua and {Lian}, Jianhui and {Cai}, Yongzhi and {Zhang}, Yangwei and {Liu}, Xiaowei},
        title = "{Early-phase Simultaneous Multiband Observations of the Type II Supernova SN 2024ggi with Mephisto}",
      journal = {ApJL},
         year = 2024,
        month = aug,
       volume = {971},
       number = {1},
          eid = {L2},
        pages = {L2},
          doi = {10.3847/2041-8213/ad62f7},
       adsurl = {https://ui.adsabs.harvard.edu/abs/2024ApJ...971L...2C}
}

@ARTICLE{Schlafly2011,
       author = {{Schlafly}, Edward F. and {Finkbeiner}, Douglas P.},
        title = "{Measuring Reddening with Sloan Digital Sky Survey Stellar Spectra and Recalibrating SFD}",
      journal = {\apj},
         year = 2011,
        month = aug,
       volume = {737},
       number = {2},
          eid = {103},
        pages = {103},
          doi = {10.1088/0004-637X/737/2/103},
       adsurl = {https://ui.adsabs.harvard.edu/abs/2011ApJ...737..103S}
}

@ARTICLE{Lang+2010,
       author = {{Lang}, Dustin and {Hogg}, David W. and {Mierle}, Keir and {Blanton}, Michael and {Roweis}, Sam},
        title = "{Astrometry.net: Blind Astrometric Calibration of Arbitrary Astronomical Images}",
      journal = {\aj},
         year = 2010,
        month = may,
       volume = {139},
       number = {5},
        pages = {1782-1800},
          doi = {10.1088/0004-6256/139/5/1782},
       adsurl = {https://ui.adsabs.harvard.edu/abs/2010AJ....139.1782L}
}

@INPROCEEDINGS{Bertin2006,
       author = {{Bertin}, E.},
        title = "{Automatic Astrometric and Photometric Calibration with SCAMP}",
    booktitle = {Astronomical Data Analysis Software and Systems XV},
         year = 2006,
       editor = {{Gabriel}, C. and {Arviset}, C. and {Ponz}, D. and {Enrique}, S.},
       series = {Astronomical Society of the Pacific Conference Series},
       volume = {351},
        month = jul,
        pages = {112},
       adsurl = {https://ui.adsabs.harvard.edu/abs/2006ASPC..351..112B}
}

@INPROCEEDINGS{Tody1986,
       author = {{Tody}, Doug},
        title = "{The IRAF Data Reduction and Analysis System}",
    booktitle = {Instrumentation in astronomy VI},
         year = 1986,
       editor = {{Crawford}, David L.},
       series = {Society of Photo-Optical Instrumentation Engineers (SPIE) Conference Series},
       volume = {627},
        month = jan,
        pages = {733},
          doi = {10.1117/12.968154},
       adsurl = {https://ui.adsabs.harvard.edu/abs/1986SPIE..627..733T}
}

@INPROCEEDINGS{1993ASPC...52..173T,
       author = {{Tody}, Doug},
        title = "{IRAF in the Nineties}",
    booktitle = {Astronomical Data Analysis Software and Systems II},
         year = 1993,
       editor = {{Hanisch}, R.~J. and {Brissenden}, R.~J.~V. and {Barnes}, J.},
       series = {Astronomical Society of the Pacific Conference Series},
       volume = {52},
        month = jan,
        pages = {173},
       adsurl = {https://ui.adsabs.harvard.edu/abs/1993ASPC...52..173T}
}

@ARTICLE{2006PASP..118..146P,
       author = {{Patat}, Ferdinando and {Romaniello}, Martino},
        title = "{Error Analysis for Dual-Beam Optical Linear Polarimetry}",
      journal = {\pasp},
         year = 2006,
        month = jan,
       volume = {118},
       number = {839},
        pages = {146-161},
          doi = {10.1086/497581},
archivePrefix = {arXiv},
       eprint = {astro-ph/0509153},
 primaryClass = {astro-ph},
       adsurl = {https://ui.adsabs.harvard.edu/abs/2006PASP..118..146P}
}

@ARTICLE{2017MNRAS.464.4146C,
       author = {{Cikota}, Aleksandar and {Patat}, Ferdinando and {Cikota}, Stefan and {Faran}, Tamar},
        title = "{Linear spectropolarimetry of polarimetric standard stars with VLT/FORS2}",
      journal = {\mnras},
         year = 2017,
        month = feb,
       volume = {464},
       number = {4},
        pages = {4146-4159},
          doi = {10.1093/mnras/stw2545},
archivePrefix = {arXiv},
       eprint = {1610.00722},
 primaryClass = {astro-ph.IM},
       adsurl = {https://ui.adsabs.harvard.edu/abs/2017MNRAS.464.4146C}
}

@ARTICLE{2020ApJ...902...46Y,
       author = {{Yang}, Yi and {Hoeflich}, Peter and {Baade}, Dietrich and {Maund}, Justyn R. and {Wang}, Lifan and {Brown}, Peter. J. and {Stevance}, Heloise F. and {Arcavi}, Iair and {Burke}, Jamison and {Cikota}, Aleksandar and {Clocchiatti}, Alejandro and {Gal-Yam}, Avishay and {Graham}, Melissa. L. and {Hiramatsu}, Daichi and {Hosseinzadeh}, Griffin and {Howell}, D. Andrew and {Jha}, Saurabh W. and {McCully}, Curtis and {Patat}, Ferdinando and {Sand}, David. J. and {Schulze}, Steve and {Spyromilio}, Jason and {Valenti}, Stefano and {Vink{\'o}}, J{\'o}zsef and {Wang}, Xiaofeng and {Wheeler}, J. Craig and {Yaron}, Ofer and {Zhang}, Jujia},
        title = "{The Young and Nearby Normal Type Ia Supernova 2018gv: UV-optical Observations and the Earliest Spectropolarimetry}",
      journal = {\apj},
         year = 2020,
        month = oct,
       volume = {902},
       number = {1},
          eid = {46},
        pages = {46},
          doi = {10.3847/1538-4357/aba759},
archivePrefix = {arXiv},
       eprint = {1903.10820},
 primaryClass = {astro-ph.HE},
       adsurl = {https://ui.adsabs.harvard.edu/abs/2020ApJ...902...46Y}
}

@software{Bertin2010,
       author = {{Bertin}, Emmanuel},
        title = "{SWarp: Resampling and Co-adding FITS Images Together}",
 howpublished = {Astrophysics Source Code Library, record ascl:1010.068},
         year = 2010,
        month = oct,
          eid = {ascl:1010.068},
archivePrefix = {ascl},
       eprint = {1010.068},
       adsurl = {https://ui.adsabs.harvard.edu/abs/2010ascl.soft10068B}
}

@ARTICLE{Bertin1996,
       author = {{Bertin}, E. and {Arnouts}, S.},
        title = "{SExtractor: Software for source extraction.}",
      journal = {\aaps},
         year = 1996,
        month = jun,
       volume = {117},
        pages = {393-404},
          doi = {10.1051/aas:1996164},
       adsurl = {https://ui.adsabs.harvard.edu/abs/1996A&AS..117..393B}
}

@ARTICLE{Huang2024ApJS,
       author = {{Huang}, Bowen and {Yuan}, Haibo and {Xiang}, Maosheng and {Huang}, Yang and {Xiao}, Kai and {Xu}, Shuai and {Zhang}, Ruoyi and {Yang}, Lin and {Niu}, Zexi and {Gu}, Hongrui},
        title = "{A Comprehensive Correction of the Gaia DR3 XP Spectra}",
      journal = {\apjs},
         year = 2024,
        month = mar,
       volume = {271},
       number = {1},
          eid = {13},
        pages = {13},
          doi = {10.3847/1538-4365/ad18b1},
       adsurl = {https://ui.adsabs.harvard.edu/abs/2024ApJS..271...13H}
}

@ARTICLE{Xiao2023ApJS,
       author = {{Xiao}, Kai and {Yuan}, Haibo and {L{\'o}pez-Sanjuan}, C. and {Huang}, Yang and {Huang}, Bowen and {Beers}, Timothy C. and {Xu}, Shuai and {Wang}, Yuanchang and {Yang}, Lin and {Alcaniz}, Jailson and {Galarza}, Carlos Andr{\'e}s and {Angulo De La Fuente}, Raul E. and {Cenarro}, A.~J. and {Crist{\'o}bal-Hornillos}, David and {Dupke}, Renato A. and {Ederoclite}, Alessandro and {Hern{\'a}ndez-Monteagudo}, Carlos and {Mar{\'\i}n-Franch}, Antonio and {Moles}, Mariano and {Sodr{\'e}}, Laerte and {V{\'a}zquez Rami{\'o}}, H{\'e}ctor and {Varela L{\'o}pez}, Jes{\'u}s},
        title = "{J-PLUS: Photometric Recalibration with the Stellar Color Regression Method and an Improved Gaia XP Synthetic Photometry Method}",
      journal = {\apjs},
         year = 2023,
        month = dec,
       volume = {269},
       number = {2},
          eid = {58},
        pages = {58},
          doi = {10.3847/1538-4365/ad0645},
       adsurl = {https://ui.adsabs.harvard.edu/abs/2023ApJS..269...58X}
}

@ARTICLE{Flewelling2020,
       author = {{Flewelling}, H.~A. and {Magnier}, E.~A. and {Chambers}, K.~C. and {Heasley}, J.~N. and {Holmberg}, C. and {Huber}, M.~E. and {Sweeney}, W. and {Waters}, C.~Z. and {Calamida}, A. and {Casertano}, S. and {Chen}, X. and {Farrow}, D. and {Hasinger}, G. and {Henderson}, R. and {Long}, K.~S. and {Metcalfe}, N. and {Narayan}, G. and {Nieto-Santisteban}, M.~A. and {Norberg}, P. and {Rest}, A. and {Saglia}, R.~P. and {Szalay}, A. and {Thakar}, A.~R. and {Tonry}, J.~L. and {Valenti}, J. and {Werner}, S. and {White}, R. and {Denneau}, L. and {Draper}, P.~W. and {Hodapp}, K.~W. and {Jedicke}, R. and {Kaiser}, N. and {Kudritzki}, R.~P. and {Price}, P.~A. and {Wainscoat}, R.~J. and {Chastel}, S. and {McLean}, B. and {Postman}, M. and {Shiao}, B.},
        title = "{The Pan-STARRS1 Database and Data Products}",
      journal = {\apjs},
         year = 2020,
        month = nov,
       volume = {251},
       number = {1},
          eid = {7},
        pages = {7},
          doi = {10.3847/1538-4365/abb82d},
       adsurl = {https://ui.adsabs.harvard.edu/abs/2020ApJS..251....7F}
}

@software{Becker2015,
       author = {{Becker}, Andrew},
        title = "{HOTPANTS: High Order Transform of PSF ANd Template Subtraction}",
 howpublished = {Astrophysics Source Code Library, record ascl:1504.004},
         year = 2015,
        month = apr,
          eid = {ascl:1504.004},
archivePrefix = {ascl},
       eprint = {1504.004},
       adsurl = {https://ui.adsabs.harvard.edu/abs/2015ascl.soft04004B}
}

@ARTICLE{Drake2009,
       author = {{Drake}, A.~J. and {Djorgovski}, S.~G. and {Mahabal}, A. and {Beshore}, E. and {Larson}, S. and {Graham}, M.~J. and {Williams}, R. and {Christensen}, E. and {Catelan}, M. and {Boattini}, A. and et al.},
        title = "{First Results from the Catalina Real-Time Transient Survey}",
      journal = {\apj},
         year = 2009,
        month = may,
       volume = {696},
       number = {1},
        pages = {870-884},
          doi = {10.1088/0004-637X/696/1/870},
archivePrefix = {arXiv},
       eprint = {0809.1394},
 primaryClass = {astro-ph},
       adsurl = {https://ui.adsabs.harvard.edu/abs/2009ApJ...696..870D}
}

@ARTICLE{Brown2013,
       author = {{Brown}, T.~M. and {Baliber}, N. and {Bianco}, F.~B. and {Bowman}, M. and {Burleson}, B. and {Conway}, P. and {Crellin}, M. and {Depagne}, {\'E}. and {De Vera}, J. and {Dilday}, B. and et al.},
        title = "{Las Cumbres Observatory Global Telescope Network}",
      journal = {\pasp},
         year = 2013,
        month = sep,
       volume = {125},
       number = {931},
        pages = {1031},
          doi = {10.1086/673168},
archivePrefix = {arXiv},
       eprint = {1305.2437},
 primaryClass = {astro-ph.IM},
       adsurl = {https://ui.adsabs.harvard.edu/abs/2013PASP..125.1031B}
}

@ARTICLE{Lawrence2007,
       author = {{Lawrence}, A. and {Warren}, S.~J. and {Almaini}, O. and {Edge}, A.~C. and {Hambly}, N.~C. and {Jameson}, R.~F. and {Lucas}, P. and {Casali}, M. and {Adamson}, A. and {Dye}, S. and et al.},
        title = "{The UKIRT Infrared Deep Sky Survey (UKIDSS)}",
      journal = {\mnras},
         year = 2007,
        month = aug,
       volume = {379},
       number = {4},
        pages = {1599-1617},
          doi = {10.1111/j.1365-2966.2007.12040.x},
archivePrefix = {arXiv},
       eprint = {astro-ph/0604426},
 primaryClass = {astro-ph},
       adsurl = {https://ui.adsabs.harvard.edu/abs/2007MNRAS.379.1599L}
}

@software{Bradley2016,
       author = {{Bradley}, Larry and {Sipocz}, Brigitta and {Robitaille}, Thomas and {Tollerud}, Erik and {Deil}, Christoph and {Vin{\'\i}cius}, Z{\`e} and {Barbary}, Kyle and {G{\"u}nther}, Hans Moritz and {Bostroem}, Azalee and {Droettboom}, Michael and et al.},
        title = "{Photutils: Photometry tools}",
 howpublished = {Astrophysics Source Code Library, record ascl:1609.011},
         year = 2016,
        month = sep,
          eid = {ascl:1609.011},
archivePrefix = {ascl},
       eprint = {1609.011},
       adsurl = {https://ui.adsabs.harvard.edu/abs/2016ascl.soft09011B}
}

@ARTICLE{Liu2025,
       author = {{Liu}, Liang-Duan and {Zhang}, Yu-Hao and {Yu}, Yun-Wei and {Du}, Ze-Xin and {Li}, Jing-Yao and {Wu}, Guang-Lei and {Dai}, Zi-Gao},
        title = "{TransFit: An Efficient Framework for Transient Light-curve Fitting with Time-dependent Radiative Diffusion}",
      journal = {\apj},
         year = 2025,
        month = oct,
       volume = {992},
       number = {1},
          eid = {20},
        pages = {20},
          doi = {10.3847/1538-4357/adfed6},
archivePrefix = {arXiv},
       eprint = {2505.13825},
 primaryClass = {astro-ph.HE},
       adsurl = {https://ui.adsabs.harvard.edu/abs/2025ApJ...992...20L}
}

@ARTICLE{Schmidt1968,
       author = {{Schmidt}, Maarten},
        title = "{Space Distribution and Luminosity Functions of Quasi-Stellar Radio Sources}",
      journal = {\apj},
         year = 1968,
        month = feb,
       volume = {151},
        pages = {393},
          doi = {10.1086/149446},
       adsurl = {https://ui.adsabs.harvard.edu/abs/1968ApJ...151..393S}
}

@ARTICLE{Sun2015ApJ,
       author = {{Sun}, Hui and {Zhang}, Bing and {Li}, Zhuo},
        title = "{Extragalactic High-energy Transients: Event Rate Densities and Luminosity Functions}",
      journal = {\apj},
         year = 2015,
        month = oct,
       volume = {812},
       number = {1},
          eid = {33},
        pages = {33},
          doi = {10.1088/0004-637X/812/1/33},
       adsurl = {https://ui.adsabs.harvard.edu/abs/2015ApJ...812...33S}
}

@ARTICLE{Sun2022ApJ,
       author = {{Sun}, Hui and {Liu}, He-Yang and {Pan}, Hai-Wu and {Liu}, Zhu and {Alp}, Dennis and {Hu}, Jingwei and {Li}, Zhuo and {Zhang}, Bing and {Yuan}, Weimin},
        title = "{Luminosity Function and Event Rate Density of XMM-Newton-selected Supernova Shock Breakout Candidates}",
      journal = {\apj},
         year = 2022,
        month = mar,
       volume = {927},
       number = {2},
          eid = {224},
        pages = {224},
          doi = {10.3847/1538-4357/ac5328},
       adsurl = {https://ui.adsabs.harvard.edu/abs/2022ApJ...927..224S}
}
\bibliographystyle{sciencemag}

%
%
%
%
%
%


\section*{Acknowledgments}
This work is based on data obtained with the Einstein Probe, a space mission supported by the Strategic Priority Program on Space Science of the Chinese Academy of Sciences, in collaboration with the European Space Agency, the Max-Planck-Institute for extraterrestrial Physics (Germany), and the Centre National d'Études Spatiales (France).
Based in part on observations made with the Nordic Optical Telescope, owned in collaboration by the University of Turku and Aarhus University, and operated jointly by Aarhus University, the University of Turku, and the University of Oslo (respectively representing Denmark, Finland, and Norway), the University of Iceland, and Stockholm University at the Observatorio del Roque de los Muchachos, La Palma, Spain, of the Instituto de Astrofisica de Canarias. The NOT data were obtained under program ID P72-811 and P73-504.
This work is based in part on observations obtained with the Multi-channel Photometric Survey Telescope (Mephisto), which is developed at and operated by the South Western Institute for Astronomy Research of Yunnan University (SWIFAR-YNU), funded by the “Yunnan University Development Plan for World-Class University” and the “Yunnan University Development Plan for World-Class Astronomy Discipline.”
The Wide Field Survey Telescope (WFST) is a joint facility of the University of Science and Technology of China, Purple Mountain Observatory.
Based on observations made with the Thai Robotic Telescopes under program ID TRTToO\_2025004, TRTToO\_2025005, and TRTC13B\_003, which is operated by the National Astronomical Research Institute of Thailand (Public Organization).
The Australia Telescope Compact Array is part of the Australia Telescope National Facility (https://ror.org/05qajvd42) which is funded by the Australian Government for operation as a National Faclity managed by CSIRO. We acknowledge the Gomero people as the Traditional Owners of the Observatory site.
The MeerkAT telescope is operated by the South African Radio Astronomy Observatory, which is a facility of the National Research Foundation, an agency of the Department of Science and Innovation. This work has made use of the ``MPIfRS-band receiver system" designed, constructed and maintained by funding of the MPI f\"{u}r Radio astronomy and the Max-Planck-Society.
The National Radio Astronomy Observatory and Green Bank Observatory are facilities of the U.S. National Science Foundation operated under cooperative agreement by Associated Universities, Inc.
We thank the staff of the GMRT that made these observations possible.  GMRT is run by the National Centre for Radio Astrophysics of the Tata  Institute of Fundamental Research. These observations were conducted under uGMRT joint ToO proposal 49\_076 (PI D. Eappachen) 49\_117 (PI  Giulia Gianfagna). G.C.A. thanks the Indian National Science Academy (INSA) for support under their Senior Scientist Programme.
We acknowledge the support of the staff of the Wuhan University 1\,m telescope. 
%
e-MERLIN is a National Facility operated by the University of Manchester at Jodrell Bank Observatory on behalf of STFC.
%
%

Observations with the SAO RAS telescopes are supported by the Ministry of Science and Higher Education of the Russian Federation.

We acknowledge the support of the staff of the Xinglong 2.16\,m and 80\,cm telescopes. This work was partially supported by National Astronomical Observatories, Chinese Academy of Sciences.
KAIT and its ongoing operation were made possible by donations from Sun Microsystems, Inc., the Hewlett-Packard Company, AutoScope Corporation, Lick Observatory, the U.S. National Science Foundation, the University of California, the Sylvia \& Jim Katzman Foundation, and the TABASGO Foundation.
A major upgrade of the Kast spectrograph on the Shane 3\,m telescope at Lick Observatory, led by Brad Holden, was made possible through gifts from the Heising-Simons Foundation,       
William and Marina Kast, and the University of California Observatories. 
Research at Lick Observatory is partially supported by a generous gift from Google. 
We appreciate the assistance of the staff at Lick Observatory.
A.V.F.’s research group at UC Berkeley acknowledges financial assistance from Gary and Cynthia Bengier, Clark and Sharon Winslow, Alan Eustace and Kathy Kwan (W.Z. is a Bengier-Winslow-Eustace Specialist in Astronomy), Timothy and Melissa Draper, Briggs and Kathleen Wood, Ellyn and Alan Seelenfreund (T.G.B. is Draper-Wood-Seelenfreund Specialist in Astronomy), and numerous other donors. 
\paragraph*{Funding:}
This work is supported by the National Key R\&D Program of China No. 2025YFF0511100, the National Natural Science Foundation of China (grant Nos. 12393811, 12303047, 12573049, the National Key Research and Development Program of China (No. 2024YFA1611600), Top Team Project (grant No.202305AT350002). 
JPZ and BZ acknowledge the Research Talent Hub for ITF project (RTH-ITF) (Project No. PiH/270/25GS) from the Innovation and Technology Commission of Hong Kong SAR. 
P.G.J.~is supported by the European Union (ERC, Starstruck, 101095973, PI Jonker). Views and opinions expressed are however those of the author(s) only and do not necessarily reflect those of the European Union or the European Research Council Executive Agency. Neither the European Union nor the granting authority can be held responsible for them.
Strategic Priority Research Program of the Chinese Academy of Sciences, Grant No. XDB0550100. 
TA acknowledge the Shanghai Oriental Talent Project.
GG, GB, LP and ALT acknowledge support by ASI (Italian Space Agency) through the Contract no. 2019-27-HH.0. 
\paragraph*{Author contributions:}
W.Y. has been serving as Principal Investigator of the Einstein Probe project since the mission proposal phase. W.Y., X.-F. W., and B.Z. initiated and coordinated this study and led the discussions. 

Y.L., H.-Q.C., and R.-D.L. processed and analysed the WXT data. Y.L., C.-K.L., Y.-H.Y., and R.-D.L. processed and analysed the FXT data. H.-Q.C., H.-W.P., and Y.L. contributed to the development of WXT data analysis software and its calibration. S.-M.J. and C.-K.L. contributed to the development of FXT data analysis software and its calibration. H.S. and J.-W.H. contributed to the estimate of the event rate. Q.-J.H., Z.-C.Z., D.-Y.L., and X.M. were the transient advocates on 21 March 2026, and contributed to the discovery and preliminary analysis of the event.

D.X. coordinated the optical data reduction and analysis. D. X., X. L., J. F., Z.-P. Z., K. C., S.-Y. F., L. B.-H., S.-Q. J., W.-X. L., K. N., S. T., and Z. F. contributed to the TRT, ALT, and NOT observations. G.-W. D., Q.-R. W., Y. F., X.-K. L., X.-W. L. and Y. P. contributed to the Mephisto observations. J.-A. J, M.-X. C, Z.-Q. J, and Z.-L. X. contributed to the WFST observations. W.-K.Z, A.F., and T.B. contributed KAIT and Shane observations. A.S.P., N.S.P., A.V.V., I.A.Z., A.S.M., E.V.K., M.V.E., O.S., V.G., and V.V. contributed to the Russian multiple telescopes. R.-D.L and X. F contributed to the WHUT observations. B.W., R.-Z. Li, and J.M. contributed to the GMG observations. J.-J.J., Z.F., H.W., H.-Y. M., Y. Z., and M.H. contributed to the Xinglong-2.16m and XL-80 observations. Y.Y. contributed to the VLT, KAIT, and Shane observations.
 
T.A. coordinated the radio observations. T.A., A.L. W., X.F.W., J.J.G., Y. Q.L., S.B. Z., X.Y. W, C.Y. D. contributed to the ATCA observations. A.-L.W., T.A., X.-F.W., J.-J.G., X.-Y.W., C.-Y.D., and Z.-W.L. contributed to the MeerKAT observations. G. G., L. P., G.B., and A.-L.T. contributed to the e-MERLIN observations. A.B., D.E., A.B., V.B., G.C. A., D.K.S., and A.P.S. contributed to the uGMRT observations. 

Y.-W.Y., J.-P.Z., and Z.L. led the theoretical investigation of the event. Y.-H.Z., G.-L.W., L.-D.L., K.-R.N., J.-S.Z., and B.W. contributed to the theoretical investigation of the event.

H-W. P, C. B, Z.-M. C, J.-X. H, J.-W. H, S.-M. J, Z.-X. L, S.-L. S, C. Z, X.-J. S, Y.-H. Z, Y.-H. C, C.-Z. C, J. G, D.-W. H, M.-H. H, G. J, C.-K. L, C. Z. L, H.-Q. L, L.-M. S, H.-Y. T, H.-T. X., and C.-B. X contributed to the hardware and software development and the science operations of the Einstein Probe mission. J.-N, K. N, S.-N. Z, B. C, H. F, C.-C. J, P. B, A. R, N. R, and J. S contributed to the science management and collaboration of the Einstein Probe project.

All authors joined various kinds of discussions at different stages of the work.

W.Y., Y.-W. Y., Y.L., D.X., T.A., Y.Y., J.-P.Z., L.-D. L., H.S., R.-D. L., X.-F. W., and B.Z drafted the manuscript with the help from all authors.
\paragraph*{Competing interests:}
The authors declare that they have no competing interests.
\paragraph*{Data and materials availability:}
The materials and data used in this paper will be made available after paper publication. Requests can be addressed to the corresponding authors. All EP data (including WXT and FXT) will be made publicly available after one-year proprietary period, at https://ep.bao.ac.cn. 


\subsection*{Supplementary materials}
Materials and Methods\\
Figures S1 to S4\\
Tables S1 to S5\\
References \textit{(51-\arabic{enumiv})}\\ 


\newpage


\renewcommand{\thefigure}{S\arabic{figure}}
\renewcommand{\thetable}{S\arabic{table}}
\renewcommand{\theequation}{S\arabic{equation}}
\renewcommand{\thepage}{S\arabic{page}}
\setcounter{figure}{0}
\setcounter{table}{0}
\setcounter{equation}{0}
\setcounter{page}{1} 


\begin{center}
\section*{Supplementary Materials for\\ \scititle}

\author{
    Weimin~Yuan$^{\ast\dagger}$,
    Qiu-Ju~Huang$^{\dagger}$,
    Jin-Ping~Zhu$^{\dagger}$\and
    Yun-Wei~Yu$^{\ast}$,
    Dong~Xu$^{\ast}$,
    Chen~Zhang$^{\ast}$\and
    Zhuo~Li,
    Yuan~Liu,
    Tao~An\and
    Giulia~Gianfagna,
    Weikang~Zheng,
    Guowang~Du\and
    Xing~Liu,
    Ji-An~Jiang,
    Johan~P.U.~Fynbo\and
    Alexei~S.~Pozanenko,
    Junjie~Jin,
    Yi~Yang\and
    Jinsong~Deng,
    Hui~Sun,
    Guang-Lei~Wu\and
    Yu-Hao~Zhang,
    Bao~Wang,
    Yu~Wang\and
    Xiang-Yu~Wang,
    Bin-Bin~Zhang,
    Yong~Chen\and
    Yonghe~Zhang,
    Bo~Wang,
    Xiaofeng~Wang\and
    Xuefeng~Wu,
    Zigao~Dai,
    Jie~An\and
    G.C.~Anupama,
    Arvind~Balasubramanian,
    Congying~Bao\and
    Aru~Beri,
    Varun~Bhalerao,
    Thomas~G.~Brink\and
    Gabriele~Bruni,
    Minxuan~Cai,
    Zhiming~Cai\and
    Krittapas~Chanchaiworawit,
    Yehai~Chen,
    Huaqing~Cheng\and
    Bertrand~Cordier,
    Chenzhou~Cui,
    Weiwei~Cui\and
    Cuiyuan~Dai,
    D.~Eappachen,
    M.~V.~Eselevich\and
    Xiao~Fan,
    Zhou~Fan,
    Yuan~Fang\and
    Hua~Feng,
    Alexei~V.~Filippenko,
    Shaoyu~Fu\and
    He~Gao,
    Jinjun~Geng,
    Vitaly~Goranskij\and
    Ju~Guan,
    Dawei~Han,
    Jinxin~Hao\and
    Linbo~He,
    Min~He,
    Jingwei~Hu\and
    Maohai~Huang,
    Shumei~Jia,
    Ziqing~Jia\and
    Shuaiqing~Jiang,
    Chichuan~Jin,
    Ge~Jin\and
    Peter~Jonker,
    E.~V.~Klunko,
    Albert~K.~H.~Kong\and
    Chengkui~Li,
    Dongyue~Li,
    Rui-Zhi~Li\and
    Wenxiong~Li,
    Run-Duo~Liang,
    Zhixing~Ling\and
    Congzhan~Liu,
    Huaqiu~Liu,
    Liangduan~Liu\and
    Xiangkun~Liu,
    Xiaowei~Liu,
    Yuanqi~Liu\and
    Zhengwei~Liu,
    Fangjun~Lu,
    Jirong~Mao\and
    Xuan~Mao,
    A.~S.~Moskvitin,
    Haiyang~Mu\and
    Kirpal~Nandra,
    Jan-Uwe~Ness,
    Kangrui~Ni\and
    Kanthanakorn~Noysena,
    Paul~O'Brien,
    Haiwu~Pan\and
    Yu~Pan,
    N.S.~Pankov,
    Luigi~Piro\and
    J.~Quirola-V{\'a}squez,
    Arne~Rau,
    Nanda~Rea\and
    D.K.~Sahu,
    Aditya~Pawan~Saikia,
    Jeremy~Sanders\and
    Liming~Song,
    Olga~Spiridonova,
    Ning-Chen~Sun\and
    Shengli~Sun,
    Xiaojin~Sun,
    Yuyin~Tan\and
    Aishwarya~Linesh~Thakur,
    Samaporn~Tinyanont,
    Valery~Vlasyuk\and
    A.V.~Volnova,
    Ailing~Wang,
    Hong~Wu\and
    Qianrui~Wu,
    Haitao~Xu,
    Zelin~Xu\and
    Changbin~Xue,
    Yi-Han~Iris~Yin,
    I.~A.~Zaznobin\and
    Jia-Sen~Zhang,
    Shuang-Nan~Zhang,
    Songbo~Zhang\and
    Yu~Zhang,
    Zipei~Zhu,
    Zecheng~Zou\and
    Bing~Zhang
}

\end{center}

\subsubsection*{This PDF file includes:}
Materials and Methods\\
Figures S1 to S4\\
Tables S1 to S5\\

\newpage


\subsection*{Materials and Methods}

\subsubsection*{Einstein Probe (EP) observations and data reduction}

EP260321a triggered the Wide-field X-ray Telescope (WXT) onboard processing unit at 2026-03-21T12:30:18 UTC during a survey observation (ObsID 11900654211) with a signal-to-noise ratio of 7.0. This source was also detected in the preceding survey observation (ObsID 11900654465) and the autonomous follow-up observation (ObsID 01709259023). Thus, WXT observations covered the entire onset and decay phase of this transient with a total exposure time of $\sim$3100 s.

The X-ray events were processed and calibrated using the dedicated data-reduction software and calibration database (CALDB) developed for the WXT instrument (Liu et al., in prep.). The CALDB was constructed based on the results of ground calibration experiments\cite{Cheng2025}, following procedures previously applied to a WXT prototype\cite{cheng2024ExA}. Photon positions were transformed into celestial coordinates in the J2000 reference frame. The energy of each event was determined using the bias and gain parameters provided in the CALDB. After identifying and flagging bad and flaring pixels, single-, double-, triple-, and quadruple-pixel events without anomalous flags were retained to produce the cleaned event file.

The 0.4--2.0 keV image was generated from the cleaned event list (Figure \ref{fig:Xraylc}). Source and background light curves, as well as time-resolved spectra, were extracted from a circular region with a radius of $9'$ and an annular region with inner and outer radii of $18'$ and $36'$, respectively. The peak count rate in the source region was only 0.01 counts per frame, indicating that the pile-up effect in the WXT data was negligible. 

An autonomous follow-up observation was carried out with the Follow-up X-ray Telescope (FXT)\cite{FXT2025} and lasted for two orbits with an exposure time of 4267 s. The FXT-A and FXT-B were working in the partial-window and full-frame modes, respectively. The Level 1 event files were processed using the FXT data-analysis software (v1.30) together with the corresponding CALDB (v1.30). The reduction procedure included particle-event identification, pulse-invariant (PI) conversion, grade assignment and selection (grade $\leq 12$), flagging of bad and hot pixels, and the screening of good time intervals based on housekeeping data. This pipeline produced cleaned event files. For FXT-A, source and background light curves, as well as time-resolved spectra, were extracted from a circular region with a radius of $1'$ and from two adjacent circular regions, each with a radius of $3'$, respectively. For FXT-B, the source region was an annulus with inner and outer radii of $20''$ and $1'$ to mitigate pile-up in the full-frame mode. The corresponding response files were then generated with the point-spread function correction.
 In addition to the autonomous follow-up observation, FXT conducted six observations to monitor the long-term X-ray behavior of EP260321a (Extended Data Table \ref{fxt_obs_log}). No significant detection was found in these observations.

\subsubsection*{X-ray spectral analysis}

Since there is no significant signal above 2.0 keV, the X-ray spectral analysis was performed over 0.4--2.0 keV and 0.3--2.0 keV for WXT and FXT in XSPEC, respectively. The spectra of FXT-A and FXT-B were generated separately and fitted simultaneously. The W-statistic  (\texttt{cstat} in XSPEC)\cite{Cash1979} was employed in deriving the best-fit spectral parameters. The integrated and time-resolved spectra were rebinned to at least 10 and 3 counts per bin, respectively.
The integrated spectra of the WXT ($[T_0, T_0+2200~\rm{s}]$) and FXT ($[T_0+1065~\rm{s}$, $ T_0+2359~\rm{s}]$, i.e., the first orbit of the autonomous follow-up observation) can be well described by an absorbed blackbody model \textit{tbabs*ztbabs*zbbody}, where the first and second components respectively account for the Galactic absorption $N_{\rm{H,\,Gal}}$ and the intrinsic absorption $N_{\rm H}$, and the third component is a blackbody function in the source rest frame. The column density of the Galactic absorption in the direction of EP260321a was fixed at \(2.64 \times 10^{20}\,\mathrm{cm}^{-2}\) (Ref. \cite{Willingale2013}) and the redshift was fixed at 0.0344. If the $N_{\rm H}$  of FXT is linked with that of WXT, the best-fit parameters are $N_{\rm H} = 8.4^{+1.6}_{-1.5} \times 10^{20}$\,$\rm cm^{-2}$, $kT_{\rm obs} = 124_{-6}^{+7}$ eV (WXT), and $112.7_{-2.1}^{+2.1}$ eV (FXT) with an acceptable statistic W-stat/(d.o.f.) $= 245.7/231$. The goodness of fit is not significantly improved with independent $N_{\rm H}$. If the integrated spectrum was fitted by an absorbed power-law model \textit{tbabs*ztbabs*zpowerlaw}, the  goodness of fit is much poorer with a statistic of W-stat/(d.o.f.) $= 272.2/231$, and the best-fit photon index ($\sim10$) is also unphysical (right panel of Figure  \ref{fig:Xrayspectra}). The difference of the Bayesian Information Criterion\cite{Schwarz1978} ($\Delta BIC=26.5$) indicates strong evidence against the absorbed power-law model. A cutoff power law could fit the integrated spectrum acceptably with a cutoff energy of $\sim$120 eV and a photon index of $\sim$$-2$, which effectively mimics a blackbody function. 

\begin{figure}[h]
\centering
\begin{overpic}[width=0.45\textwidth]{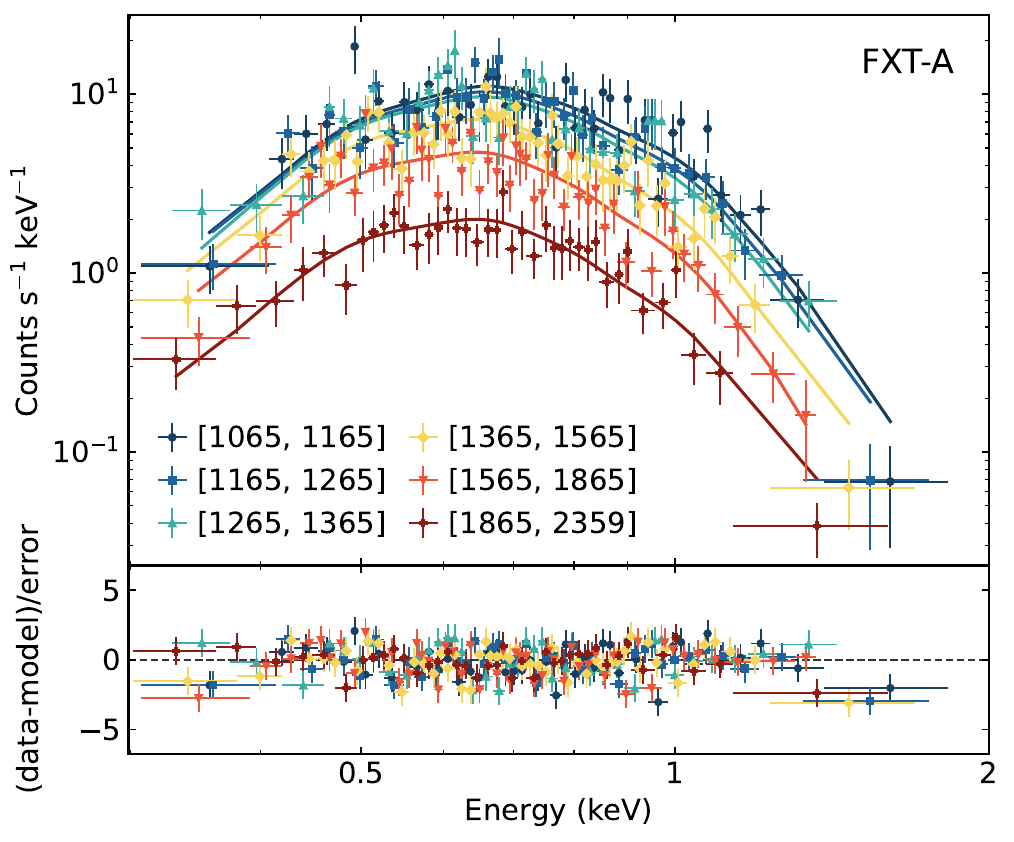}\put(-2,79){\textbf{(A)}} 
\end{overpic}
\begin{overpic}[width=0.45\textwidth]{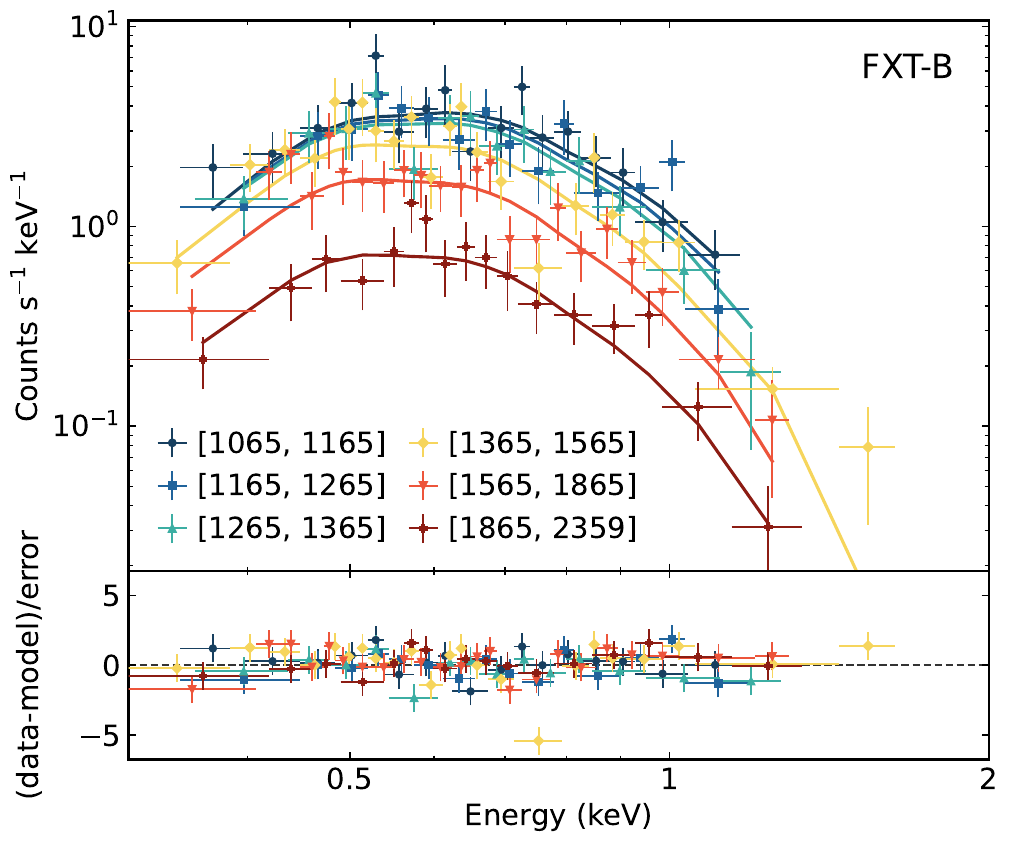}\put(-2,79){\textbf{(B)}} 
\end{overpic}
\begin{overpic}[width=0.45\textwidth]{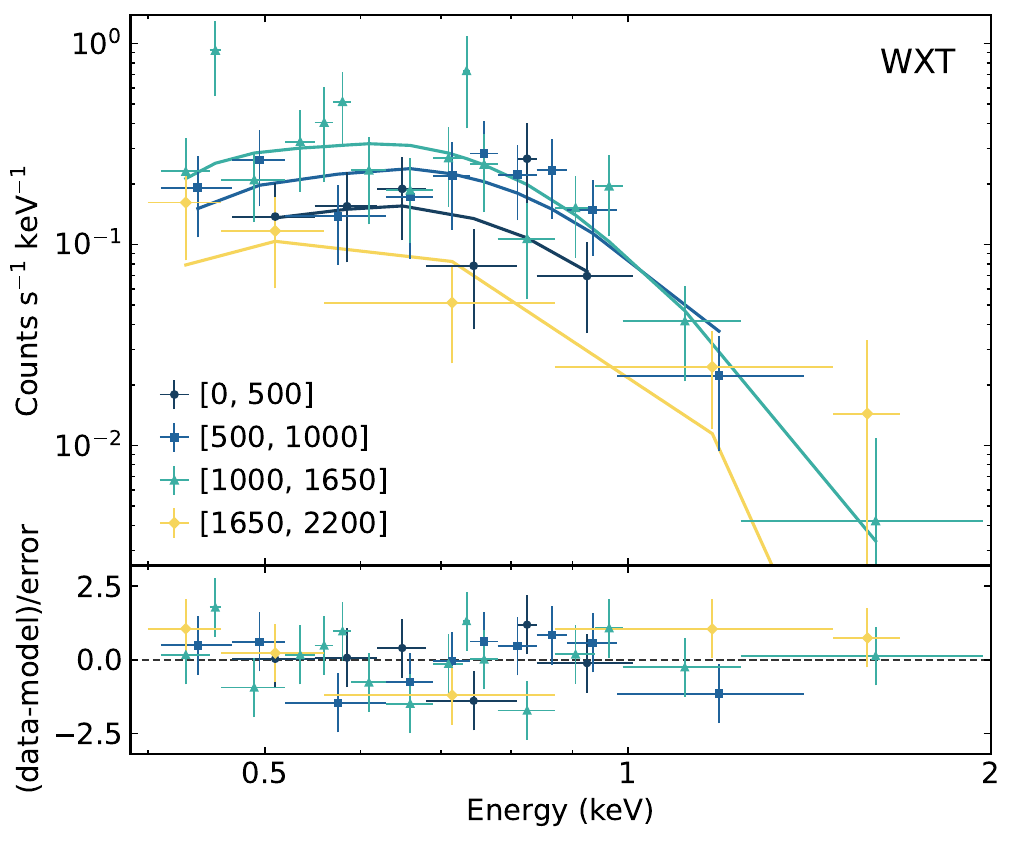}\put(-2,79){\textbf{(C)}} 
\end{overpic}
\caption{{\bf The evolution of the time-resolved WXT and FXT spectra.} The data points are shown with $1\sigma$ uncertainty and the lines represent the best-fit absorbed blackbody model. The data points of WXT {(A)} and FXT {(B, C)} spectra are rebinned to $2\sigma$ and $3\sigma$ significance, respectively, for clarity. The numbers in the square brackets of the legend represent the time intervals (in seconds) relative to $T_0 = \mathrm{2026\mbox{-}03\mbox{-}21T12{:}16{:}08}$ UTC. The best-fit spectral parameters are shown in Extended Table  \ref{ep_sliced_spec}.}
\label{fig:epspecseg}
\end{figure}

To investigate the possible spectral evolution, the WXT observation was divided into four segments; the FXT observation during the first orbit of the autonomous follow-up observation was divided into six segments of each FXT module (12 FXT segments in total). The resulting 16 spectra were simultaneously fitted by an absorbed blackbody model \textit{tbabs*ztbabs*zbbody}. The $N_{\rm H}$ values of all segments were linked assuming no significant variation of the intrinsic absorption. The blackbody temperatures of FXT-A and FXT-B in the same segment were also linked. The spectra of all segments are consistent with the absorbed blackbody model. The best-fit $N_{\rm H}$ is $7.9_{-1.6}^{+1.7} \times 10^{20}$\,$\rm cm^{-2}$. The blackbody temperature slightly decreased from $\sim$135 eV to $\sim$110 eV  (Extended Table  \ref{ep_sliced_spec}; Extended Data Figure \ref{fig:epspecseg}).

In the second orbit of the autonomous follow-up observation ($[T_0+5137~\rm{s},$ $T_0+8111~\rm{s}]$), EP260321a was still well detected with a significance of $\sim$5. The flux was estimated to be $(7.1\pm2.5)\times10^{-14}~{\rm erg~cm^{-2}~s^{-1}}$ in 0.4--2.0 keV by assuming an absorbed blackbody with $N_{\rm H} = 8 \times 10^{20}$\,$\rm cm^{-2}$ and $kT_{\rm obs}=100$ eV. Additionally, the $3\sigma$ upper limits for the nondetection in the other FXT observations were derived under the same spectral shape (Extended Table  \ref{fxt_obs_log}).

\subsubsection*{Optical and near-infrared photometric observations} 

Our ground-based optical and near-infrared photometric observations were conducted using multiple facilities and are still ongoing. The current results are summarized in Extended Table \ref{tab:phot}. The 1.6\,m Mephisto telescope began imaging the field 36 min after $T_0$, operating simultaneously in either the $ugi$ or $vrz$ bands\cite{Yang2024, Chen2024}. 
The Jinshan project commenced continuous monitoring in the $grizJ$ bands using the 1.0\,m 100A/100B/100C and 0.5\,m 50A telescopes starting at $T_0 + 1.8$ hr. The 0.76\,m Katzman Automatic Imaging Telescope (KAIT; Filippenko et al. 2001) observed the field in the $BVRI$ and {\it Clear} bands from $T_0$ + 16.5 hr. Further optical observations were carried out with the 2.5\,m Wide Field Survey Telescope (WFST) in the $ugr$ bands starting at $T_0$ + 2.04 days. Additional photometric data were obtained with the Tsinghua-NAOC 0.8\,m telescope (TNT)\cite{Wang2008}, the 0.7\,m telescope of the Thai Robotic Telescope network (TRT) located at the Spring Brook Remote Obsrevatory (SBO) and the Cerro Tololo Inter-American Observatory (CTO), the SAO RAS 1.0\,m Zeiss-1000 Optical Telescope (Zeiss-1000) and the 0.5\,m AS-500/2 telescope, the 2.56\,m Nordic Optical Telescope (NOT), the 2.6\,m Shajn Telescope (ZTSh) and the 50-cm Maksutov telescope (MTM-500) of the Crimean Astrophysical Observatory, and the 1.5\,m AZT-33IK telescope of the Mondy Observatory.  Along the line of sight of EP260321a/SN 2026gzf, the Galactic extinction is $E_{B-V} = 0.021$\, mag\cite{Schlafly2011}.

All photometric images were calibrated following standard procedures, including bias subtraction, dark-current subtraction (for CMOS detectors), and flat-field correction. We used \textit{astrometry.net}\cite{Lang+2010} and \texttt{SCAMP}\cite{Bertin2006} for astrometric calibration, \texttt{IRAF}\cite{Tody1986} and \texttt{SWarp}\cite{Bertin2010} for image stacking. Aperture photometry on the reduced images was carried out with \texttt{SExtractor}\cite{Bertin1996}. Mephisto $uvgriz$ photometry was calibrated against the synthetic photometry derived from the recalibrated Gaia BP/RP (XP) spectra\cite{Huang2024ApJS, Xiao2023ApJS}. The Sloan $griz$ (as well as $g'r'i'z'$) photometry was calibrated against nearby reference stars from the Pan-STARRS DR2 catalog in the AB magnitude system\cite{Flewelling2020}. For the Johnson-Cousins filters, the \textit{BVRcIc}-band photometry in the Vega system was calibrated using magnitudes converted from the Sloan system (\url{https://www.sdss.org/dr12/algorithms/sdssUBVRITransform/#Lupton}). To subtract the host contribution, we used \texttt{HOTPANTS}\cite{Becker2015} to perform image subtraction. As template images we used multiband stack images from PanSTARRS-1 (PS1) survey, unfiltered coadded images from the Catalina Sky Survey\cite{Drake2009}, $BV$-band archival images from the Las Cumbres Observatory (LCO) network\cite{Brown2013}, and a $J$-band archival image from the UKIRT Infrared Deep Sky Survey\cite{Lawrence2007} (UKIDSS). Images from Mephisto and WFST were subtracted using their own archival templates. \texttt{Photutils}\cite{Bradley2016} was then used to perform aperture photometry on the difference image to obtain the transient magnitude. 

\subsubsection*{Optical spectroscopic observations} 

To investigate the ejecta properties and chemical evolution of SN\,2026gzf, we conducted a multiepoch spectroscopic campaign spanning from $T_{0}$ + 2.7 days to $T_{0}$ + 56.4 days. 19 high-quality spectra were obtained, with their detailed observational parameters listed in the Extended Table~\ref{tab:opt_spec}. The resulting spectral sequence is illustrated in the left panel of Extended Data Figure~\ref{fig:lc_bolometric}, providing a comprehensive view of the transient's spectral transformation. All spectroscopic data were reduced using standard procedures within IRAF\cite{Tody1986, 1993ASPC...52..173T}, including bias subtraction, flat-fielding, and optimal extraction. Wavelength and flux calibrations were performed using daily HeNe comparison-lamp exposures and spectrophotometric standard stars, respectively.

Low-resolution optical spectra were acquired using the ALFOSC spectrograph on the 2.56\,m Nordic Optical Telescope (NOT) between 2026 March 26 and May 16; 13 reliable spectra were obtained by the following procedures. We employed Grism \#4 in conjunction with a $1.3''$ slit, yielding a typical spectral resolution of $R \equiv \lambda/\Delta \lambda \approx 300$ and a usable wavelength coverage of 3800--9200 \AA. We performed multiple successive exposures at each epoch, allowing for robust removal of cosmic rays during the image-stacking process. The spectroscopic data were reduced using standard IRAF-based pipelines, including bias subtraction, flat-field correction, and one-dimensional spectral extraction. Wavelength calibration was anchored to HeNe comparison-lamp spectra. Flux calibration was achieved using spectrophotometric standard stars (e.g., Feige 34 or HD 84937) observed during the same nights at similar airmasses, with their reference spectrum taken from the Oke catalog\cite{Oke+1990}.

Follow-up spectroscopy was also conducted using the Kast double spectrograph mounted on the 3\,m Shane telescope at Lick Observatory; 3 spectra were acquired between 2026 March 24 and 2026 March 27. The observations utilized a dual-beam configuration with the d55 dichroic mirror, typically employing a 600/4310 grating for the blue channel and a 300/7500 grating for the red channel. This setup provided broad and continuous wavelength coverage from $\sim 3500$ {\AA} to 10,500 \AA. The long $2''$-wide slit was aligned at the parallactic angle (Filippenko 1982) to minimize differential light losses caused by atmospheric dispersion. A complete log of the Kast observations, including exposure times and specific instrumental setups for each epoch, is detailed in Extended Table~\ref{tab:opt_spec}.

Additional optical spectra were obtained using the BFOSC instrument mounted on the Xinglong 2.16\,m telescope of the National Astronomical Observatories, Chinese Academy of Sciences (NAOC)\cite{Fan+2016}. These observations were carried out between 2026 March 26 and April 15, yielding two high-quality spectra. We utilized the Grism 4 (G4), which, combined with a $2.3''$ slit, provided  wavelength coverage of 3600–9000 \AA. A detailed journal of the BFOSC observations is included in Extended Table~\ref{tab:opt_spec}.

A high-signal-to-noise ratio optical spectrum was obtained on 2026 April 6 using the FORS2 (Focal Reducer and low dispersion Spectrograph 2) instrument mounted on the 8.2\,m Very Large Telescope (VLT) UT1 (Antu) at ESO's Paranal Observatory\cite{Appenzeller+1998}. We utilized the 300V grism in combination with a $1.0''$ slit, providing wavelength range of 3410--9330 \AA. 
Observations were carried out in the Polarimetric Multi-Object Spectroscopy (PMOS) mode; 8 integrations were obtained, each with an exposure time of 480 s.   
We preprocessed the raw data collected at each retarder plate angle and extracted the ordinary and extraordinary beams using standard methods in IRAF. Wavelength calibration was conducted with a root-mean-square accuracy of $\sim0.2$\,\AA. 
Details of the FORS2/PMOS data reduction and the derivation of the Stokes parameters can be found in the FORS2 Spectropolarimetry Cookbook and Reflex Tutorial\footnote{\url{ftp://ftp.eso.org/pub/dfs/pipelines/instruments/fors/fors-pmos-reflex-tutorial-1.3.pdf}} and Refs~\cite{2017MNRAS.464.4146C, 2020ApJ...902...46Y}, following the procedures described in Ref~\cite{2006PASP..118..146P}. 

It is clear that there are not $\rm [N\,II]$ emission lines in the spectral sequence, while prominent Balmer and He I emission lines are present, which reveals that SN\,2026gzf is located at a place of extremely low metallicity. Interestingly, the survey images from PanSTARRS and Legacy Survey show that the location of SN\,2026gzf is on top of a bright blue knot of the host galaxy, indicating that the SN\,2026gzf location is actually very close to highly star-forming regions.

\subsubsection*{Supernova bolometric light curve} 

A bolometric light curve for the supernova was constructed using NOT and Shane spectra, combined with multiband photometry from ALT (\textit{g, r, i, z, J} bands) and Mephisto (\textit{u, v, z} bands), utilizing temporal interpolation where necessary. Spectral fluxes were scaled to match the multiband photometry. When available, these were complemented by effective fluxes at $3460\, \text{\AA}$ from the Mephisto \textit{u} band and those at 12,200\,$\text{\AA}$ from the ALT \textit{J} band. Before integration, the spectra were extrapolated to assumed zero-flux points at $3000\, \text{\AA}$ and 13,500\,$ \text{\AA}$. By comparing the spectral energy distribution (SED) of the multiband photometry before and after host-galaxy subtraction, we estimated a host contribution of $(2.5\pm 0.7) \times 10^{-13}\, \text{erg}\,\text{s}^{-1}\,\text{cm}^{-2}$, which represents approximately $8\% \text{--} 20\%$ of the total flux depending on the epoch. This contribution was then subtracted from the spectrum-integrated bolometric fluxes. Galactic and host-galaxy reddening were not taken into account, as their contributions are negligible for this supernova.

\begin{figure}[h]
\centering
\begin{overpic}[width=0.4\textwidth]{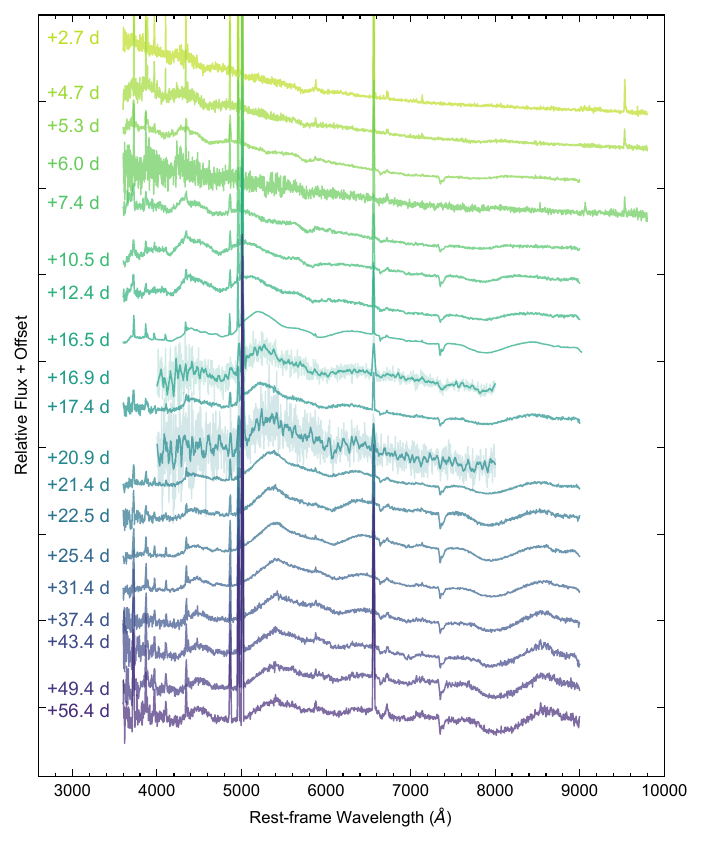}
        \put(-5,95){\textbf{(A)}}
\end{overpic}
\begin{overpic}[width=0.5\textwidth]{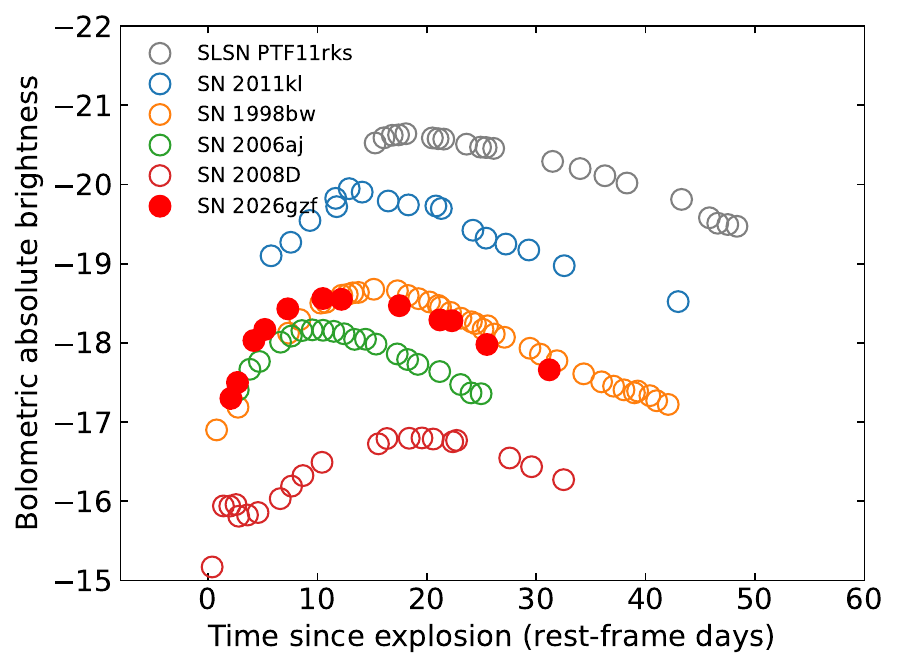}
        \put(-2,71){\textbf{(B)}}
\end{overpic}
\caption{\textbf{The spectral evolution and bolometric light curve of SN~2026gzf.} {(A) The Spectra of SN~2026gzf from +2.7 d to +56.4 d. (B) Comparison of the bolometric light curve of SN~2026gzf with those of representative SNe Ic-BL and superluminous SNe. } The spectrum-integrated absolute bolometric magnitude of SN~2026gzf is represented by solid circles, plotted alongside the quasibolometric counterparts of representative normal-luminosity, luminous, and superluminous stripped-envelope supernovae. Relative to the other two supernovae with reported X-ray SBO detections, SN~2026gzf rises rapidly to a peak magnitude of $M_{\rm bol}\approx -18.6$ mag, peaking brighter and slightly later than SN~2006aj, but $\sim 10$ days earlier than the normal-luminosity SN~2008D. The overall shape of the light curve is similar to that of the well-modelled prototypical broad-lined supernova SN~1998bw, although it peaks earlier, suggesting a somewhat smaller $M_{\rm ej}^3/E$ value. These properties place SN~2026gzf among rapidly evolving luminous supernovae, rather than superluminous ones like PTF11rks.}
\label{fig:lc_bolometric}
\end{figure}

The results were converted to absolute magnitudes by adopting a distance modulus of $\mu=35.84$ mag. To account for minor epoch differences between the spectroscopic and photometric data, effective epochs and corresponding uncertainties were assigned to each bolometric point. Magnitude uncertainties include contributions from photometric measurements, host-galaxy subtraction, and spectral flux scaling (fixed at $5\%$). Spectra from days $+2.7$ and $+4.7$ were adopted for epochs $+2.2$ and $+4.2$ days, respectively. Similarly, Mephisto photometry from day $+6.2$ was utilized for days $+4.2$ and $+5.2$. While (quasi-)bolometric magnitudes derived from host-subtracted multiband photometric SEDs generally underestimate the spectrum-integrated results by $0\text{--}0.15$ mag across most epochs, the discrepancy reaches $\sim 0.3$ mag at days $+2.1$ and $+2.7$. We conservatively adopted mean values for these two earliest epochs, with assigned uncertainties of $0.15$ mag. The final bolometric light curve, converted to the supernova rest frame, is shown in the right panel of Extended Data Figure \ref{fig:lc_bolometric}.

\subsubsection*{SBO from a CSM shell }

The peak timescale of an SBO signal is determined by both photon diffusion and the light-travel time. If the peak is dominated by photon diffusion, the peak luminosity is comparable to the instantaneous injected luminosity at the SBO,  $L_{\rm p,X}\approx 2\pi r_{\rm SBO}^2\rho_{\rm CSM}v_{\rm sh}^3$. The diffusion timescale $t_{\rm diff}\approx \kappa_{\rm X}\rho_{\rm CSM}\Delta r^2/c$ is approximately equal to the dynamical timescale $t_{\rm dyn}\approx \Delta r/v_{\rm sh}$, both of order $t_{\rm p,X}$, where $\kappa_{\rm X}$ is the opacity, $\Delta r$ is the width of the material outside the SBO radius between the SBO radius and outer boundary, and $c$ is the speed of light. Thus, the SBO radius can be expressed as $r_{\rm SBO}\approx \sqrt{\kappa_{\rm X} L_{\rm p}t_{\rm p,X}/(2\pi cv_{\rm sh})}$. As described in the main text, the spectrum of EP260321a suggests  thermal-like emission, implying that the shock cannot be far from local thermal equilibrium, with $v_{\rm sh}\lesssim0.1\,c$. Furthermore, as the SBO emission is powered by the conversion of the kinetic energy of the outer ejecta, the inferred radiated energy places another constraint on the shock velocity. 
For a typical broken-power-law structure of supernova ejecta\cite{Chevalier1989}, the available kinetic energy within the outer ejecta above velocity $v_{\rm sh}$ can be expressed as $E_{\rm kin}(>v_{\rm sh}) \propto E_{\rm kin} ({v_{\rm sh}}/{v_{\rm tr}})^{-5}$, where $v_{\rm tr}$ is the transition velocity. By equating $E_{\rm kin}(>v_{\rm sh})$ to $\sim E_{\rm th,X}$, we obtain
\begin{equation}
    v_{\rm sh}\sim0.09\,c\left(\frac{E_{\rm kin}}{10^{51}\,{\rm erg}}\right)^{7/10}\left(\frac{M_{\rm ej}}{1\,M_\odot}\right)^{-1/2}\left(\frac{E_{\rm th,X}}{10^{48}\,{\rm erg}}\right)^{-1/5},
\end{equation}
where $M_{\rm ej}$ and $E_{\rm kin}$ are the total mass and kinetic energy of the supernova ejecta, respectively. This implies that the shock velocity cannot be very low, even allowing for reasonable variations in $E_{\rm kin}$ and $M_{\rm ej}$. Then, the SBO radius determined by photon diffusion can be estimated as
\begin{equation}
\begin{split}
    r_{\rm SBO}\approx \sqrt{\frac{\kappa_{\rm X} L_{\rm p}t_{\rm p,X}}{2\pi cv_{\rm sh}}}\sim 270\,R_\odot\left(\frac{\kappa_{\rm X}}{0.2\,{\rm cm}^2\,{\rm g}^{-1}}\right)^{1/2}\left(\frac{L_{\rm p}}{10^{45}\,{\rm erg}\,{\rm s}^{-1}}\right)^{1/2}\left(\frac{t_{\rm p,X}}{700\,{\rm s}}\right)^{1/2}\left(\frac{v_{\rm sh}}{0.09\,c}\right)^{-1/2}.
\end{split}
\end{equation}
Alternatively, if the observed peak timescale of EP260321a is attributed to the light-travel effect, the corresponding SBO radius is inferred to be 
\begin{equation}
    r_{\rm SBO}\lesssim r_{\rm out}\approx ct_{\rm p,X}\sim 300\,R_\odot\left(\frac{t_{\rm p,X}}{700\,{\rm s}}\right).
\end{equation}
Thus, both the light-travel effect and photon diffusion constrain the SBO radius of EP260321a to be $\sim300\,R_\odot$.

For the light-curve and temperature modeling, we then consider a radiation-mediated shock propagating with a velocity of $v_{\rm sh}$ through a CSM shell confined in a range of $r_{\rm in} \lesssim r \lesssim r_{\rm out}$, with a wind-like density profile $\rho_{\rm CSM}(r)= Kr^{-2}$. 
The normalization coefficient of the density profile is given by $K=M_{\rm CSM}/[4\pi(r_{\rm out}-r_{\rm in})]$, with $M_{\rm CSM}$ representing the CSM shell mass. 
At the moment of SBO, where the time zero ($t=0$) is defined, the optical depth outside the SBO radius satisfies
\begin{equation}
    \tau(t=0) \approx \int_{r_{\rm SBO}}^{r_{\rm out}}\kappa_{\rm X}\rho_{\rm CSM} dr =\kappa_{\rm X} K\left(\frac{1}{r_{\rm SBO}}-\frac{1}{r_{\rm out}}\right)\approx \frac{c}{v_{\rm sh}},
\end{equation}
which is contributed by the unshocked CSM. 
As the shock propagates outward, photons trapped in the inner shocked CSM can gradually diffuse out, thus producing cooling emission\cite{Nakar2010,Piro2021}. We define the layer of the diffusing shocked CSM with optical depth $\tau \approx c / v_{\rm sh}$ as the {\em breakout shell}. 
Then, we calculate the mass $M_{\rm diff}$ of the diffusing shocked CSM and the unshocked CSM (if the shock has not yet swept up all the CSM) outside the breakout shell at different times $t$. The diffusion of photons for the entire shocked CSM can be determined by $M_{\rm diff}\approx M_{\rm CSM}$, thus obtaining a characteristic diffusion timescale $t_{\rm diff,c}$ for the entire shocked CSM. 

The initial energy of the breakout shell at time $t$ is given by $E_{\rm int,0}\approx M_{\rm diff}v_{\rm sh}^2/2$. 
Since the CSM is very thin, the internal energy can be approximately treated as constant, i.e., $E_{\rm int}\approx E_{\rm int,0}$, when $r_{\rm sh}\lesssim r_{\rm out}$. Once the shocked material expands beyond the outer edge of the CSM ($r_{\rm sh}\gtrsim r_{\rm out}$), adiabatic expansion leads to an evolution of the internal energy as $E_{\rm int}\propto r^{-1}$. The internal energy can therefore be estimated as $E_{\rm int}\approx M_{\rm diff}v_{\rm sh}^2r_{\rm out}/(2r_{\rm sh})$. Then, we can model the source luminosity lightcurve of the cooling emission as
\begin{equation}
\label{equ:Lightcurve}
    L \approx \begin{cases}
    E_{\rm int}\over t+(r_{\rm out}-r_{\rm SBO})/v_{\rm sh}, & \text{if}\;t\leq t_{\rm diff,c}, \\
    {E_{\rm int}\over t_{\rm diff,c}+(r_{\rm out}-r_{\rm SBO})/v_{\rm sh}}e^{-[(t-t_{\rm diff,c})^2/t_{\rm diff,c}^2]/2}, & \text{if}\;t> t_{\rm diff,c},
\end{cases}
\end{equation}
where $(r_{\rm out}-r_{\rm SBO})/v_{\rm sh}$ is the diffusion timescale for the breakout shell, and the exponential decline describes the rapid fading after most trapped thermal photons in the shocked CSM have diffused out, followed by radiative and adiabatic cooling\cite{Piro2021,Zhu2025}.

Under local thermal equilibrium, the temperature is determined by balancing the radiation energy density (dominated by the luminosity $u\propto L\tau/cr_{\rm sh}^2$) with the kinetic energy density: $T_{\rm th}\approx \left(u/{a}\right)^{1/4}$, 
where $a$ is the radiation constant. When the shock velocity $v_{\rm sh}/c\gtrsim0.05$ occurs, there is not enough time to form a local photon-electron thermal equilibrium, and blackbody radiation cannot build up\cite{Nakar2010,Katz2010}. In such a case, photons and electrons are in Compton equilibrium at a higher temperature $T$. One can define a thermal coupling coefficient, $\eta$, as the ratio of the photon density required for thermal equilibrium to the number of photons produced by free-free emission over the available photon-production timescale in the breakout shell, thereby assessing whether local thermal equilibrium can be established\cite{Nakar2010,Katz2010}. 
If $\eta\lesssim1$, the radiation is in local thermal equilibrium and the observed temperature is $T_{\rm obs}=T_{\rm th}$; while if $\eta\gtrsim1$, $T_{\rm obs}$ can be modified by Comptonization as $T_{\rm obs}=T_{\rm th}\eta^2/\xi(T_{\rm obs})^2$, where $\xi(T_{\rm obs})$ 
represents photons produced at lower frequencies and Comptonized to a temperature of $T$. 
In our case, $T_{\rm obs}$ at the early cooling phase can exceed the thermal equilibrium temperature $T_{\rm th}$ by a factor of $\sim 3$--4, suggesting that the emission is not in full local thermal equilibrium but remains partially thermalized. We therefore define a color correction factor, $f_{\rm col}=T_{\rm obs}/T_{\rm th}$, to account for the deviation of the source temperature from the blackbody temperature expected under complete thermal equilibrium, which is used in the main text. The X-ray spectrum is expected to show a Wien peak and is conventionally approximated as a diluted blackbody spectrum characterized by $T_{\rm obs}$ when modeling the X-ray light curve, an assumption that is also supported by the observed spectral properties of EP260321a. 


The observed light curve and temperature are modified by the geometric light-travel-time effect, whereby photons emitted simultaneously from different locations on a spherical surface reach the observer at different times. 
Then, we use this model to simultaneously fit the X-ray light curve and temperature of EP260321a with a total of five free parameters: outer boundary $r_{\rm out}$, ratio between inner and outer boundary $r_{\rm in}/r_{\rm out}$, shock dimensionless velocity $\beta_{\rm sh}=v_{\rm sh}/c$, CSM mass $M_{\rm CSM}$, and time of shock breakout relative to the first observed data point $t_{\rm first}$. We perform Markov Chain Monte Carlo fitting with {\em emcee} to model the X-ray light curve and temperature of EP260321a, with the resulting fits shown in Figure \ref{fig:Xraylc}. 
The median values with $1\sigma$ regions of the key parameters are $r_{\rm out}= 324.5^{+2.7}_{-3.6}\,R_\odot$, $r_{\rm in}/r_{\rm out}=0.807^{+0.003}_{-0.003}$ (corresponding to $r_{\rm in}=261.7^{+2.2}_{-2.7}\,R_\odot$), $\beta_{\rm sh}=0.0894^{+0.0003}_{-0.0003}$, $\log_{10}(M_{\rm CSM}/M_\odot)=-3.88^{+0.01}_{-0.01}$, and $t_{\rm first}=142.1^{+5.4}_{-8.8}\,{\rm s}$.

\subsubsection*{Supernova light-curve modeling}

\begin{figure}[h]
\centering
\includegraphics[width=0.8\textwidth]{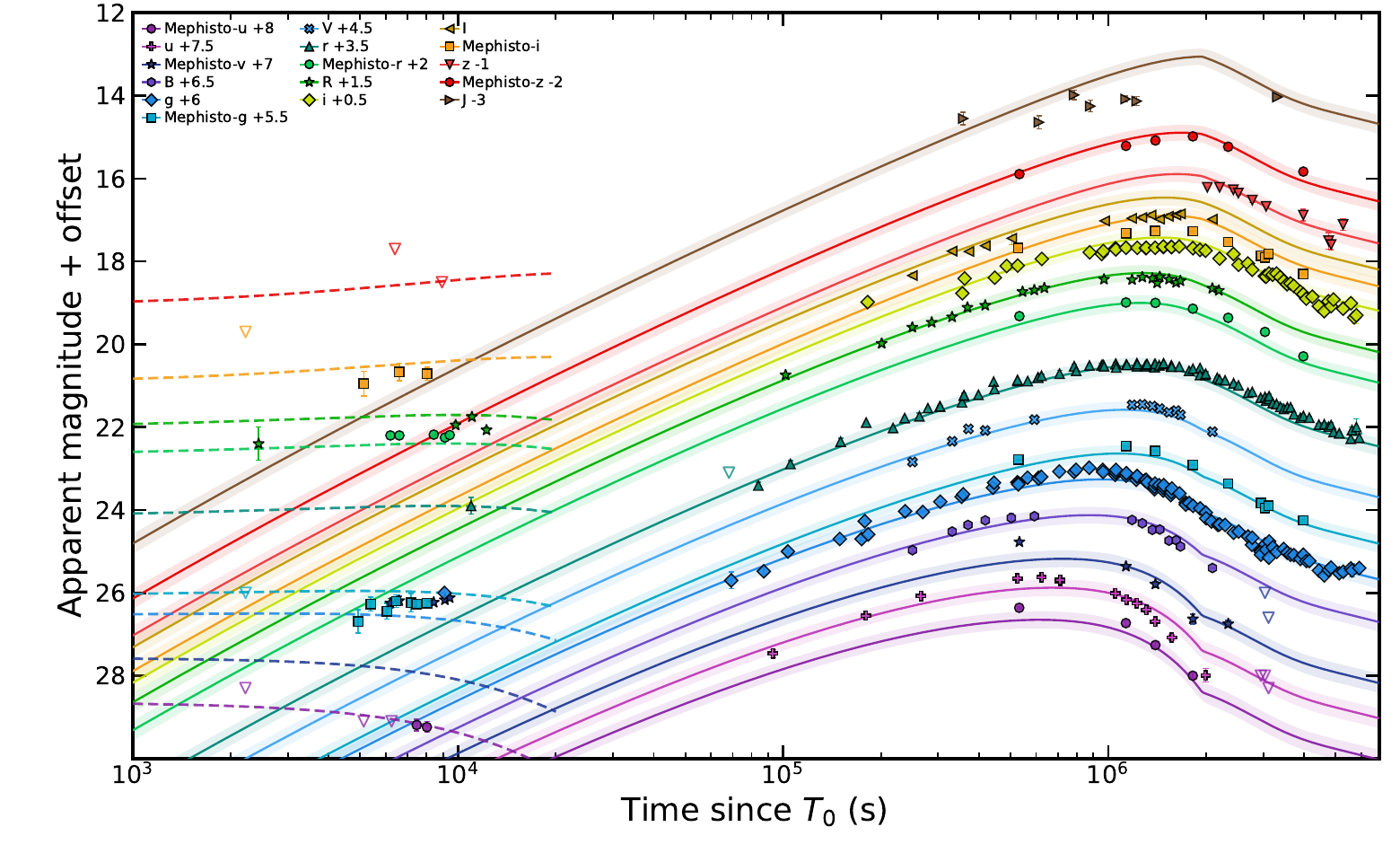}
\caption{{\bf Fits of SN 2026gzf with a magnetar engine}. The solid lines give the theoretical light curves from the magnetar engine model for following parameters: initial spin period $P_{\rm i}=1.60^{+0.09}_{-0.08}$ ms, dipolar magnetic field strength $B_{\rm p}=4.39^{+0.21}_{-0.18}\times10^{15}$ G, ejecta mass $2.38^{+0.20}_{-0.20}~M_{\odot}$, $^{56}$Ni mass $0.26^{+0.01}_{-0.01}~
M_{\odot}$, and kinetic energy $2.48^{+1.00}_{-0.97}\times10^{51}$ erg. The dashed light curves are obtained by assuming blackbody emission with an empirical power-law-evolving temperature and photospheric radius.}
\label{fig:SNfit_comparison}
\end{figure}

The fitting results presented in Figure \ref{fig:SNlightcurves} suggest that the traditional radioactive model can provide a plausible explanation for SN~2026gzf, but a high nickel-to-ejecta mass ratio and extreme $^{56}\rm Ni$ mixing are required, which somewhat reduces the feasibility of the model. Therefore, here we try to model the supernova in alternative scenarios --- in particular, by invoking other energy sources. 
As usually suggested for unusual supernova phenomena\cite{Kasen2010ApJ...717..245K,Yu2015ApJ...806L...6Y,Zhu2026}, a powerful central engine (e.g., a millisecond magnetar) could be formed in SN 2026gzf, injecting energy into the supernova ejecta during its spin-down. 
In our calculations, as well for the radioactive model, the monochromatic luminosities in different filters are obtained by considering blackbody spectra for the supernova emission. However, for short-wavelength emission (i.e., Mephisto-$v$, Mephisto-$u$, and $u$), an additional phenomenological suppression factor such as $f_{\rm UV}=\left[\min\left(1,{\lambda_{\rm rest}/4500~{\rm \AA}}\right)\right]^3$ is invoked to match the observational data\cite{Nicholl2017,Yan2018ApJ...858...91Y}, where \(\lambda_{\rm rest}\) is the rest-frame wavelength of each filter. The fitting results with a magnetar engine are shown in Extended Data Figure~\ref{fig:SNfit_comparison}. As shown, the shallow increase of SN 2026gzf has been well explained, because the $t^{-2}$ behavior of the energy release from the magnetar naturally determines a logarithmic increase\cite{Yu2015ApJ...806L...6Y}. Moreover, the inferred parameter values are also consistent with previous work for other SNe Ic‑BL in the magnetar engine model\cite{Srinivasaragavan2025,Zhu2025MNRAS.544L.139Z,Zhu2026}. Nevertheless, in order to account for the tail of the supernova emission, $0.26\,M_{\odot}$ of $^{56}$Ni is still required.

In addition, owing to the potentially complicated CSM environment, one may suggest that SN 2026gzf is powered by an interaction of the supernova ejecta with the CSM. 
However, this model inevitably requires the presence of a highly optically thick CSM shell, which would determine a large optical depth for soft X-rays due to photoionization or electron scattering, and thus can in principle obscure the previous SBO X-ray emission. In principle, the potential anisotropy of the supernova explosion could still help circumvent this difficulty. The more serious challenge to the CSM-interaction model is that it is actually difficult for the nearly static CSM to generate sufficiently broad lines to match the spectral observations of SN~2026gzf. 

The optical data at $\lesssim 10^4$ s always exceed the supernova early emission, regardless of the power scenario. The empirical blackbody fit of the early data gives a blackbody radius $\sim 5\times10^{14}\rm cm$. This makes it difficult to explain with the supernova emission as well as the cooling of the SBO, no matter which energy mechanism is invoked. Alternatively, this early optical bump is likely to arise from another emission region, which is of a transrelativistic velocity. Here, we tentatively describe the early optical emission by using an evolving blackbody component parameterized as
\begin{equation}
T_{\rm BB}=T_0\left(\frac{\hat{t}_{\rm }}{t_{\rm ref}}\right)^{-\alpha_T},\qquad
u(\hat{t}_{\rm })= \Gamma\beta
=
u_0
\left(
\frac{\hat{t}_{\rm }}{t_{\rm ref}}
\right)^{-\alpha_u},
\end{equation}
where the dynamical time $\hat{t}_{\rm }$ is defined to relate to the observer's time $t$ by 
\begin{equation}
\hat{t}_{\rm }
=
\frac{t}{(1+z)(1-\beta)}
+
{1-\beta_{\rm ej}\over 1-\beta}{R_{\rm SBO}\over v_{\rm ej}}. 
\end{equation}
The best fit, as presented by the dashed lines in Extended Data Figure \ref{fig:SNfit_comparison}, gives
$u_0\approx 0.89$, $\alpha_u\approx 0.14$, $T_0\approx 1.83\times10^4\rm K$ , and
$\alpha_T\approx 0.61$ for $t_{\rm ref}=10^4$ s, corresponding to $\beta\approx0.67$ and $\Gamma\approx1.34$ at $t_{\rm ref}=10^4$ s. Here, the Mephisto-$v$ band obviously cannot be included by the blackbody description. This fitting result implies that a subrelativistic cocoon could provide a plausible explanation for the early optical bump, as the bump must not be accounted for by the late cooling of the nonrelativistic SBO. A powerful central engine as suggested above may help to understand the existence of such a subrelativistic cocoon, since a relativistic jet can be driven by the engine even though the jet is finally choked in the progenitor.

\subsubsection*{Radio observations and constraints}

\begin{figure}[h]
\centering
\begin{overpic}[width=0.45\textwidth]{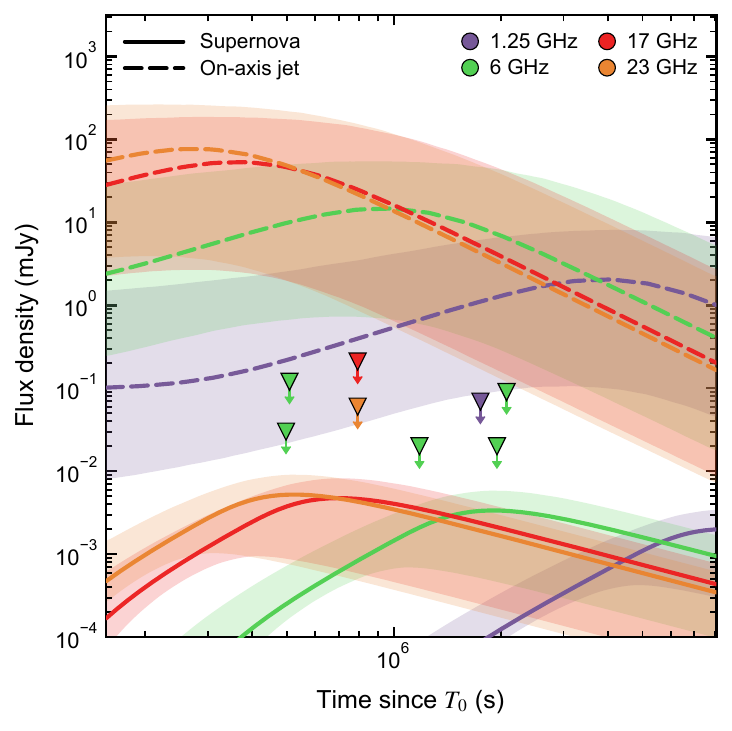}
        \put(0,95){\textbf{(A)}} 
    \end{overpic}
\begin{overpic}[width=0.45\textwidth]{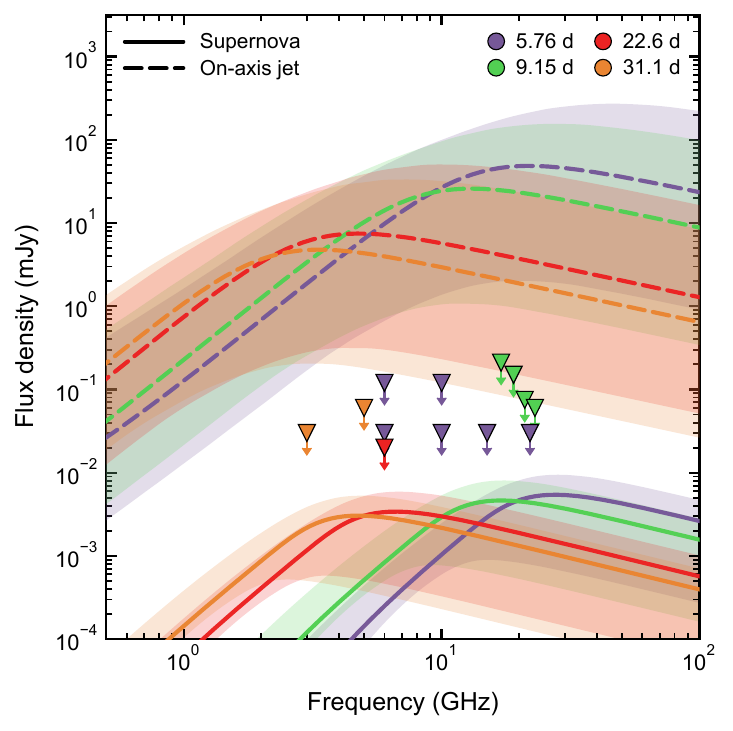}
        \put(0,95){\textbf{(B)}} 
    \end{overpic}
\caption{ \textbf{Radio upper limits (downward triangles) compared with model predictions.} 
\zjp{(A)} Radio light curves at 1.25, 6, 17, and 23 GHz, shown as a function of time since $T_0$. Solid curves illustrate the expected emission from supernova ejecta–CSM interaction, while dashed curves indicate the radio afterglow of on-axis jets. The shaded regions arises from the typical uncertainties of the model parameters. \zjp{(B)} Theoretical radio spectra at representative observing epochs of 5.76, 9.15, 22.6, and 31.1 days. 
}
\label{fig:Radio}
\end{figure}

We compiled radio follow-up observations of EP260321a/SN~2026gzf from our collaboration and public GCN Circulars, as summarized in Extended Data Table~\ref{tab:radio_obs}. Times are measured relative to the WXT onset, $T_0=\texttt{2026-03-21T12:16:08}$ UTC; for observations with reported time ranges, the listed epoch is the midpoint of the interval. The observations cover $0.7$--$23$~GHz from $5.1$ to $31.1$~d after $T_0$. No secure radio counterpart is detected. Upper limits were measured from the local root-mean square (RMS) at the optical position of SN~2026gzf in the final images. MeerKAT observations at 1.28 and 3.0~GHz were reduced with the \textsc{OXKAT} pipeline \cite{2020ascl.soft09003H}, while ATCA observations were calibrated and imaged with standard CASA procedures. The uGMRT Band-5 public nondetection was reduced with the SPAM pipeline as reported in GCN~44227 (Ref. \cite{GCN44227}). e-MERLIN 5\,GHz C-band observations, reduced in CASA with the e-MERLIN pipeline, also yielded nondetections at $t_{\rm obs}=13.489$ and $30.989$~d, with RMS values of 22 and $12~\mu$Jy beam$^{-1}$ and corresponding $5\sigma$ limits of $<110$ and $<60~\mu$Jy beam$^{-1}$, respectively. Public VLA and ATCA constraints are included from GCN Circulars\cite{GCN44229,GCN44239,GCN44357,GCN44403}. A low-significance VLA feature reported at $\sim13.6$~d is not considered a reliable counterpart after subsequent follow-up/reinspection\cite{GCN44239,GCN44357}, revealing no significant variability. These limits, as presented in Extended Data Figure~\ref{fig:Radio}, can provide a useful constraint on the radio emission arising from interactions of supernova ejecta and, in particular, relativistic jets with CSM.

We further examined the radio constraints by comparing two representative external-shock processes, which arise from the propagation of supernova ejecta or a GRB-like relativistic jet into a wind-like CSM. 
In both scenarios, the density profile of the wind-like CSM is assumed to be \(\rho=A r^{-2}\) with a normalization \(A_*\equiv A/(5\times10^{11}\ {\rm g\ cm^{-1}})\), where \(A_*=1\) approximately corresponds to \(\dot{M}=10^{-5}\,M_\odot\,{\rm yr^{-1}}\) for \(v_{\rm w}=1000~{\rm km~s^{-1}}\). 
Specifically, for the supernova radio emission, we varied the ejecta velocity over 
\(\beta_{\rm ej}=v_{\rm ej}/c\in[0.03-0.08]\), together with the microphysical parameters 
\(\log \epsilon_{\rm e}\in[-1.75,-0.25]\) and 
\(\log \epsilon_{\rm B}\in [-2,-1]\). 
For the jet afterglow emission, we scan the initial Lorentz factor 
\(\Gamma_0\in[10,300]\), the isotropic-equivalent kinetic energy 
\(\log E_{\rm iso}/{\rm erg}\in[51,53]\), 
\(\log \epsilon_{\rm e}\in[-1.25,-0.25]\), and 
\(\log \epsilon_{\rm B}\in[-4,-0.5]\). 
For each scenario, \(10^4\)  parameter combinations were sampled. 
The shaded regions in  Extended Data Figure~\ref{fig:Radio} show the $1\sigma$ range of the predicted flux density at each time or frequency of observation, while the solid and dashed curves denote the representative curves. To be specific, the representative parameter values are taken as 
\(\Gamma_0=100\), \(E_{\rm iso}=10^{52}\,{\rm erg}\), 
\(\epsilon_{\rm e}=0.1\), and \(\epsilon_{\rm B}=0.01\) for the jet external shock, which are broadly consistent with GRB afterglow observations\cite{Granot2002,Santana2014,Chandra2012}, while for the supernova external shock, we adopted
\(\beta_{\rm ej}=0.1\), \(\epsilon_{\rm e}=0.1\), and \(\epsilon_{\rm B}=0.1\) (Refs. \cite{Chevalier1998,Weiler2002,Chevalier2006}).

The observed radio upper limits are consistent with the low fluxes expected from the ejecta-CSM interaction. 
In contrast, a standard energetic GRB-like afterglow in a comparable wind environment would generally produce substantially brighter radio emission and can exceed the observational limits over a broad region of the scanned parameter space. 
These radio nondetections therefore disfavor the presence of a luminous, on-axis, GRB-like relativistic jet associated with EP260321a/SN~2026gzf. 
However, they should not be interpreted as ruling out all possible relativistic or engine-driven outflows, such as
an off-axis jet of a viewing angle $\theta_{\rm v}\gtrsim 45^{\circ}$ and, in particular, a deeply choked jet.
As a result, a weak sub-relativistic cocoon could appear on the line of sight, which contributes to radio emission of  flux\cite{Piran2013MNRAS.430.2121P} of $F_\nu\approx 1.8\times10^{-5}\,{\rm mJy}(\beta/0.6)^{31/4}(t_{\rm obs}/10\,{\rm day})^3(\nu/10\,{\rm GHz})^{-3/4}$, where $p=2.5$, $\epsilon_{\rm e}=0.1$, and $\epsilon_{\rm B}=0.01$ are adopted and a typical density of $n=1\rm cm^{-3}$ is used because the cocoon has moved into the interstellar medium rather than staying in the wind-like CSM. Obviously, this flux is safely below the current upper limits for fiducial parameters.


\subsubsection*{Event-rate density} 

We estimate the local event-rate density of X-ray shock breakout $\rho_{0,\rm SBO}$ from the single detection of EP260321a in the systematic WXT survey, using the $V_{\rm max}$ method \cite{Schmidt1968} as implemented for systematic study of high-energy transients\cite{Sun2015ApJ,Sun2022ApJ}. For EP260321a at $z = 0.034$ (luminosity distance $D_L \approx 150$ $\rm  Mpc$), we compute the maximum volume within which the source would have been detectable given its luminosity light curve and the WXT’s sensitivity limit. The field of view of WXT ($\Omega_{\rm WXT}$) is 3600 square degrees. The total exposure time ($T_{\rm WXT}$) of WXT by March 2026 is $\sim 34$ million seconds, indicating a survey efficiency of $\sim 50\%$. The WXT simulation indicates that an EP260321a-like event can be well detected by WXT up to a luminosity distance of 250 Mpc with signal-to-noise ratio above 7. We obtain an effective volume-time $\mathcal{M} = V_{\rm max}T_{\rm WXT} \approx 0.006$ $\rm Gpc^{3}\,yr$, where $V_{\rm max}$ is the monitored maximum volume weighted by the density evolution and
time dilation, 
\begin{equation}
V_{\rm max} = \int_{0}^{z_{\rm max}} \frac{\Omega_{\rm WXT}}{4\pi}\,\frac{f(z)}{(1+z)} \frac{dV(z)}{dz} dz.
\label{eq:V'max}
\end{equation}
The comoving volume is given by
\begin{equation}
\frac{dV(z)}{dz}=\frac{c}{{\rm H}_0}\frac{4\pi D_L(z)^2}{(1+z)^2[\Omega_M(1+z)^3+\Omega_{\Lambda}]^{1/2}}\, ,
\end{equation} 
where $D_L(z)$ is the luminosity distance at the corresponding redshift $z$. 
The $f(z)$ factor is model dependent. By assuming that core-collapse supernovae trace the star-formation history, we adopt the analytical formula from Ref. \cite{Yuksel2008ApJ},
\begin{equation}
f(z)=\left[ (1+z )^{3.4\eta}+\left( \frac{1+z}{5000}\right) ^{-0.3\eta}+\left( \frac{1+z}{9}\right) ^{-3.5\eta} \right] ^{\frac{1}{\eta}},
\label{eq:sfh}
\end{equation}
where $\eta=-10$.

With a single event detected by WXT, the event-rate density of X-ray shock breakout is 
\begin{equation}
\rho_{0, \rm  SBO} \geq \frac{1}{\mathcal{M}} \approx 170_{-140}^{+390}\ \rm{Gpc}^{-3}\ {yr}^{-1},   
\end{equation}
The error represents 1$\sigma$ uncertainty, which is calculated from small-sample statistics\cite{Gehrels1986ApJ}. This value should be regarded as a lower limit on the event-rate density owing to the incompleteness of the identification and classification of EP transients, possibly due to the lack of timely follow-up observations of some EP transients.

For comparison, following the estimate of XRO 080109 \cite{Soderberg2008Natur.453..469S} and extending the \textit{Swift} operation time to date, we re-estimate the event-rate density from the serendipitous detection of XRO 080109 to be $1.0_{-0.8}^{+2.3}\times 10^4$ $\rm{Gpc}^{-3}\ {yr}^{-1}$. The systematic WXT survey yields a lower rate density from EP260321a, yet with an order of magnitude higher luminosity. Both estimates are based on single events and thus remain sensitive to Poisson fluctuations. The higher luminosity of EP260321a compared to XRO 080109, together with their different rate densities, also implies a luminosity-dependent evolution of the SBO rate. To explore this possibility, we make use of EP260321a, together with XRO 080109 and XRF 060218, to probe the X-ray SBO luminosity function following the method in Refs. \cite{Sun2015ApJ, Sun2022ApJ}. The luminosity function $\Phi(L)$ is defined as the event-rate density per logarithmic luminosity interval (in units of $\rm Gpc^{-3}\, yr^{-1}\, dex^{-1}$). A bin width of $\Delta \log L = 1$ dex is adopted in the calculation. Assuming a power-law form for the luminosity function, these three sources are described by a single power-law distribution with a slope of $1.3 \pm 0.4$ ($1\sigma$), as shown in the right panel of Figure~\ref{fig:ComXrayLC}. 

\clearpage

\begin{table*}
    \centering
    \small
    \begin{threeparttable}
    \caption{\textbf{FXT monitoring observation list.}}\label{fxt_obs_log}%
    \begin{tabular}{ccccc}
    \hline
        ObsID & Start Time (UTC) &  End Time (UTC) & Exp. Time (s) & Upper Limit\footnotemark[2] \\ \hline 
        06800001291\footnotemark[1] & 2026-03-21T21:22:38 & 2026-03-21T22:58:30 & 2974 &$<4.4$ \\ 
        06800001293 & 2026-03-22T06:55:06 & 2026-03-22T11:45:26 & 8321 &$<2.9$\\ 
        06800001297 & 2026-03-23T18:05:31 & 2026-03-23T22:54:25 & 8894 &$<1.5$\\ 
        06800001298 & 2026-03-24T10:05:25 & 2026-03-24T14:52:59 & 8849 &$<1.3$\\ 
        06800001346 & 2026-04-09T14:15:17 & 2026-04-09T15:51:08 & 2962 &$<3.7$\\ 
        06800001347 & 2026-04-10T06:09:48 & 2026-04-10T09:25:33 & 4977 &$<1.8$\\ 
        \hline
    \end{tabular}
    \begin{tablenotes}
    \footnotesize 
    \item[1]{The FXT-A was working in the partial-window mode with the thin filter in ObsID 06800001291. Other observations were performed in the full-frame mode with the thin filter.}
    \item[2]{The FXT-A/B combined $3\sigma$ upper limits (0.4--2.0 keV) are in units of $10^{-14}~{\rm erg~cm^{-2}~s^{-1}}$.}
    \end{tablenotes}
    \end{threeparttable}
\end{table*}

\begin{table*}
\centering
\small
\begin{threeparttable}
\caption{\textbf{Spectral parameters of the time-resolved WXT and FXT spectra.} All errors represent the 1$\sigma$ uncertainties.}\label{ep_sliced_spec}%
\begin{tabular}{cccc}
\hline
Time interval (s) \tnote{1}& $kT_{\rm obs}$ (eV) & Unabsorbed flux\tnote{2} & W-stat/dof \\
\hline
\multicolumn{4}{l}{\textbf{WXT segments}} \\
\hline
$[0,500]$ & $132_{-21}^{+30}$ & $1.19_{-0.23}^{+0.28}$ & $6.0/9$ \\
$[500, 1000]$ & $136_{-12}^{+14}$ & $1.85_{-0.28}^{+0.32}$ & $18.9/19$ \\
$[1000, 1650]$ & $125_{-9}^{+10}$ & $1.07_{-0.14}^{+0.16}$ & $38.6/29$ \\
$[1650, 2200]$ & $114_{-22}^{+26}$ & $0.35_{-0.08}^{+0.10}$ & $14.3/11$ \\
\hline
\multicolumn{4}{l}{\textbf{FXT-A/B segments}} \\
\hline
$[1065,1165]$ & $122.4_{-3.3}^{+3.4}$ & $1.40_{-0.12}^{+0.14}$ & $132.0/152$ \\
$[1165,1265]$ & $119.6_{-3.3}^{+3.4}$ & $1.32_{-0.11}^{+0.13}$ & $133.8/139$ \\
$[1265,1365]$ & $117.1_{-3.4}^{+3.5}$ & $1.25_{-0.11}^{+0.13}$ & $112.1/133$ \\
$[1365,1565]$ & $110.5_{-2.8}^{+2.8}$ & $0.96_{-0.09}^{+0.10}$ & $189.0/165$ \\
$[1565,1865]$ & $107.8_{-2.8}^{+2.9}$ & $0.64_{-0.06}^{+0.07}$ & $167.1/168$ \\
$[1865,2359]$ & $107.9_{-3.0}^{+3.2}$ & $0.27_{-0.03}^{+0.03}$ & $105.8/138$ \\
\hline
\end{tabular}
\begin{tablenotes}
\footnotesize 
\item[1] Time intervals are relative to $T_0 = \mathrm{2026\mbox{-}03\mbox{-}21T12{:}16{:}08}$ UTC.
\item[2] Flux values (0.4--2.0 keV) and their 1$\sigma$ errors are given in units of $10^{-10}\,\mathrm{erg\,cm^{-2}\,s^{-1}}$.
\end{tablenotes}
\end{threeparttable}
\end{table*}

\clearpage

\begin{longtable}{lccccc}
\caption{Photometric Table} \label{tab:phot} \\
\toprule
Time (day) & Band & Magnitude & Error & System & Telescope \\
\midrule
\endfirsthead
\caption[]{Photometric Table} \\
\toprule
Time (day) & Band & Magnitude & Error & System & Telescope \\
\midrule
\endhead
\midrule
\multicolumn{6}{r}{Continued on next page} \\
\midrule
\endfoot
\bottomrule
\endlastfoot
1.078 & u & 19.91 & 0.03 & AB & WFST \\
2.078 & u & 19.03 & 0.02 & AB & WFST \\
3.077 & u & 18.56 & 0.02 & AB & WFST \\
6.089 & u & 18.13 & 0.02 & AB & WFST \\
6.089 & u & 18.15 & 0.02 & AB & WFST \\
7.217 & u & 18.11 & 0.02 & AB & WFST \\
7.218 & u & 18.11 & 0.02 & AB & WFST \\
8.239 & u & 18.21 & 0.04 & AB & WFST \\
8.240 & u & 18.18 & 0.04 & AB & WFST \\
12.190 & u & 18.46 & 0.03 & AB & WFST \\
13.179 & u & 18.64 & 0.03 & AB & WFST \\
14.181 & u & 18.70 & 0.03 & AB & WFST \\
15.153 & u & 18.88 & 0.04 & AB & WFST \\
16.157 & u & 19.15 & 0.03 & AB & WFST \\
17.158 & u & 19.17 & 0.04 & AB & WFST \\
18.151 & u & 19.50 & 0.03 & AB & WFST \\
22.122 & u & 20.05 & 0.05 & AB & WFST \\
23.094 & u & 20.32 & 0.07 & AB & WFST \\
24.129 & u & 20.45 & 0.12 & AB & WFST \\
25.129 & u & 20.54 & 0.09 & AB & WFST \\
26.167 & u & 20.73 & 0.07 & AB & WFST \\
29.186 & u & 21.24 & 0.14 & AB & WFST \\
0.105 & g & 20.00 & 0.10 & AB & ALT100B \\
1.115 & g & 19.55 & 0.11 & AB & WFST \\
1.729 & g & 18.70 & 0.02 & AB & TRT-CTO \\
2.064 & g & 18.27 & 0.03 & AB & ALT100B \\
2.115 & g & 18.51 & 0.05 & AB & WFST \\
2.746 & g & 18.03 & 0.02 & AB & TRT-CTO \\
3.114 & g & 18.00 & 0.03 & AB & WFST \\
3.523 & g & 17.80 & 0.01 & AB & TRT-CTO \\
4.100 & g & 17.68 & 0.02 & AB & XL-80 \\
4.146 & g & 17.62 & 0.01 & AB & ALT100B \\
5.114 & g & 17.46 & 0.01 & AB & ALT100B \\
5.157 & g & 17.34 & 0.01 & AB & XL-80 \\
6.084 & g & 17.33 & 0.01 & AB & WHUT \\
6.126 & g & 17.33 & 0.02 & AB & WFST \\
6.126 & g & 17.35 & 0.02 & AB & WFST \\
6.537 & g & 17.22 & 0.02 & AB & TRT-CTO \\
7.070 & g & 17.19 & 0.02 & AB & WFST \\
7.070 & g & 17.21 & 0.02 & AB & WFST \\
7.206 & g & 17.20 & 0.03 & AB & ALT100B \\
8.191 & g & 17.07 & 0.03 & AB & ALT100B \\
9.093 & g & 17.08 & 0.05 & AB & XL-80 \\
9.214 & g & 17.03 & 0.03 & AB & ALT100B \\
10.121 & g & 16.98 & 0.02 & AB & ALT100B \\
11.143 & g & 17.03 & 0.03 & AB & ALT100B \\
11.166 & g & 17.06 & 0.02 & AB & WFST \\
12.162 & g & 17.15 & 0.01 & AB & WFST \\
12.201 & g & 17.09 & 0.01 & AB & WHUT \\
12.234 & g & 17.03 & 0.03 & AB & ALT100B \\
12.162 & g & 17.13 & 0.01 & AB & WFST \\
13.167 & g & 17.11 & 0.01 & AB & ALT100B \\
13.141 & g & 17.15 & 0.01 & AB & WFST \\
14.121 & g & 17.24 & 0.02 & AB & WFST \\
14.192 & g & 17.19 & 0.02 & AB & ALT100B \\
14.209 & g & 17.18 & 0.01 & AB & WHUT \\
15.041 & g & 17.28 & 0.01 & AB & XL-80 \\
15.085 & g & 17.28 & 0.02 & AB & WFST \\
15.144 & g & 17.27 & 0.01 & AB & WHUT \\
15.264 & g & 17.33 & 0.02 & AB & ALT100B \\
16.083 & g & 17.50 & 0.01 & AB & XL-80 \\
16.086 & g & 17.41 & 0.02 & AB & WFST \\
16.129 & g & 17.36 & 0.01 & AB & WHUT \\
16.614 & g & 17.40 & 0.02 & AB & TRT-CTO \\
17.086 & g & 17.50 & 0.02 & AB & WFST \\
17.102 & g & 17.46 & 0.01 & AB & WHUT \\
17.155 & g & 17.40 & 0.03 & AB & ALT100B \\
18.050 & g & 17.61 & 0.02 & AB & WFST \\
19.051 & g & 17.69 & 0.03 & AB & WFST \\
18.107 & g & 17.57 & 0.01 & AB & WHUT \\
18.164 & g & 17.48 & 0.01 & AB & ALT100B \\
19.178 & g & 17.60 & 0.02 & AB & ALT100B \\
20.050 & g & 17.78 & 0.03 & AB & WFST \\
20.051 & g & 17.83 & 0.03 & AB & WFST \\
20.535 & g & 17.83 & 0.01 & AB & TRT-CTO \\
21.116 & g & 17.90 & 0.02 & AB & ALT100B \\
22.203 & g & 17.96 & 0.02 & AB & ALT100B \\
23.112 & g & 18.06 & 0.02 & AB & ALT100B \\
23.164 & g & 18.14 & 0.04 & AB & WFST \\
24.057 & g & 18.23 & 0.04 & AB & WFST \\
25.200 & g & 18.30 & 0.04 & AB & WFST \\
25.471 & g & 18.35 & 0.01 & AB & NOT \\
26.504 & g & 18.36 & 0.01 & AB & TRT-CTO \\
28.156 & g & 18.55 & 0.05 & AB & ALT100B \\
29.180 & g & 18.52 & 0.05 & AB & ALT100B \\
31.171 & g & 18.64 & 0.05 & AB & ALT100B \\
32.063 & g & 18.74 & 0.06 & AB & WFST \\
32.129 & g & 18.67 & 0.03 & AB & ALT100B \\
34.092 & g & 18.87 & 0.07 & AB & WFST \\
34.117 & g & 18.90 & 0.07 & AB & WFST \\
34.151 & g & 18.96 & 0.08 & AB & WFST \\
35.178 & g & 18.83 & 0.09 & AB & ALT100B \\
35.419 & g & 18.96 & 0.02 & AB & NOT \\
36.137 & g & 19.15 & 0.11 & AB & WFST \\
36.217 & g & 18.76 & 0.08 & AB & ALT100B \\
37.168 & g & 18.90 & 0.08 & AB & ALT100B \\
38.097 & g & 18.93 & 0.08 & AB & WFST \\
38.107 & g & 19.02 & 0.08 & AB & WFST \\
38.186 & g & 19.01 & 0.08 & AB & ALT100B \\
40.143 & g & 18.94 & 0.06 & AB & ALT100B \\
41.146 & g & 19.03 & 0.06 & AB & ALT100B \\
42.098 & g & 19.17 & 0.09 & AB & WFST \\
42.128 & g & 19.08 & 0.07 & AB & ALT100B \\
43.132 & g & 19.09 & 0.07 & AB & ALT100B \\
43.143 & g & 19.21 & 0.09 & AB & WFST \\
45.155 & g & 19.18 & 0.08 & AB & ALT100B \\
46.128 & g & 19.08 & 0.06 & AB & ALT100B \\
47.134 & g & 19.23 & 0.10 & AB & ALT100B \\
48.135 & g & 19.23 & 0.03 & AB & ALT100B \\
51.478 & g & 19.46 & 0.04 & AB & TRT-CTO \\
53.478 & g & 19.58 & 0.04 & AB & TRT-CTO \\
55.134 & g & 19.45 & 0.04 & AB & ALT100C \\
56.140 & g & 19.39 & 0.05 & AB & ALT100C \\
59.473 & g & 19.53 & 0.06 & AB & TRT-CTO \\
61.158 & g & 19.49 & 0.10 & AB & ALT100C \\
64.563 & g & 19.41 & 0.12 & AB & TRT-CTO \\
65.136 & g & 19.46 & 0.16 & AB & ALT100B \\
68.545 & g & 19.40 & 0.10 & AB & TRT-CTO \\
0.127 & r & 20.40 & 0.20 & AB & ALT100B \\
1.740 & r & 18.85 & 0.08 & AB & TRT-CTO \\
2.040 & r & 18.64 & 0.06 & AB & WFST \\
2.085 & r & 18.39 & 0.03 & AB & ALT100B \\
2.739 & r & 18.28 & 0.02 & AB & TRT-CTO \\
3.040 & r & 18.18 & 0.05 & AB & WFST \\
3.227 & r & 18.04 & 0.01 & AB & GMG \\
3.515 & r & 18.00 & 0.02 & AB & FORS2 \\
3.516 & r & 18.00 & 0.02 & AB & TRT-CTO \\
4.111 & r & 17.90 & 0.02 & AB & XL-80 \\
4.167 & r & 17.74 & 0.02 & AB & ALT100B \\
4.168 & r & 17.69 & 0.03 & AB & WFST \\
4.618 & r & 17.72 & 0.01 & AB & FORS2 \\
5.142 & r & 17.40 & 0.01 & AB & ALT100B \\
5.170 & r & 17.58 & 0.01 & AB & XL-80 \\
6.085 & r & 17.36 & 0.02 & AB & WFST \\
6.092 & r & 17.36 & 0.01 & AB & WHUT \\
6.545 & r & 17.39 & 0.02 & AB & TRT-CTO \\
7.042 & r & 17.27 & 0.02 & AB & WFST \\
7.042 & r & 17.28 & 0.02 & AB & WFST \\
7.217 & r & 17.26 & 0.03 & AB & ALT100B \\
8.203 & r & 17.21 & 0.08 & AB & ALT100B \\
9.097 & r & 17.03 & 0.05 & AB & XL-80 \\
9.228 & r & 17.14 & 0.04 & AB & ALT100B \\
10.137 & r & 17.01 & 0.02 & AB & ALT100B \\
11.050 & r & 17.01 & 0.05 & AB & XL-80 \\
11.158 & r & 16.99 & 0.02 & AB & ALT100B \\
11.243 & r & 17.01 & 0.02 & AB & WFST \\
12.209 & r & 16.98 & 0.01 & AB & WHUT \\
12.217 & r & 16.97 & 0.04 & AB & ALT100B \\
12.249 & r & 17.03 & 0.02 & AB & WFST \\
13.152 & r & 16.99 & 0.01 & AB & ALT100B \\
13.213 & r & 16.98 & 0.02 & AB & WFST \\
14.181 & r & 16.96 & 0.03 & AB & ALT100B \\
14.207 & r & 17.00 & 0.01 & AB & WFST \\
14.216 & r & 16.98 & 0.01 & AB & WHUT \\
15.045 & r & 17.03 & 0.01 & AB & XL-80 \\
15.151 & r & 16.99 & 0.01 & AB & WHUT \\
15.253 & r & 16.98 & 0.03 & AB & ALT100B \\
16.048 & r & 16.98 & 0.01 & AB & WFST \\
16.087 & r & 17.05 & 0.01 & AB & XL-80 \\
16.137 & r & 16.97 & 0.01 & AB & WHUT \\
16.609 & r & 17.01 & 0.01 & AB & TRT-CTO \\
17.109 & r & 16.98 & 0.01 & AB & WHUT \\
17.143 & r & 16.96 & 0.02 & AB & ALT100B \\
18.083 & r & 17.01 & 0.01 & AB & WFST \\
18.115 & r & 17.01 & 0.01 & AB & WHUT \\
18.143 & r & 17.04 & 0.01 & AB & ALT100B \\
19.084 & r & 17.00 & 0.02 & AB & WFST \\
19.166 & r & 17.04 & 0.01 & AB & ALT100B \\
20.527 & r & 17.14 & 0.01 & AB & TRT-CTO \\
21.104 & r & 17.09 & 0.03 & AB & ALT100B \\
22.191 & r & 17.07 & 0.02 & AB & ALT100B \\
22.192 & r & 17.21 & 0.02 & AB & WFST \\
23.057 & r & 17.21 & 0.02 & AB & WFST \\
23.100 & r & 17.21 & 0.05 & AB & ALT100B \\
25.059 & r & 17.28 & 0.02 & AB & WFST \\
25.474 & r & 17.35 & 0.01 & AB & NOT \\
26.496 & r & 17.37 & 0.01 & AB & TRT-CTO \\
28.144 & r & 17.44 & 0.02 & AB & ALT100B \\
29.134 & r & 17.49 & 0.01 & AB & ALT100A \\
31.159 & r & 17.67 & 0.02 & AB & ALT100B \\
32.117 & r & 17.65 & 0.02 & AB & ALT100B \\
34.089 & r & 17.75 & 0.03 & AB & WFST \\
35.159 & r & 17.77 & 0.03 & AB & ALT100B \\
35.422 & r & 17.86 & 0.01 & AB & NOT \\
36.128 & r & 17.78 & 0.04 & AB & WFST \\
36.192 & r & 17.75 & 0.05 & AB & ALT100B \\
37.141 & r & 17.92 & 0.05 & AB & ALT100B \\
37.160 & r & 17.84 & 0.04 & AB & WFST \\
38.070 & r & 17.96 & 0.04 & AB & WFST \\
38.080 & r & 17.94 & 0.03 & AB & WFST \\
38.159 & r & 17.94 & 0.05 & AB & ALT100B \\
39.071 & r & 18.00 & 0.04 & AB & WFST \\
39.202 & r & 18.01 & 0.07 & AB & ALT100B \\
40.197 & r & 18.06 & 0.05 & AB & ALT100B \\
41.074 & r & 18.06 & 0.04 & AB & WFST \\
41.099 & r & 18.07 & 0.04 & AB & WFST \\
41.172 & r & 18.02 & 0.04 & AB & ALT100B \\
42.141 & r & 18.14 & 0.04 & AB & WFST \\
42.144 & r & 18.02 & 0.04 & AB & ALT100B \\
43.074 & r & 18.13 & 0.04 & AB & WFST \\
43.146 & r & 18.19 & 0.04 & AB & ALT100B \\
44.074 & r & 18.24 & 0.06 & AB & WFST \\
45.167 & r & 18.27 & 0.05 & AB & ALT100B \\
46.078 & r & 18.20 & 0.05 & AB & WFST \\
46.142 & r & 18.23 & 0.04 & AB & ALT100B \\
47.104 & r & 18.24 & 0.04 & AB & WFST \\
48.077 & r & 18.29 & 0.05 & AB & WFST \\
48.121 & r & 18.27 & 0.02 & AB & ALT100B \\
50.095 & r & 18.37 & 0.05 & AB & WFST \\
51.482 & r & 18.44 & 0.03 & AB & TRT-CTO \\
53.486 & r & 18.43 & 0.03 & AB & TRT-CTO \\
55.140 & r & 18.49 & 0.05 & AB & ALT100B \\
56.161 & r & 18.45 & 0.05 & AB & ALT100B \\
57.148 & r & 18.62 & 0.04 & AB & ALT100B \\
59.480 & r & 18.65 & 0.04 & AB & TRT-CTO \\
64.569 & r & 18.78 & 0.10 & AB & TRT-CTO \\
65.121 & r & 18.57 & 0.08 & AB & ALT100B \\
67.146 & r & 18.50 & 0.20 & AB & ALT100B \\
68.552 & r & 18.76 & 0.10 & AB & TRT-CTO \\
2.107 & i & 18.48 & 0.10 & AB & ALT100B \\
4.121 & i & 18.27 & 0.03 & AB & XL-80 \\
4.188 & i & 17.91 & 0.05 & AB & ALT100B \\
5.185 & i & 17.89 & 0.02 & AB & XL-80 \\
5.639 & i & 17.61 & 0.03 & AB & TRT-CTO \\
6.099 & i & 17.60 & 0.01 & AB & WHUT \\
7.233 & i & 17.44 & 0.03 & AB & ALT100B \\
10.157 & i & 17.28 & 0.04 & AB & ALT100B \\
11.054 & i & 17.32 & 0.10 & AB & XL-80 \\
11.180 & i & 17.25 & 0.05 & AB & ALT100B \\
12.193 & i & 17.18 & 0.05 & AB & ALT100B \\
12.216 & i & 17.21 & 0.02 & AB & WHUT \\
13.129 & i & 17.17 & 0.05 & AB & ALT100B \\
14.224 & i & 17.17 & 0.02 & AB & WHUT \\
15.120 & i & 17.18 & 0.02 & AB & ALT100B \\
15.159 & i & 17.17 & 0.02 & AB & WHUT \\
16.144 & i & 17.17 & 0.01 & AB & WHUT \\
17.117 & i & 17.15 & 0.02 & AB & WHUT \\
17.126 & i & 17.14 & 0.03 & AB & ALT100B \\
18.122 & i & 17.15 & 0.02 & AB & WHUT \\
18.126 & i & 17.15 & 0.03 & AB & ALT100B \\
19.149 & i & 17.14 & 0.02 & AB & ALT100B \\
21.087 & i & 17.18 & 0.03 & AB & ALT100B \\
22.174 & i & 17.22 & 0.05 & AB & ALT100B \\
23.076 & i & 17.24 & 0.03 & AB & ALT100B \\
25.477 & i & 17.43 & 0.01 & AB & NOT \\
28.127 & i & 17.32 & 0.04 & AB & ALT100B \\
29.151 & i & 17.58 & 0.04 & AB & ALT100B \\
31.142 & i & 17.55 & 0.03 & AB & ALT100B \\
32.100 & i & 17.70 & 0.03 & AB & ALT100B \\
35.117 & i & 17.85 & 0.10 & AB & ALT100B \\
35.426 & i & 17.87 & 0.01 & AB & NOT \\
36.170 & i & 17.79 & 0.08 & AB & ALT100B \\
37.113 & i & 17.82 & 0.08 & AB & ALT100B \\
38.128 & i & 17.79 & 0.07 & AB & ALT100B \\
39.174 & i & 17.88 & 0.11 & AB & ALT100B \\
40.170 & i & 17.97 & 0.06 & AB & ALT100B \\
41.115 & i & 18.05 & 0.05 & AB & ALT100B \\
42.107 & i & 18.03 & 0.05 & AB & ALT100B \\
43.113 & i & 18.09 & 0.08 & AB & ALT100B \\
45.138 & i & 18.23 & 0.07 & AB & ALT100B \\
46.109 & i & 18.29 & 0.06 & AB & ALT100B \\
47.114 & i & 18.40 & 0.06 & AB & ALT100B \\
49.151 & i & 18.35 & 0.05 & AB & ALT100B \\
51.497 & i & 18.58 & 0.10 & AB & TRT-CTO \\
53.495 & i & 18.70 & 0.10 & AB & TRT-CTO \\
55.118 & i & 18.46 & 0.12 & AB & ALT100B \\
56.140 & i & 18.58 & 0.10 & AB & ALT100B \\
57.126 & i & 18.42 & 0.08 & AB & ALT100B \\
61.125 & i & 18.64 & 0.10 & AB & ALT100B \\
64.579 & i & 18.51 & 0.13 & AB & TRT-CTO \\
66.145 & i & 18.85 & 0.15 & AB & ALT100B \\
67.142 & i & 18.80 & 0.20 & AB & ALT100B \\
23.366 & z & 17.21 & 0.01 & AB & NOT \\
25.481 & z & 17.21 & 0.01 & AB & NOT \\
28.098 & z & 17.27 & 0.05 & AB & ALT100A \\
29.110 & z & 17.35 & 0.04 & AB & ALT100A \\
32.136 & z & 17.52 & 0.06 & AB & ALT50A \\
35.429 & z & 17.67 & 0.01 & AB & NOT \\
46.131 & z & 17.88 & 0.14 & AB & ALT50A \\
55.156 & z & 18.50 & 0.20 & AB & ALT100C \\
56.161 & z & 18.59 & 0.12 & AB & ALT100C \\
61.138 & z & 18.11 & 0.14 & AB & ALT100C \\
0.026 & Mephisto-u & $>$20.3 & - & AB & Mephisto \\
0.059 & Mephisto-u & $>$21.1 & - & AB & Mephisto \\
0.072 & Mephisto-u & $>$21.1 & - & AB & Mephisto \\
0.086 & Mephisto-u & 21.19 & 0.14 & AB & Mephisto \\
0.093 & Mephisto-u & 21.24 & 0.13 & AB & Mephisto \\
6.164 & Mephisto-u & 18.36 & 0.02 & AB & Mephisto \\
13.124 & Mephisto-u & 18.73 & 0.05 & AB & Mephisto \\
16.173 & Mephisto-u & 19.26 & 0.07 & AB & Mephisto \\
21.084 & Mephisto-u & 20.00 & 0.07 & AB & Mephisto \\
34.162 & Mephisto-u & $>$20.0 & - & AB & Mephisto \\
35.146 & Mephisto-u & $>$20.0 & - & AB & Mephisto \\
36.041 & Mephisto-u & $>$20.3 & - & AB & Mephisto \\
0.072 & Mephisto-v & 19.25 & 0.12 & AB & Mephisto \\
0.076 & Mephisto-v & 19.19 & 0.11 & AB & Mephisto \\
0.097 & Mephisto-v & 19.23 & 0.05 & AB & Mephisto \\
0.105 & Mephisto-v & 19.17 & 0.04 & AB & Mephisto \\
0.109 & Mephisto-v & 19.12 & 0.04 & AB & Mephisto \\
6.172 & Mephisto-v & 17.77 & 0.04 & AB & Mephisto \\
13.130 & Mephisto-v & 18.36 & 0.05 & AB & Mephisto \\
16.175 & Mephisto-v & 18.79 & 0.07 & AB & Mephisto \\
21.091 & Mephisto-v & 19.63 & 0.12 & AB & Mephisto \\
27.088 & Mephisto-v & 19.75 & 0.08 & AB & Mephisto \\
35.156 & Mephisto-v & $>$19.0 & - & AB & Mephisto \\
36.050 & Mephisto-v & $>$19.6 & - & AB & Mephisto \\
0.026 & Mephisto-g & $>$20.5 & - & AB & Mephisto \\
0.057 & Mephisto-g & 21.19 & 0.28 & AB & Mephisto \\
0.062 & Mephisto-g & 20.77 & 0.17 & AB & Mephisto \\
0.070 & Mephisto-g & 20.94 & 0.19 & AB & Mephisto \\
0.075 & Mephisto-g & 20.70 & 0.15 & AB & Mephisto \\
0.083 & Mephisto-g & 20.74 & 0.22 & AB & Mephisto \\
0.086 & Mephisto-g & 20.77 & 0.08 & AB & Mephisto \\
0.093 & Mephisto-g & 20.75 & 0.08 & AB & Mephisto \\
6.136 & Mephisto-g & 17.28 & 0.01 & AB & Mephisto \\
13.124 & Mephisto-g & 16.96 & 0.01 & AB & Mephisto \\
16.173 & Mephisto-g & 17.07 & 0.01 & AB & Mephisto \\
21.084 & Mephisto-g & 17.41 & 0.01 & AB & Mephisto \\
27.083 & Mephisto-g & 17.86 & 0.01 & AB & Mephisto \\
34.162 & Mephisto-g & 18.33 & 0.06 & AB & Mephisto \\
35.146 & Mephisto-g & 18.45 & 0.06 & AB & Mephisto \\
36.041 & Mephisto-g & 18.40 & 0.05 & AB & Mephisto \\
46.124 & Mephisto-g & 18.75 & 0.02 & AB & Mephisto \\
0.072 & Mephisto-r & 20.20 & 0.10 & AB & Mephisto \\
0.076 & Mephisto-r & 20.20 & 0.10 & AB & Mephisto \\
0.097 & Mephisto-r & 20.18 & 0.06 & AB & Mephisto \\
0.105 & Mephisto-r & 20.25 & 0.06 & AB & Mephisto \\
0.109 & Mephisto-r & 20.19 & 0.05 & AB & Mephisto \\
6.172 & Mephisto-r & 17.32 & 0.01 & AB & Mephisto \\
13.130 & Mephisto-r & 16.99 & 0.01 & AB & Mephisto \\
16.175 & Mephisto-r & 17.00 & 0.01 & AB & Mephisto \\
21.091 & Mephisto-r & 17.14 & 0.01 & AB & Mephisto \\
27.088 & Mephisto-r & 17.36 & 0.01 & AB & Mephisto \\
35.156 & Mephisto-r & 17.70 & 0.07 & AB & Mephisto \\
46.128 & Mephisto-r & 18.29 & 0.01 & AB & Mephisto \\
0.026 & Mephisto-i & $>$19.7 & - & AB & Mephisto \\
0.059 & Mephisto-i & 20.95 & 0.30 & AB & Mephisto \\
0.076 & Mephisto-i & 20.67 & 0.21 & AB & Mephisto \\
0.093 & Mephisto-i & 20.71 & 0.16 & AB & Mephisto \\
6.120 & Mephisto-i & 17.68 & 0.01 & AB & Mephisto \\
13.124 & Mephisto-i & 17.32 & 0.01 & AB & Mephisto \\
16.173 & Mephisto-i & 17.26 & 0.01 & AB & Mephisto \\
21.086 & Mephisto-i & 17.27 & 0.01 & AB & Mephisto \\
27.083 & Mephisto-i & 17.53 & 0.01 & AB & Mephisto \\
34.162 & Mephisto-i & 17.86 & 0.04 & AB & Mephisto \\
35.146 & Mephisto-i & 17.90 & 0.05 & AB & Mephisto \\
36.045 & Mephisto-i & 17.82 & 0.03 & AB & Mephisto \\
46.124 & Mephisto-i & 18.30 & 0.02 & AB & Mephisto \\
0.074 & Mephisto-z & $>$19.7 & - & AB & Mephisto \\
0.103 & Mephisto-z & $>$20.5 & - & AB & Mephisto \\
6.172 & Mephisto-z & 17.89 & 0.02 & AB & Mephisto \\
13.130 & Mephisto-z & 17.21 & 0.01 & AB & Mephisto \\
16.175 & Mephisto-z & 17.08 & 0.01 & AB & Mephisto \\
21.091 & Mephisto-z & 16.98 & 0.01 & AB & Mephisto \\
27.088 & Mephisto-z & 17.23 & 0.03 & AB & Mephisto \\
46.128 & Mephisto-z & 17.83 & 0.04 & AB & Mephisto \\
2.894 & B & 18.47 & 0.05 & Vega & KAIT \\
3.834 & B & 18.02 & 0.05 & Vega & KAIT \\
4.288 & B & 17.86 & 0.01 & Vega & ZTSh \\
4.848 & B & 17.75 & 0.04 & Vega & KAIT \\
5.826 & B & 17.69 & 0.12 & Vega & KAIT \\
6.864 & B & 17.65 & 0.04 & Vega & KAIT \\
13.711 & B & 17.74 & 0.04 & Vega & KAIT \\
14.741 & B & 17.82 & 0.04 & Vega & KAIT \\
15.828 & B & 17.98 & 0.04 & Vega & KAIT \\
16.685 & B & 17.97 & 0.05 & Vega & KAIT \\
17.809 & B & 18.24 & 0.06 & Vega & KAIT \\
18.732 & B & 18.23 & 0.07 & Vega & KAIT \\
19.315 & B & 18.38 & 0.01 & Vega & Zeiss-1000 \\
24.227 & B & 18.90 & 0.01 & Vega & Zeiss-1000 \\
2.895 & V & 18.34 & 0.03 & Vega & KAIT \\
3.834 & V & 17.84 & 0.03 & Vega & KAIT \\
4.303 & V & 17.54 & 0.01 & Vega & ZTSh \\
4.844 & V & 17.58 & 0.03 & Vega & KAIT \\
6.864 & V & 17.32 & 0.03 & Vega & KAIT \\
13.712 & V & 16.96 & 0.02 & Vega & KAIT \\
14.742 & V & 16.95 & 0.01 & Vega & KAIT \\
15.829 & V & 16.99 & 0.01 & Vega & KAIT \\
16.686 & V & 17.05 & 0.02 & Vega & KAIT \\
17.810 & V & 17.14 & 0.02 & Vega & KAIT \\
18.732 & V & 17.10 & 0.02 & Vega & KAIT \\
19.317 & V & 17.20 & 0.01 & Vega & Zeiss-1000 \\
24.231 & V & 17.61 & 0.01 & Vega & Zeiss-1000 \\
0.028 & R & 20.90 & 0.40 & Vega & TRT-SBO \\
0.114 & R & 20.45 & 0.07 & Vega & Mondy \\
0.128 & R & 20.25 & 0.06 & Vega & Mondy \\
0.142 & R & 20.57 & 0.08 & Vega & Mondy \\
1.178 & R & 19.24 & 0.01 & Vega & Mondy \\
2.327 & R & 18.48 & 0.05 & Vega & Astrosib-500 \\
2.891 & R & 18.09 & 0.02 & Vega & KAIT \\
3.316 & R & 17.97 & 0.01 & Vega & Astrosib-500 \\
3.831 & R & 17.84 & 0.02 & Vega & KAIT \\
4.271 & R & 17.61 & 0.01 & Vega & ZTSh \\
4.845 & R & 17.56 & 0.02 & Vega & KAIT \\
6.311 & R & 17.23 & 0.01 & Vega & MTM-500 \\
6.861 & R & 17.19 & 0.02 & Vega & KAIT \\
7.366 & R & 17.14 & 0.02 & Vega & Zeiss-1000 \\
11.239 & R & 16.93 & 0.01 & Vega & MTM-500 \\
13.713 & R & 16.94 & 0.01 & Vega & KAIT \\
14.738 & R & 16.88 & 0.01 & Vega & KAIT \\
15.830 & R & 16.91 & 0.01 & Vega & KAIT \\
16.374 & R & 17.02 & 0.04 & Vega & Astrosib-500 \\
16.682 & R & 16.87 & 0.01 & Vega & KAIT \\
17.811 & R & 16.93 & 0.02 & Vega & KAIT \\
18.729 & R & 17.01 & 0.02 & Vega & KAIT \\
19.306 & R & 16.97 & 0.01 & Vega & Zeiss-1000 \\
24.214 & R & 17.15 & 0.01 & Vega & Zeiss-1000 \\
25.353 & R & 17.20 & 0.01 & Vega & Zeiss-1000 \\
2.892 & I & 18.34 & 0.06 & Vega & KAIT \\
3.832 & I & 17.75 & 0.04 & Vega & KAIT \\
4.317 & I & 17.76 & 0.01 & Vega & ZTSh \\
4.846 & I & 17.62 & 0.05 & Vega & KAIT \\
5.851 & I & 17.44 & 0.15 & Vega & KAIT \\
11.288 & I & 17.02 & 0.02 & Vega & MTM-500 \\
13.674 & I & 16.96 & 0.05 & Vega & KAIT \\
14.705 & I & 16.94 & 0.06 & Vega & KAIT \\
15.695 & I & 16.88 & 0.06 & Vega & KAIT \\
16.675 & I & 16.98 & 0.07 & Vega & KAIT \\
17.742 & I & 16.91 & 0.06 & Vega & KAIT \\
18.715 & I & 16.88 & 0.06 & Vega & KAIT \\
19.312 & I & 16.85 & 0.01 & Vega & Zeiss-1000 \\
24.223 & I & 16.98 & 0.01 & Vega & Zeiss-1000 \\
4.146 & J & 17.55 & 0.15 & Vega & ALT100C \\
7.087 & J & 17.64 & 0.16 & Vega & ALT100C \\
9.091 & J & 16.98 & 0.11 & Vega & ALT100C \\
10.191 & J & 17.25 & 0.14 & Vega & ALT100C \\
13.093 & J & 17.08 & 0.07 & Vega & ALT100C \\
14.159 & J & 17.13 & 0.11 & Vega & ALT100C \\
38.369 & J & 17.03 & 0.01 & Vega & NOT \\
0.713 & clear & 20.04 & 0.07 & Vega & KAIT \\
0.862 & clear & 19.57 & 0.06 & Vega & KAIT \\
2.893 & clear & 18.35 & 0.01 & Vega & KAIT \\
3.833 & clear & 18.19 & 0.01 & Vega & KAIT \\
3.861 & clear & 17.78 & 0.01 & Vega & KAIT \\
4.847 & clear & 17.71 & 0.01 & Vega & KAIT \\
4.875 & clear & 17.30 & 0.02 & Vega & KAIT \\
5.852 & clear & 17.37 & 0.03 & Vega & KAIT \\
6.863 & clear & 17.20 & 0.01 & Vega & KAIT \\
13.714 & clear & 16.89 & 0.01 & Vega & KAIT \\
14.740 & clear & 17.06 & 0.01 & Vega & KAIT \\
15.818 & clear & 17.08 & 0.01 & Vega & KAIT \\
15.827 & clear & 16.94 & 0.01 & Vega & KAIT \\
16.684 & clear & 16.98 & 0.01 & Vega & KAIT \\
17.803 & clear & 16.90 & 0.01 & Vega & KAIT \\
17.813 & clear & 17.02 & 0.01 & Vega & KAIT \\
18.731 & clear & 17.09 & 0.01 & Vega & KAIT \\
18.813 & clear & 17.12 & 0.01 & Vega & KAIT \\
\end{longtable}

\begin{table*}
\centering
\small
\caption{Log of spectroscopy of SN\,2026gzf.}\label{tab:opt_spec}
\begin{tabular}{ccccccc}
\toprule
UTC Date & $\Delta t$ & Range & Airmass  &  Total Exp. & Telescope  & Instrument \\
 & (day) &({\AA})&    & (s) & \\
\hline
 Mar. 24 & 2.7 & 3800--10000 & 1.4  &  6000 & Shane & Kast   \\
 Mar. 26 & 4.7 & 3800--10000 & 1.0  &  3600 & Shane & Kast   \\
 Mar. 26 & 5.3 & 3800--9200 & 1.3  &  1800 & NOT & ALFOSC   \\
 Mar. 27 & 6.0 & 3800--10000 & 1.3  &  3600 & Shane & Kast   \\
 Mar. 28 & 7.4 & 3800--9200 & 1.2  &  2400 & NOT & ALFOSC   \\
 Mar. 31 & 10.5& 3800--9200 & 1.3  &  2400 & NOT & ALFOSC   \\
 Apr. 02 & 12.4 & 3800--9200 & 1.1  &  2400 & NOT & ALFOSC   \\
 Apr. 06 & 16.5 & 3410--9330 & 1.2  &  480$\times$8 & VLT & FORS2   \\ 
 Apr. 07 & 16.9 & 4000--9000 & 1.6  &  3600 & Xinglong 2.16\,m & BFOSC   \\
 Apr. 07 & 17.4  & 3800--9200 & 1.1  &  2400 & NOT & ALFOSC   \\
 Apr. 11 & 20.9 & 4000--9000 & 1.6  &  3600 & Xinglong 2.16\,m & BFOSC   \\
 Apr. 11 & 21.4 & 3800--9200 & 1.1  &  2400 & NOT & ALFOSC   \\
 Apr. 12 & 22.5 & 3800--9200 & 1.9  &  2400 & NOT & ALFOSC   \\
 Apr. 15 & 25.4 & 3800--9200 & 1.2  &  2400 & NOT & ALFOSC   \\
 Apr. 21 & 31.4 & 3800--9200 & 1.1  &  2400 & NOT & ALFOSC   \\
 Apr. 27 & 37.4 & 3800--9200 & 1.2  &  2400 & NOT & ALFOSC   \\
 May. 03 & 43.4 & 3800--9200 & 1.3  &  2400 & NOT & ALFOSC   \\
 May. 09 & 49.4 & 3800--9200 & 1.3  &  2400 & NOT & ALFOSC   \\
 May. 16 & 56.4 & 3800--9200 & 1.3  &  2400 & NOT & ALFOSC   \\
\bottomrule
\end{tabular}
\end{table*}

\begin{table*}[t]
\centering
\small
\caption{Radio observations of EP260321a/SN~2026gzf. Epochs are computed relative to $T_0=\texttt{2026-03-21T12:16:08}$ UTC. For observations with reported UTC ranges, the listed epoch is the midpoint. Upper limits are $3\sigma$ unless otherwise stated.}
\label{tab:radio_obs}
\begin{tabularx}{\textwidth}{@{}l p{3.1cm} c c X p{2.1cm}@{}}
\toprule
Facility & Epoch / UTC range & $\Delta t$ (d) & $\nu$ (GHz) & Result & Obs. code / ref. \\
\midrule
uGMRT & 2026-03-26 15:00--16:00 & 5.135 & 0.7 & rms $=120~\mu$Jy; $<360~\mu$Jy & 49\_076/49\_114; GCN~44227\cite{GCN44227}, this work \\

uGMRT & 2026-03-26 18:00--19:00 & 5.260 & 1.1 & rms $=90~\mu$Jy; $<270~\mu$Jy & 49\_076/49\_114; GCN~44227\cite{GCN44227}, this work \\

VLA & 2026-03-27 05:12--07:42 & 5.758 & 6, 10, 15, 22 & $<30~\mu$Jy beam$^{-1}$ at each band & GCN~44229\cite{GCN44229} \\

ATCA & 2026-03-27 09:00--15:00 & 5.889 & 6.0, 10.0 & rms $\simeq40~\mu$Jy; $<120~\mu$Jy & CX595; this work \\

ATCA & 2026-03-30 10:30--21:30 & 9.155 & 17, 19, 21, 23 & $<210,150,75,60~\mu$Jy at 17, 19, 21, 23~GHz & GCN~44403\cite{GCN44403} \\

ATCA & 2026-04-02 10:30--14:00 & 11.999 & 20.0 & rms $\simeq32~\mu$Jy; $<96~\mu$Jy & C3735; this work \\

e-MERLIN & 2026-04-01 04:05 -- 04-02  02:04 \& 2026-04-06 19:25 -- 04-07  07:30  & 13.489 & 5.0 & rms $=22~\mu$Jy; $5\sigma<110~\mu$Jy & CY21232; this work \\

VLA & 2026-04-04 04:03--05:03 & 13.658 & 6.0 & low-significance feature; rms $=5~\mu$Jy; $4\sigma\lesssim20~\mu$Jy & GCN~44239\cite{GCN44239} \\

MeerKAT & 2026-04-10 18:17--19:04 & 20.267 & 1.28 & rms $\simeq23~\mu$Jy; $<69~\mu$Jy & DDT-20260331; this work \\

VLA & 2026-04-13 01:20--02:20 & 22.565 & 6.0 & no confirmed counterpart; rms $=5~\mu$Jy; $4\sigma\lesssim20~\mu$Jy & GCN~44357\cite{GCN44357} \\

MeerKAT & 2026-04-13 16:02--16:49 & 23.173 & 3.0 & rms $\simeq11~\mu$Jy; $<33~\mu$Jy & DDT-20260331; this work \\

ATCA & 2026-04-14 09:00--16:00 & 24.010 & 6.0, 10.0 & rms $\simeq30~\mu$Jy; $<90~\mu$Jy & C3735; this work \\

e-MERLIN & 2026-04-20 16:10--2026-04-21 06:51 & 30.989 & 5.0 & rms $=12~\mu$Jy; $5\sigma<60~\mu$Jy & CY21232; this work \\

MeerKAT & 2026-04-21 14:08--15:26 & 31.105 & 3.0 & rms $\simeq8.8~\mu$Jy; $<26~\mu$Jy & DDT-20260331; this work \\
\bottomrule
\end{tabularx}
\end{table*}






\end{document}